\documentclass[11pt]{article}

\usepackage[hmargin=2cm,vmargin=3cm,bindingoffset=0.5cm]{geometry}

\usepackage{palatino}
\usepackage{xcolor}
\usepackage{graphicx}
\usepackage{amsfonts}
\usepackage{amssymb}
\usepackage{url}
\usepackage{amsmath}
\usepackage{amscd}
\usepackage{latexsym}
\usepackage{dsfont}
\usepackage{graphicx}
\usepackage{latexsym}
\usepackage{amscd}
\usepackage{tikz}
\usepackage{bm}
\usepackage{pdfpages}
\usepackage{cancel}
\usepackage{empheq}
\usepackage{hyperref}
\usepackage{relsize}
\usepackage{lmodern}
\usepackage{bigints}
\usepackage{bm}
\usepackage{tikz}
\usepackage[framemethod=default]{mdframed}
\usepackage{textcomp}

\DeclareMathOperator\arctanh{arctanh}

\DeclareMathOperator{\re}{Re} 

\begin{document}


\newtheorem{theorem}{Theorem}
\newtheorem{teo}{Theorem}
\newtheorem{lem}{Lemma}
\newtheorem{prp}{Proposition}
\newtheorem{proposition}{Proposition}
\newtheorem{ass}{Assertion}
\newtheorem{assum}{Assumption}
\newtheorem{stat}{Statement}
\newtheorem{cor}{Corollary}
\newtheorem{hyp}{Hypothesis}
\newtheorem{con}{Conjecture}
\newtheorem{definition}{Definition}
\newtheorem{problem}{Problem}
\newtheorem{notat}{Notation}
\newtheorem{quest}{Question}
\newtheorem{rem}{Remark}
\newtheorem{exa}{Example}
\newtheorem{cas}{Case}
\newtheorem{claim}{Claim}
\newtheorem{com}{Comment}
\newtheorem{proof}{Proof. \hspace{1mm}}

\newcommand\overN{\stackrel{\mathclap{\small\mbox{N}}}{\cdots}}
\newcommand\overtwoN{\stackrel{\mathclap{\small\mbox{2N}}}{\cdots}}
\newcommand{\R}{\ensuremath{\mathbb{R}}}
\newcommand{\proofnsymprop}{\textit{Proofs of Theorems 6 and 7.}}
\newcommand{\Z}{\mathbb{Z}}
\newcommand{\N}{\mathbb{N}}
\newcommand{\Q}{\mathbb{Q}}
\newcommand{\K}{\mathbb{K}}
\newcommand{\T}{\mathbb{T}}
\newcommand{\n}{}
\newcommand{\m}{\medskip}
\newcommand{\D}{\displaystyle}
\newcommand{\End}{\operatorname{right }}

\newcommand\myeq{\stackrel{\mathclap{\normalfont\mbox{\tiny{N}}}}{\cdots}}

\newcommand{\erm}{{\rm e}}
\newcommand{\bdA} {{\bf  A}}
\newcommand{\bdU} {{\bf  U}}
\newcommand{\bdu} {{\bf  u}}
\newcommand{\bdv} {{\bf  v}}
\newcommand{\bdo} {{\bf \om}}
\newcommand{\bdn} {{\bf  n}}
\newcommand{\bdx} {{\bf  x}}
\newcommand{\bde} {{\bf  e}}
\newcommand{\bdR} {{\bf  R}}
\newcommand{\bdO} {{\bf \Om}}
\newcommand{\cinfty} {\hbox{\rm C}^\infty}
\newcommand{\diver} {{\rm div} \, }
\newcommand{\ov} {\overline}
\newcommand{\sech} {\hbox{\rm sech}\,}
\newcommand{\Ro}{{\rm R}}
\newcommand{\Real}{{\rm Re}}
\newcommand{\Imag}{{\rm Im}}
\newcommand{\Si}{{\rm S}}
\newcommand{\Rc} {{\cal R}}
\newcommand{\Oc} {{\cal O}}
\newcommand{\oc} {\hbox{\scriptsize $\cal O$}}
\newcommand{\calA}{{\cal A}}
\newcommand{\calL}{{\cal L}}
\newcommand{\calG}{{\cal G}}
\newcommand{\calF}{{\cal F}}
\newcommand{\calR}{{\cal R}}
\newcommand{\ap} {\alpha}
\newcommand{\bt} {\beta}
\newcommand{\dt} {\delta}
\newcommand{\Dt} {\Delta}
\newcommand{\ep} {\epsilon}
\newcommand{\Gm} {\Gamma}
\newcommand{\gm} {\gamma}
\newcommand{\Lb} {\Lambda}
\newcommand{\lb} {\lambda}
\newcommand{\om} {\omega}
\newcommand{\Om} {\Omega}
\newcommand{\sg} {\sigma}
\newcommand{\Sg} {\Sigma}
\newcommand{\vp} {\varphi}
\newcommand{\Th} {\Theta}
\newcommand{\defi}{\stackrel{\rm def}{=}}
\newcommand{\0} {\underline 0}
\newcommand{\ptl} {\partial}
\def\P{{\mathbb P}}
\def\S{{\mathbb S}}
\def\Z{{\mathbb Z}}
\def\Xm{{\mathbb X}}
\def\R{{\mathbb R}}
\def\Tm{{\mathbb T}}
\def\N{{\mathbb N}}
\def\C{{\mathbb C}}
\def\H{{\mathbb H}}
\def\D{{\mathbb D}}
\newcommand{\Id}{{\rm 1}\!{\rm l}}
\newcommand{\bproof}{\noindent {\it Proof. }}
\newcommand{\eproof}{\hfill$\Box$}
\newcommand{\nd} {\noindent}

\newcommand{\htheta}{\hat{\theta}}
\newcommand{\diagg}{\mathrm{diag}}
\newcommand{\hJ}{\hat{J}}

\newcommand{\proofn}{Proof.}

\newcommand{\qed}{\hfill $\blacksquare$}


\newtheorem{notation}{Notation}
\renewcommand{\thenotation}{} 
\newtheorem{terminology}{Terminology}
\renewcommand{\theterminology}{} 

\newcommand\dee{\partial}

\def\I{\mathrm{i}}
\def\Log{{\rm Log\,}}
\def\D{{\mathbb D}}
\def\R{{\mathbb R}}
\def\C{{\mathbb C}}
\def\P{{\mathbb P}}
\def\Z{{\mathbb Z}}
\def\N{{\mathbb N}}
\def\DD{{\mathcal{D}}}
\def\OO{{mathcal{O}}}
\def\Wilde{}
\def\spect#1{\rho(#1)}


\begin{center}
{\Large{\bf
On the interplay between  vortices and harmonic flows: \\ Hodge decomposition of Euler's equations in 2d }}
\\ \, \\
Clodoaldo Grotta-Ragazzo\footnote{\noindent
Partially supported by FAPESP grant 2016/25053-8; ORCID: 0000-0002-4277-4173.} 
 (ragazzo@usp.br) 
\\
Instituto de Matematica e Estatistica, Universidade de S\~ao Paulo\\
R. do Mat\~ao  1010,
S\~ao Paulo - SP, 05508-090, Brazil
\\ \, \\
 Bj\"orn Gustafsson 
(gbjorn@kth.se)
\\
Department of Mathematics,   KTH Royal Institute of Technology\\ SE-100 44 Stockholm, Sweden  \\ \, \\
 Jair Koiller (jairkoiller@gmail.com) \\
Departamento de Matem\'atica, Universidade Federal de Juiz de Fora\\Via Local, 880 - S\~ao Pedro,
Juiz de Fora - MG, 36036-900, Brazil 
\end{center}

\begin{abstract} 
    Let  $\Sigma$  be  a compact manifold without boundary  whose first homology is nontrivial. Hodge decomposition of  the  incompressible Euler's equation in terms of 1-forms  yields a coupled PDE-ODE system. The $L^2$-orthogonal components are a `pure' vorticity flow  and  a potential flow (harmonic, with the dimension of the homology).     In this paper we focus on  $N$  point vortices on a compact Riemann surface without boundary of genus $g$,   with a  metric  chosen in the conformal class.  
 The phase space has finite dimension  $2N+ 2g$.   We compute a surface of section for  the motion of a single vortex ($N=1$) on a torus ($g=1$) with a non-flat metric, that shows typical features of non-integrable 2-dof Hamiltonians. In contradistinction, for   flat tori the harmonic part is constant. Next, we   turn to hyperbolic surfaces ($ g \geq 2$),  having constant curvature -1,    with discrete symmetries. Fixed points of involutions  yield vortex crystals in the Poincar\'e disk.    Finally we consider  multiply connected planar domains.  The image method due to
 Green and Thomson is viewed  in the Schottky  double. The Kirchhoff-Routh hamiltonian given in  C.C. Lin's celebrated theorem is recovered by  Marsden-Weinstein  reduction from $2N+2g$ to $2N$.  
 The relation between  the electrostatic Green function and the hydrodynamical Green function is clarified.
 A number of questions are suggested.
 \end{abstract}

\begin{center}  keywords:    vortex motion, Riemann surfaces, Hodge decomposition \\
76B47, 76M60, 34C23, 37E35  \\ \, \\ \, \\
 $$
 $$
 {\it Dedicated to the memory of Alexey Borisov}
 \end{center}

\newpage

{\footnotesize

\tableofcontents

}

\newpage

$$
$$

\section{INTRODUCTION} 
\vspace{1mm}

\subsection*{``Let there be a vault between the waters to separate water from water"\\
(Genesis 1:6,   New International Version).}

\bigskip \bigskip

This paper\footnote{Extended version of   talks given by one of the authors (JK)  at two conferences in memory
of Alexey Borisov:\\ GDIS 2022, Zlatibor, Serbia, June 5-11 2022 and November 22-December 3, 2021, Steklov Mathematical Institute.}  continues   recent work    by two of the co-authors,   \cite{Gustafsson, Ragazzo, Ragazzo0}  about  $N$-vortex systems on a closed surface $\Sigma$ of genus $g \geq1$. They called attention   (also noticed in \cite{Bogatskii1, Bogatskii}) that the phase space must be $2N+2g$ dimensional.       In any  dimension,  if   the manifold has nontrivial first homology,   two incompressible vector fields with the same vorticity 2-form will differ by a harmonic field \cite{MarsdenWeinstein}. 
     
Thus it seems natural to apply Hodge theory  \cite{Hodge}, that  makes possible to separate  \textit{orthogonally} what could be called a  `pure'  vorticity  term  from a uniquely defined harmonic complement.  It turns out, for better or for  worse, that the  Euler equations are coupled. 

For compact boundaryless surfaces of genus $g$ the harmonic forms are $2g$-dimensional.  Here we will show \textit{how} the  harmonic component   \textit{interacts } with the  $N$ point vortices. 
  The dynamics lifts  to the  universal cover, containing infinitely many   replicas of a fundamental domain. In  \cite{KoillerBoatto}  there was  an attempt to extend to closed surfaces C. C. Lin's approach  \cite{Lin-a}  for planar domains,  via the  Green function of the Laplacian.    However   the equations  were  \textit{incomplete}, missing   the harmonic contribution\footnote{How much this impacts the evolution will be     discussed in the final section.}. 
   
A (quite reasonable) misunderstanding may arise in the case of multiply connected \textit{planar domains}, because    in the seminal work by C.C. Lin the Hodge decomposition \textit{was not used}.    One may ask, how come the phase space in his approach has only dimension $2N$ and not $2N+2g$?  As one of the referees sensibly remarked,
in  Lin's  setting each point vortex ``carries an harmonic field external to it"\footnote{So to speak, it is like the snail that carries its shell while   it moves.}. Indeed, if a domain has $g+1$ boundaries,  the  \textit{hydrodynamical Green function} 
\cite{Gustafsson-1979-b, Flucher-Gustafsson-1997-a, Flucher-1999-a}
has  $g$ circulations built-in.  Moreover, there is no distributed  vorticity in the background.  

To fit in the framework of this paper, a closed boundaryless surface is constructed by taking the Schottky double of the planar domain, that we call a ``pancake". If $N$ vortices are moving in the front face,  there are $N$ other  symmetric counter  vortices
on the backside, implementing   the image method of Green and Thomson.  We will show that the Hodge decomposition in the double    leads quite elegantly to  Lin's result  via Marsden-Weinstein reduction \cite{Marsden}.   
   
 In fact, with a natural choice of homology basis, $g$ among the
 circulations vanish  because of the involution  of the double.  Furthermore, there are additional $g$  independent constants of motion, circulations around the inner boundary curves (the external circulation is  a dependent quantity).

 There is   a finer  decomposition (Friedrichs-Morrey)  for manifolds with boundary 
  \cite{Friedrichs-1955-a, Schwarz-1995-a,  Morrey-a}. Such decompositions seem to be useful in the study of   circulatory/heart  flows \cite{Raza-a, Saqr-a,  Razaf-a}. Implementation for 2d found, eg. in \cite{Poelke-a}, and for 3d  in \cite{Desbrun-a}.
As we  were about to send  the revision, Boris Khesin called our attention to  the  computational geometry paper \cite{SanDiego},  congenial to ours\footnote{``Fluid cohomology" (nice title!). It includes codes for  the numerical implementations both in 2d and 3d
and a beautiful (and impressive)  video demonstration, \url{https://yhesper.github.io/fc23/fc23.html} . }.

\newpage

   \subsection*{Results and organization of the paper}  
   \vspace{1mm}

 Section \ref{general} reviews the \textit{Hodge  decomposition} of 1-forms on  manifolds without boundary.
 An incompressible Euler flow splits in  a unique way into $L^2$-orthogonal components,  `pure vorticity' (lacking a better name) and harmonic (present when the first homology is nontrivial).
  Section   \ref{surfaces} reviews the main results of \cite{Gustafsson, Ragazzo},  that give  the complete  equations for point vortices coupled with the harmonic flows on \textit{closed surfaces without boundary with genus $g \geq 1.$} In section \ref{flatnonflattori} 
we present  simulations of a single vortex on  a non-flat torus.  Flat tori are exceptional: the harmonic part is constant.  
 Section \ref{crystals} contains  a study  in the Poincar\'e disk with the hyperbolic metric. We show that vortex crystals are abundant, specially those related to hyperbolic tesselations. More precisely, to  fixed points of discrete symmetries on a  Fuchsian group. Sections \ref{scho}-\ref{hydrogreen0} discuss the  Schottky double of a planar domain.
We re-interpret C.C.Lin's theorem  in terms of Marsden-Weinstein reduction.
Section  \ref{questions} lists a number of questions, hoping to  attract interest to `vorticists', as well as geometers and analysts.  For instance,   quantifiying by how much  the incomplete equations    (discarding the  harmonic component)  are off  the mark.
The Appendices aim to make the paper self-contained.  

$$ $$

  \section {HODGE DECOMPOSITION OF EULER's EQUATION} \label{general} 

\vspace{1mm}  

It is well known since   V. Arnold famous 1966 paper   \cite{Arnold} (for an introductory   review see \cite{Modin1}) that  Euler's equation
makes sense for Riemannian manifolds of  any dimension,  as  the right reduction of the geodesic  equations on the infinite dimensional group of  volume preserving diffeomorphisms $\phi_t \in {\rm SDiff}(\Sigma),  $   with  the metric
  \begin{equation}
 T(v) = 1/2 \, \int_{\Sigma} \,\langle v,v \rangle \, \mu,\, {\rm where}\,\, v(x,t) = \partial/\partial t_{|t=0} \, \phi(t,x) \,\,,\,\,\, \mu = {\rm volume}\,{\rm form}
  \end{equation}
  to the Lie algebra $ v \in {\rm sDiff}(\Sigma) $ of divergence free vector fields ($ v || \partial \Sigma $ at boundaries).

 It seems  more convenient to
work  in the dual $  {\rm sDiff(\Sigma)}^*.$ Recall the musical isomorphism  between vector fields ($v$)  and 1-forms ($\nu$) on a Riemannian
manifold $(\Sigma^n, \langle , \rangle)$ 
\begin{equation}   \nu = v^\flat   \,\,\,\, \leftrightarrow\,\,\,\,  v  = \nu^{\sharp} ,  \,\,\,\, v^\flat(X) = \,\langle v, X \rangle .
\end{equation}

\begin{prp}  \label{ArnoldKhesin}
 Euler's equations are given by 
\begin{equation} \label{Euler}
 \frac{\partial \nu}{\partial t}  + L_{\nu^\sharp}\, \nu =   db,    \,\,{\rm with}\,\, b =  \frac{1}{2} |v|^2 - p  \end{equation}
where $L$ is the Lie derivative operator  and $b$ is called the Bernoulli function.
\end{prp}
In this  nice dual formulation,   $\nu^\sharp$ is the only object that changes with the metric\footnote{We refer  to  Arnold-Khesin \cite{ArnoldKhesin},   a comprehensive treatise about  the topological-geometrical approach to Euler's equations. Our final section, with  questions,   mentions  recent work  in 3d  manifolds published by  two groups, of  Boris Khesin and of  Eva Miranda,  with their respective coauthors.}.

\newpage

\subsection{Hodge decomposition} \label{hodgesubsection}
\vspace{0.5mm}

 Instrumental for our viewpoint is the  Hodge decomposition\footnote{See  Appendix A with basic informations on Hodge theory and  and Appendix B on general versions of the  well known Biot-Savart formula in three dimensions. We should not confuse  Helmholtz-Hodge decomposition  with  Ladyzhenskaya's  for simply connected domains (on irrotational and  solenoidal components), used in Chorin's projection method \cite{Chorin}.} for the  1-forms $\nu = v^{\flat}$: 
\begin{equation} \label{nu}  \nu =   df  \oplus \delta \Psi \oplus \eta  .
\end{equation}
where $ \delta   $ is the codifferential operator.   $ \,\,   \eta $  is harmonic  ($d\eta = 0, \delta \eta = 0$).
 The  2-form $\Psi = \Psi_{\omega} $  can be   recovered (uniquely up to $ {\rm Kernel} \, \delta$)  from the  {\it vorticity} 2-form\footnote{Aristotle already had the intuition that vorticity is what drives fluid motion. 
    ``Vortices are  sinews and muscles of fluid motions'" (K\"uchemann, 1965), see \cite{Kuch, Saffman}.  In two dimensions, we will use  the greek letter $\psi = \psi_{\omega}$  to denote the stream function of $\omega$  as a $0$-form, which is the traditional usage, and denote the corresponding $2$-form
stream function by $\Psi$. Then $\Psi=\star \psi$.}
  \begin{equation} \label{omega}   \omega = d \nu .
    \end{equation}

 The \textit{divergence}  of a   vector field $v$ is defined as the $n$-form $ {\rm div} \, v =  d( \star (v^\flat))$,  where $\star$ is the Hodge operator.   Clearly, saying that $v$ is divergence free is the same as saying $v \in {\rm sDiff(\Sigma)}$. 
 \vspace{1mm}

\begin{lem} Let
$\Sigma$ be a compact Riemannian manifold.      If $v$ is  divergence free, then in the Hodge decomposition of $\nu = v^{\flat}$ there is  no component $  df $. 
\end{lem}
 \proofn   We must show
$$ \mathlarger{\int}_{\mathsmaller{\Sigma}} \langle {\rm grad} f , v \rangle  \, \mu  =   \mathlarger{\int}_{\mathsmaller{\Sigma}} df(v) \, \mu   = 0\,,\,\,\, \forall f \in C^{\infty}(\Sigma)\,\,,  \,\, \forall v \in {\rm sDiff(\Sigma)}\,\, \,\, \rm{(recall}\,\, \mu =  \rm{volume}\, {\rm form)}.  $$ 
Check first the identity $    df(v)\, \mu   =  df \wedge (\star v^\flat)  $.  Since 
$   d(f \star v^\flat) = df \wedge \star v^\flat + f \cancel{d(\star v^\flat)},$  
by Stokes theorem
$$ \mathlarger{\int}_{\mathsmaller{\Sigma}} \, df(v) \, \mu  = \mathlarger{\int}_{\mathsmaller{\Sigma}} \, d(f \star v^\flat) = 0 ,  $$
because  we are assuming  there is no boundary.
Actually the result also holds with boundaries:   $ v || \partial \Sigma$  implies that the boundary integrals vanish.
 \qed   \vspace{2mm}

 This result  was assumed obvious in Marsden and Weinstein \cite{MarsdenWeinstein}. An  idea suggested in that paper   and explored here is to decompose incompressible Euler's equations in terms of $(\omega, \eta)$\footnote{Boris Khesin (personal communication) warned us that, due to   resulting coupled equations,
 mathematicians have mostly stayed   with the space of $\nu$ mod  $df$.  Nonetheless, he  and  his coworkers  used Hodge  explicitly for 3d fluids \cite{Khesin3dHodge} and    implicitly in \cite{helicity3d}.  A recent survey  by Peskin and collaborators \cite{Peskin} recognizes that their  use of periodic  boundary conditions in numerical simulations  may introduce artifacts.
  }. 
  
\begin{rem}
There is a vast  literature in  Physics and Engineering exploring  Helmholtz-Hodge decomposition  (see e.g. \cite{Joseph}, \cite{Bhatia}) and
 more recently, also in Biomathematics \cite{Lefevre}. 
Among recent applications we found:  visualization
and computer graphics,  robotics,  
medical imaging and and bio-engineering.
Hodge decomposition has been also relevant in  fluids  (oceanography, geophysics and astrophysics). 
\end{rem}

  \subsection{A brief intermission for sign conventions} \label{signconv} 

Before giving the coupled equations for $(\dot{\omega},\, \dot{\eta})$ we  fix our conventions and notations. In this subject the choice of signs in the definitions may become  a  terrible curse. Authors adopt different conventions  - sometimes the same author!  

The sign we take for the geometer's Laplacian $\Delta$ (in any dimension) is such that it becomes a {\it positive definite}  operator.   The signs chosen in Appendix B for the codifferential operators $\delta^k:\Omega ^{k+1}(\Sigma) \to \Omega ^{k}(\Sigma)$ are  such that in 2 dimensions the Hodge Laplacian   $d \delta + \delta d$  is equal to the geometer's ($+ \Delta$  and not -$\Delta$). 

Most importantly the reader must keep in mind our {\it convention for circulations}.  Here we do  {\it  not} follow one of the authors previous papers \cite{Gustafsson-1979-b, Flucher-Gustafsson-1997-a}, where  Green function `periods'
 were considered, but those parameters were not really the circulations for the flow boundaries.

The line integral defining the circulation
\begin{equation}
p = \oint_{\gamma} \nu
\end{equation}
involves the direction that the curve $\gamma$ is being traversed. 
The canonical orientation  of a boundary curve  is such that when traveling along the curve  one sees the surface on the left, so in a planar domain it goes \textit{clockwise} on the inner boundaries.  When there is an external boundary $b_o$, it is run \textit{counter-clockwise.} This agrees with \cite{Marchioro} (equation (2.27) there), and certainly  many other sources.

On the other hand, for a point vortex we view its circulation and strength as positive if the flow goes counter-clockwise (increasing polar angle) seen from the vortex  itself. Another 
difference between a vortex point and a (very small) boundary component is that boundaries
are kept fixed, while vortices are free to move. 
``Bound'' vortices can be thought as degenerate boundary components, and one should decide what convention to use.

\begin{exa}  \label{exa1}  
The simplest example is the   unit disk $\D$  with a unit point vortex at the origin.
We take for stream function for the flow  the Green function  of the whole plane
\begin{equation}  \label{basicstreamf}
\psi (z)=G(z,0)=-\frac{1}{2\pi}\log |z|= -\frac{1}{4\pi}\log (x^2+y^2),\,\,\,\, z = x+i y,  \, |z| \leq 1
\end{equation}
whereby
$$
d\psi= -\frac{1}{2\pi} \frac{xdx+ydy}{x^2+y^2},\quad \star d\psi=-\frac{1}{2\pi}\frac{xdy-ydx}{x^2+y^2}.
$$
This indicates that  our  choice of sign for the flow should be
\begin{equation} \label{correctchoice}
\nu=-\star d\psi = -\frac{1}{2\pi}\frac{ydx-xdy}{x^2+y^2}.
\end{equation}
corresponding  to the vector field
\begin{equation}
v=\nu^\#=-\frac{1}{2\pi}\Big(\frac{y}{x^2+y^2}\frac{\partial}{\partial x}-\frac{x}{x^2+y^2}\frac{\partial}{\partial y}\Big) 
\end{equation}
\centerline{(for  points on the positive $y$-axis this goes in the negative $x$-direction as desired).}
\smallskip

The vorticity as a two-form is
$$
\omega=d\nu=-d\star d\psi=\delta_0 \,dx\wedge dy.  
$$
a positive point mass at the origin. 
Notice that
\begin{equation*}
\nu=\frac{1}{2\pi}d\arg z=\frac{1}{2\pi} d\theta,
\end{equation*}
where $z=|z|e^{i\theta}$. It is natural to define the circulation $p$ around a boundary component $b$ in such a way that it is positive
if the flow goes in the same direction as the orientation of the component. 

If we use this stream function in the unit disk  $\D$ as the flow domain, the boundary $b_o = \partial\D$ is run counterclockwise, so as our convention prescribes
\begin{equation}
p_o = 
 \oint_{\partial\D} \nu=1  \,\, \text{(circulation of } b).
\end{equation}

 If we have several inner  boundary components $b_1, \cdots, b_g $ and several vortices $z_1, \cdots, z_N$ with respective vortex strengths $\Gamma_j$, we should then have
 by Cauchy residue  theorem, and according to our sign convention,
 \begin{equation} \label{byresidue}
p_o =  - \sum_{j=1}^g p_j + \sum_{k=1}^N \Gamma_k . 
\end{equation}

If there are no vortices in the domain then $p_o + p_1 + \cdots p_g = 0$. 
This may be a cause for misunderstandings. Assume for example that the domain is the annulus
$$
A=\{z\in\C:\varepsilon<|z|<1\},
$$
so that $\psi$ is harmonic.  
Thus   $g=1$, $b_0=\{|z|=1\}$ (run counterclockwise), $b_1=\{| z|=\varepsilon \}$ (run clockwise).  Consider the stream
   function (\ref{basicstreamf}). We have $p_0=1$.   If $\varepsilon>0$ is very small  we are very close to the case of the unit disk and a point vortex $\Gamma = 1$ at the origin. 
   
   But now we have $p_0+p_1=0$.
Comparing $p_1$ with $\Gamma$ there is a sign change:  $p_1$ is counted positive in the other direction of  $\Gamma$.
\begin{equation} p_1=-\Gamma = -1 .
\end{equation}
\end{exa}

\vspace{0.3mm}

\subsection*{Glossary of main symbols  appearing in the paper}
\vspace{2mm}

\vspace{2mm}

\noindent $\Sigma$ will be   a 2d surface. $g$ will  denote its genus.  $s$ will be a point in $\Sigma$.\\
$s_j$ a point vortex with vorticity $\Gamma_j$.  $N$ will be the number of vortices, $\Gamma$ the total vorticity.\\

\noindent $\langle,\rangle$ a metric. $\mu$ the area form.  $V$ the area. $\Delta = d \delta + \delta d$  the Laplace-Beltrami operator. \\
$G$ its Green function for closed surfaces without boundary.\\
$R$  the corresponding Robin functions. \\
$\star$ the Hodge operator. $\delta = - \star d \star $ the codifferential operator (all degrees).\\

\noindent $\nu$ a 1-form.  $\eta$ a harmonic  1-form.
$v$ a  divergence free vector field.  \\ $\nu = v^{\flat}, \, v = \nu^{\sharp}$ the musical isomorphisms. 
$v_{s_o}^{vort} $  a pure vortex flow. $\psi$ a stream function. \\  
$\omega = d\nu $ a vorticity 2-form.   $\Psi_{\omega} $ corresponding 2-form (Biot-Savart).\\
$ \nu =  \delta\Psi_{\omega}  \oplus \eta  = \,-\star dG^\omega \oplus \eta\,\,$ (Hodge decomposition).\\

\noindent $\{ (a, b) \},  \, \{ (\alpha , \beta) \} $ homology and cohomology basis. \\
$P, Q, R $  matrices appearing in Riemann relations. \\
\noindent $U_{\gamma}$ the  multivalued function built using $G$ and a closed curve $ \gamma$. \\

\noindent $(A, B)$ circulations of $\eta$.  \\ $(\tilde{A}, \tilde{B})$ circulations of $\nu$.
$\Omega$ the symplectic form in  $(\Sigma)^N \times \Re^{2g}$\\
$ H = H_{\rm vort} +  H_{\rm harm} $ the split Hamiltonian using $(A, B)$ \\

\noindent $\D$ unit disk with Euclidian metric.  $D$  unit disk with the Poincar\'e metric (curvature -1).\\
${\cal F} $ a fundamental domain for the uniformization in $D$ of  a closed Riemann surface.\\

\noindent For bounded planar domains $\Sigma$: $b_o$ external curve, $b_i, 1\leq i \leq g$ inner boundaries.\\
\noindent $S$ = Schwarz function for the Schottky double $\hat{\Sigma}$. $I$ the involution on $\hat{\Sigma}$.\\
Three types of Green functions: $G_{{\rm electro}},\, G_{{\rm hydro}},\, G_{{\rm double}}.$\\
$u = (u_1, \cdots, u_g)$ harmonic measures.\\
$p = (p_1, \cdots, p_g)$ circulations.

\bigskip

\subsection{Coupled Euler equations in $(\omega, \eta)$}

One can sweep  ``under the rug"  the right-hand-side of Euler's equations  (\ref{Euler}), that contains the pressure in it:
\begin{equation} \label{EulerArnold}  \frac{\partial \nu}{\partial t}  + L_{\nu^\sharp}\, \nu =   0  \,\,\,\,{\rm (mod}\,\, {\rm exact)} \,\,,\,\, \nu = \delta \Psi + \eta .
\end{equation}
We now project in the   two  Hodge-orthogonal components,   the pure vorticity given by $\omega = d\delta \Psi$   and  the harmonic part  $\eta$. This  is (formally) easy to do\footnote{See section \ref{recoverpress}  for an additional discussion,
 suggested by one of the referees,  about  recovering  the pressure.}. 
 Let $i_v$ denote the contraction operator, 
 $\pi_{har}$ the projection over the harmonic component. Then
\begin{equation*}
\frac{\partial}{\partial t} (\delta \Psi)   + \frac{\partial \eta}{\partial t}
  + L_{ (\delta \Psi)^\sharp }\, \delta \Psi  + L_{\eta^\sharp}\, \delta \Psi  + \cancel{L_{(\delta \Psi)^\sharp}\, \eta}  + \cancel{L_{\eta^\sharp}\, \eta}   = 0  \,\, {\rm (mod}\,\, {\rm exact)} .
\end{equation*}
The last two terms vanish due to Cartan's recipe
$ L_{v} = i_{v} d + d i_{v}$. All  the exact terms can be removed. 

\begin{teo}  \label{coupledeq}
The coupled PDE-ODE for  ${(\omega, \eta),\,\omega \in \Omega^2(\Sigma),  \eta \in {\rm Harm}^1(\Sigma) } $ are given by.
\begin{equation} \label{omegaeta}
\begin{split}
& \frac{\partial \omega}{ \partial t}
  + d \, [ i_{(\delta \Psi)^\sharp}\, \omega  +    i_{\eta^\sharp}\, \omega ]  = 0\,\,,\,\, (\delta \Psi = \delta \Psi_{\omega})    \\
&   \frac{\partial \eta}{ \partial t}  + \pi_{har} [ i_{(\delta \Psi)^\sharp}\,  \omega    +  i_{\eta^\sharp}\,  \omega  ] = 0 .
\end{split}
\end{equation}
Furthermore,  Hodge theory implies that the total energy   splits:
 $$ H(\nu) = H(\omega, \eta) =  H_{vort}(\omega) + H_{harm}(\eta)\,\,\, \,, \,\, H_{vort}(\omega) := H(\delta \Psi) .
 $$
 Here $H_{harm}$  is a quadratic  functional giving the energy of a harmonic 1-form.
 \end{teo}

 \begin{cor} \label{cor1} If  initially there is no vorticity  $(\omega_{(t=0)} \equiv 0)$,  then
no  vorticity will appear in future time,  anywhere. Moreover, the harmonic component is steady. \\
\end{cor}

\begin{rem} Arnold  provided the Poisson structure for Euler's equations:
$$  
  \dot{\nu}  =  B(H(\nu), \nu)  \,\,\,{\rm where}\,\,\, B: Lie \times Lie^* \to  Lie^*\,\, , \,\, B(v, \nu)(\cdot) \,= \, \nu [\cdot, v].
   $$
We leave as an exercise to split $B$ in terms of $\omega, \eta$.
\end{rem}

\begin{rem} The humble wasp Encarsia Formosa knows better than Corollary \ref{cor1}.   Insects and small birds perform a  `clap-and-fling' trick (\cite{Weis}, \cite{Lighthill-a}, \cite{Kol}, \cite{Cheng}). It produces topology changes in the underlying manifold at the clap and impulsive motions at  the start of the  fling\footnote{Topology breaking by motions starting impulsively  were  already described  in Helmholtz 
1858 paper \cite{Helmholtz}.
Impulsive motions have been studied since Blasius and Lamb \cite{Telionis}.}. A vortex ring is created periodically. On a planar section one sees a pair of opposite vortices\footnote{A far-fetched analogy is the  emergence of vortex pairs
in 2-dimensional (i.e, thin)  Bose-Einstein condensates, leading to the BKT transition  and turbulence as they proliferate \cite{BKT}.}.
\textit{In this paper we will  gloss over  topology changes, 
 vortex sheets, vorticity generation at  boundaries, etc.} \cite{Moffatt}. 
\end{rem}

\vspace{4mm}

\section{POINT VORTICES AND HARMONIC FLOWS} \label{surfaces}

\vspace{1mm}

Let the Riemann surface $\Sigma$ have genus $>0$ nonzero. Assume that the vorticity  is concentrated in  $N$ 
  points.   If there are boundaries,  they will  be  analytic curves,  with slip conditions. The phase space has dimension   $2N+2g$.  Hodge decomposition extends to such singular flows.   We will define  a `pure' vortex flow in section \ref{purevort}.  \vspace{1mm}
   
   For closed Riemann surfaces we take coordinates $(A,B) \in \R^{2g}$   in the cohomology
\begin{equation}  \label{etadec}  \eta = \sum_{k=1}^g \,\left(  A_{k} {\alpha_{k}} + B_{k} {\beta_{k}} \right)  \in {\rm Harm}^1(\Sigma).
\end{equation}

The basis $\{ \alpha_{k} , \beta_{k} \} $ is dual to a corresponding homology basis $\{ a_{k} , b_{k} \}$
(next section \ref{Riemrel}).

\vspace{1mm}

As we just saw, without vorticity  any Euler flow will be stationary.  However, if vortices are present, the  harmonic component $\eta$ will
change  in time.  This corresponds to the second set of general equations (\ref{omegaeta}). In turn, the harmonic component   also drives the vortices, corresponding to the term $d i_{\eta^\sharp}\, \omega $ in   the first set.

\subsection{Harmonic flows: Riemann's   relations  \cite{FarkasKra}} \label{Riemrel}  

Recall that by  uniformization, every closed Riemann surface $\Sigma$ of genus $g \geq 1$   can be constructed by
identifying in pairs appropriate sides of a fundamental  $4g$-sided polygon inside the Poincar\'e disk.
Each one of the sides  of the polygon $a_1b_1a_1^{-1}b_1^{-1} \cdots a_gb_ga_g^{-1}b_g^{-1}$
corresponds to a representative curve $a_j\,, \, b_j, \, 1\leq j \leq g$ of a homology class of the surface.

The intersection numbers of these curves (see \cite{FarkasKra}, III.1) are
\begin{equation} \label{intnumb} 
 a_j \cdot b_k = \delta_{jk}, \, a_j \cdot a_k =   b_j \cdot b_k  = 0. 
\end{equation}

\newpage
  Let
 the dual basis be  (in the sequel we will not worry about subscripts and superscripts)
 $$ \alpha^1, \cdots , \alpha^g, \beta^1, \cdots,  \beta^g \in   {\rm Harm}^1(\Sigma). $$
\begin{equation} \label{dualbasis} \oint_{a_k} \alpha_j =  \oint_{b_k} \beta_j = \delta_{jk}\,\,,\,\, \oint_{b_k} \alpha_j =  \oint_{a_k} \beta_j = 0 \,\, .
\end{equation}
It follows that
\begin{equation} \label{dualbasis1}
\int_{\Sigma} \alpha_j \wedge \beta_k = \delta_{jk}\,,\,\, \int_{\Sigma} \alpha_j \wedge \alpha_k = \int_{\Sigma} \beta_j \wedge \beta_k = 0\, ,
\end{equation}
so one can write 
\begin{equation}  \label{etaexp}
\eta = \sum_{j=1}^g \, A_j \alpha_j + B_j \beta_j \,\,,\,\,\,{\rm with}\,\,\,\,  A_j = \oint_{a_j} \eta \,,\,\, B_j = \oint_{b_j} \eta .
\end{equation}

 \begin{lem} (See, e.g.,  \cite{FarkasKra}, III.3). Given   closed differentials $\theta, \tilde{\theta}$ on a Riemann surface,
  \begin{equation}   \mathlarger{\int}_{\Sigma} \theta \wedge \tilde{\theta} = \sum_{i=1}^g \,\left( \int_{a_i} \theta    \int_{b_i} \tilde{\theta}   -  \int_{b_i}  \theta      \int_{a_i}  \tilde{\theta}   \right).
 \end{equation}
\end{lem}
This formula is very useful.  It implies, for instance

  \begin{prp} \label{RR} (Riemann relations).
    The harmonic forms
    $ \{\star\alpha_1,\ldots,\star\alpha_g,\star\beta_1,\ldots \star\beta_g\}$
    also form a basis of the cohomology of $\Sigma$.  Moreover,
    \begin{equation}  \label{staralphabeta}
      \left(\begin{array}{c}\star \alpha\\\star\beta\end{array}\right)=
      \left(\begin{array}{ll}- R  & P \\- Q  &  R^{\dagger}\end{array}\right)
       \left(\begin{array}{c} \alpha\\\beta\end{array}\right) 
            \end{equation}
     where $P>0$, $Q>0$  (symmetric and positive definite) and  $R$ are  $g\times g$ square matrices given by
     \begin{equation}   \label{staralphabeta1}
       P_{jk}=\int_{\Sigma} \alpha_j\wedge\star \alpha_k\,,\quad
       Q_{jk}=\int_{\Sigma} \beta_j\wedge\star \beta_k\,,\quad
       R_{jk}=\int_{\Sigma} \alpha_j\wedge\star \beta_k\, . 
     \end{equation}
     \end{prp}
 See  \cite{FarkasKra} chapter 3,  or
    \cite{Gustafsson} Eqs. (5.20) and (5.21)).     Since conformal maps preserve Hodge star on one-forms,  the  change of variable formula for integrals implies that $P,Q,R$ are \textit{ invariantly defined } on a Riemann surface\footnote{An account on Riemann's discovery of the bilinear period relations can be found in \cite{Chai}.}. Transformation rules under a change of basis on ${\rm Harm}^1(\Sigma)$ are obvious. 

\begin{cor} \label{naomanjado}
\begin{equation}
\left( \begin{array}{cc}
                                            -R & P  \\
                                              -Q & R^\dagger  \\
\end{array} \right)^2=-
\left( \begin{array}{cc}
                                            I & 0  \\
                                              0 & I  \\
\end{array} \right)
\end{equation}
\end{cor}
\proofn
This is a consequence of the fact that applying Hodge star twice on a one-form just produces a minus sign. One has to notice
also that the column vector with the $\alpha$ and $\beta$ span the entire space as one moves around on the manifold. If this
were not the case, then a non-trivial linear combination of the harmonic forms would vanish, contradicting that they form a basis.
The corollary also appears as a  direct statement in Section~III.2.5 of \cite{FarkasKra}.   \qed

\subsection{`Pure' vortices: Laplace-Beltrami Green function} \label{purevort}
\vspace{1mm}

Recall that in local isothermal coordinates $ds^2 = \lambda^2 (dx^2 + dy^2) $  for a Riemannian metric $g$,  the  Laplace-Beltrami operator  for functions is given by
$$\Delta  f = - (1/\lambda^2)( f_{xx} + f_{yy}) \,\,\,\, \text{ (note the minus sign so that $\Delta$ is a positive operator)}. $$

 The Green function  for  $\Delta$ is defined by  the properties (see eg. \cite{Okikiolu}) 
\begin{equation} \label{geometerdelta}
\Delta_s G(s,r)=\dt_r(s)-V^{-1}\, , \qquad \int_{\Sigma} G(s,r)\mu(s)=0 \,\,, \,\, G(s,r) = G(r,s) ,
\end{equation}
where  $V (= {\rm Area}(\Sigma)) = \int_{\Sigma} \mu$.   In our paper $ \mu $ always denotes the  area form.  It is well known that in view of Green's identities  the meaning of  $\dt_r(s)$ is that  
$$ \int_{\Sigma} G(s,r) \Delta f(r) \mu(r)  = f(s) - \frac{1}{V} \int_{\Sigma} f \mu .
$$

The regularization of the Green's function
 at $s_o$ is  called the {\it Robin function}: 
    \begin{equation}
R(s_o)=
\lim_{\ell(s,s_o) \to 0} \left[
G(s,s_o)+\frac{1}{2\pi}\log \ell(s,s_o)\right]\,,{\rm where} \, \,\ell \, \,{\rm is}\,{\rm the} \, {\rm Riemannian} \,{\rm distance.}
\label{robin}
\end{equation}
\begin{rem} $G(r,s)$ can be interpreted as the mutual energy between the point charges
$\delta_r$ and $\delta_s$, see eg. \cite{Gustafsson, Gustafsson-2019a}. The symmetry is actually a consequence of the first two equalities, that imply
$$
G(s,r)=\int_\Sigma dG(\cdot,s)\wedge\star dG(\cdot,r),
$$
\end{rem}

 \begin{definition}
 The `pure' vortex flow determined by a vortex of intensity 1 pinned at a point $s_o$ is
  \begin{equation}
  s \mapsto  v_{s_o}^{vort}(s) = (\nu_{s_o}{^{vort}})^\sharp (s),\,\,\,\,\, \text{where} \,\,\, \nu_{s_o}{^{vort}} =: - \star d_s G(s, s_o)\, .
    \end{equation}
    \end{definition}
The choice of minus sign becomes clear from Example \ref{exa1}.  \\

Our starting point  is  the  following proposition and the two lemmas in next section \ref{mainlemmas}.\\

\begin{prp}   (see \cite{Gustafsson}) \\ \, \\
i)   Hodge  decomposition of a flow  $\nu$ containing $N$ point vortex singularities $
    (s_k, \Gamma_k), 1 \leq k \leq N$:
    \begin{equation} \label{Hodgedecompflow}
  \nu(s) = - \left[  \sum_{k=1}^N \,\Gamma_k   \star d_s G(s, s_k) \right] \oplus  \eta, \,\,\, \eta \in {\rm Harm}^1(\Sigma),\,\, s \neq s_k \, (1 \leq k \leq N).
    \end{equation}
ii)  The vorticity 2-form corresponding to the $N$ pure vortices is given by
\begin{equation}  \label{vortwoform}  \omega = \left[ \sum_{k=1}^N \, \Gamma_k \delta_{s_k} - (\Gamma/V) \right]\, \mu   
\end{equation}
iii) The decomposition (\ref{Hodgedecompflow}) is  $L^2$-orthogonal:   $\nu = \text{pure vortices flow} \oplus \text{harmonic}$.
\end{prp}  

The word `Hodge' in the proposition  and the symbol $\oplus$ mean $L^2$-orthogonality\footnote{It is not hard to show that Hodge theorem extends to   (\ref{Hodgedecompflow}) and        Euler's equation will have the singular limit discussed here
(see  \cite{Gustafsson}), and for background Appendix  B).}. The flow $\nu$  given by (\ref{Hodgedecompflow}) is undefined at the locations of the vortices. Nonetheless,  it is common knowledge  among  the vorticists community  the following regularization at the vortices themselves:
{\small
\begin{equation}  \label{regularized1}
(\dot{s}_j)^{\flat}  =  
-\star d_{s=s_j} \left[\frac{1}{2} \Gm_j R(s)+\sum_{k\ne j}^{N}\Gm_k G(s,s_k)
\right] +\,  \eta(s_j)\,\,\,, \,\,\,\,\,
\eta(s_j) = \sum_{\ell=1}^g \,\left(  A_{\ell} {\alpha_{\ell}}_{|s_j} + B_{\ell} {\beta_{\ell}}_{|s_j} \right)  .
\end{equation}
}

\begin{rem}  Why
  the factor
  $\frac{1}{2}$ in front of $R$?
The regularized
  self-velocity at the point $s_o$ is
  \[-\star \lim_{\ell(s,s_o)\to 0} d_s\underbrace{\left[
      G(s,s_o)+\frac{1}{2\pi}\log \ell(s,s_o)\right]}_{=: \tilde R(s,s_o)}=
    -\star d_s
\tilde R(s,s_o)_{s=s_o}\,,
\]
 \label{onehalf} 

The symmetry  of $G$ 
 implies $\tilde R(s,s_o)=\tilde R(s_o,s)$ and
$d_s\tilde R(s,s_o)=d_s\tilde R(s_o,s)$.
The Robin function is given by $R(s_o)= \lim_{s\to s_o}\tilde R(s,s_o)=\tilde R(s_o,s_o)$. Then
 $$d R(s_o)=d_s \tilde R(s_o,s)_{s=s_o}+d_s \tilde R(s,s_0)_{s=s_o}=
2d_s \tilde R(s,s_o)_{s=s_o} \,\,\,\, \Rightarrow \,\,\, d_s \tilde R(s,s_o)_{s=s_o}=\frac{1}{2}
dR(s_o). $$
\end{rem}

 \subsection{Two  key  lemmas (\cite{Gustafsson})} 
 \label{mainlemmas}
 \vspace{1.5mm}
\textit{The story does not end with} (\ref{regularized1}), where the harmonic part is fixed.  It was recently shown  in \cite{Gustafsson} that if  a flow $\nu(t)$  obeys Euler's equations,
 the harmonic components  $A, B$  of $\eta$  {\it must} also change.  \smallskip
 
  The following
 two results are key for deriving the ODEs for  $\eta$. The   reader must take his/hers time in filling the details of the proofs.
We may denote here $G_s(\cdot) = G(\cdot , s)$ and $d_s$ operating on $s$. 
 Define
 \begin{equation}
   U_{\gamma}(s)   =  \mathlarger{\oint}_{\gamma(r)}\star_r d_r G_{s}(r)  ,\,\, s \notin \gamma .
\end{equation}

 \begin{lem}  (``Hat trick": three Messiesque goals) \label{lem3}
Given $\Sigma$  as a Riemann surface, i.e. with only its  conformal structure, a basis of harmonic forms can be obtained from the Green function of any chosen metric in the conformal class. \\ 

\noindent i)  $U_{\gamma}(s)$ extends to all of $\Sigma$ as   a multivalued function in the sense of `ancients': it   makes a unit jump everytime $s$ crosses $\gamma$ from its left to the right side of $\Sigma$.
 \\

\noindent ii)  $dU_{\gamma}$ is a  harmonic differential in $\Sigma$. \\ 

\noindent iii)  
Let  $\{ \alpha_k ,\, \beta_k \} $  the 1-forms in  $\Sigma$ dual to $\{ a_k,,\, b_k \}$ as in (\ref{intnumb}). Then
\begin{equation}  \label{Gbasis}
dU_{\alpha_{\ell}} = d_s {\mathlarger{\mathlarger{\oint}}}_{a_{\ell}} \star_r d_r G(r,s) = \beta_{\ell} \,\,,\,\, dU_{\beta_{\ell}} = d_s \mathlarger{\oint}_{b_{\ell}} \star_r d_r G(r,s) = -\alpha_{\ell} \,.
\end{equation}
where   the line integrals done in $r$. 
\end{lem}
 
The multivaluedness disappears under differentiation. In Appendix C we show i) in detail.  We will show ii) next,  that the    
differential $d_sU_\gamma$ is harmonic. The reader is challenged to verify iii). 

\bigskip

  \begin{lem}   \label{lem2}  (``Crossing the Rubicon'')  If $\nu(s, t)$  is  the   Euler flow (\ref{Hodgedecompflow})  of  a  system of point
vortices on a closed boundaryless surface, then the rate of change of its circulation around a curve
 that avoids the singularities can be computed by
 \begin{equation}  \label{eqlem2}
  \frac{d}{dt} \, \mathlarger{\oint}_{\gamma} \nu  =  (\Gamma/V) \, \mathlarger{\oint}_{\gamma}\, \star \nu \, .
\end{equation}
where $\Gamma = \sum_{k=1}^N \Gamma_k$ is the total vorticity of the point vortices.
\end{lem}

\bigskip

\begin{rem} At first sight,  one could imagine that for any incompressible flow $\nu$, the line integral $\oint_{\gamma} \star \nu $ in the right-hand-side of (\ref{eqlem2}) should vanish.   The following simple example shows that this is not the case.
Consider  the flat torus corresponding to a rectangular lattice,
  the harmonic vector field  $\nu = \mathbf e_1$  and  the closed curve $\gamma$ tangent to the direction $\mathbf e_2$.
 The integral $ \oint_{\gamma}\, \star \nu $ is nonzero.  Although 
  $ \oint_{\gamma(t)} \nu \equiv 0,$ the validity of (\ref{eqlem2}) is saved,  because  $\Gamma = 0$.
  \end{rem}

\noindent Proof (outline)  of Lemma \ref{lem2}. 
 Recall that  if $\nu(t)$ is an Euler flow, then $\dot \nu + L_{\nu^\sharp}  \nu = {\rm exact}. $  Hence,
 $$ \frac{d}{dt} \, \oint_{\gamma} \nu =  - \oint_{\gamma}   L_v \nu  \,,\,\, v = \nu^\sharp . $$

\noindent  In  $ L_v = i_v d  + d i_v$,  the  $d i_v$ piece does not contribute to the integral over a closed curve, so
 $$ \frac{d}{dt} \, \oint_{\gamma} \nu = - \oint_{\gamma}\, i_v d\nu =  - \oint_{\gamma}\, i_v  \omega,\,\,\,{\rm with}\,\,\,  \omega = d\nu = \left[ \sum_{k=1}^N \, \Gamma_k \delta_{s_k} - (\Gamma/V) \right]\, \mu  
 $$
 where we used  (\ref{vortwoform} )and that the harmonic part of $\nu = \delta \psi \oplus \eta $ disappears when computing $d\nu$.

\noindent The lemma   follows from the following observations, that the reader should appreciate:
\vspace{2mm}

i)   For any vector field :  $ \,  i_v  \mu = \star v$, see Appendix B.

ii)
Since  in our calculations   $s \neq s_k$,   it  is readily seen that if  $s \neq s_k$ then
 $  ( i_v \delta_{s_k} \mu)(s) = 0 .$  \qed   
 
\bigskip

\noindent Proof for  ii)  on Lemma \ref{lem3}.  
In computing  
$  d_s \mathlarger{\oint}_{\gamma}\star dG(\cdot, \,s) 
$ 
recall that    $\,\, d , \,  \star,$ and  $\oint$  operate to the $\cdot$ slot.
The Laplace operator $\Delta_s$ can go inside the integral.  We just observe that in the integrand 
\begin{equation}  \Delta_{s} \star_r d_{r}  G(r ,s ) =  \star_r  d_{r}   \Delta_{s }  G(r , s) = \star_r d_{r} \left( \delta_{s} (r) -  (\frac{1}{V}) \right) = \star_r d_{r}  \delta_{s}(r) = 0  \,\, \text{for} \,\, s \neq r .  
\end{equation}

\bigskip

\begin{rem}This may be a far-fetched observation,  but in a way  the  expression (\ref{eqlem2}) resembles (somehow in reverse) Faraday's law of induction $ \oint_{\partial \Sigma} {E} \cdot d\ell  = - \frac{\mathrm{d}}{\mathrm{d}t} \int_{\Sigma}  B \cdot \mathrm{dA}  . $ 
\end{rem}

\newpage

\subsection{Equations of motion} 
\vspace{1mm}

  Felix Klein took ideal 2d-hydrodynamics as the motivation in his outstanding  introductory book about  Riemann 
  surfaces  (\cite{Kleina}, 1882)\footnote{He was  probably the first to realize that a surface with a metric is also a Riemann surface by taking an atlas of local isothermal  coordinates.}. For Klein, 
a meromorphic 1-form  $\nu$    is   interpreted as a steady \textit{ideal flow}, composed by  elementary  components:  harmonic, point vortices, sources-sinks and higher order singularities. 
  These  singularities remained \textit{static}, pinned at fixed  points on the  surface. 
  
The following 
theorem  provides  the  \textit{Euler dynamics} to  $\nu(t)$ 
 associated to the  chosen metric in the   conformal class. We take a basis   for ${\rm Harm}^1(\Sigma)$, $\,\, \{ \alpha_k , \beta_k\,,\,  1\leq k \leq g \} $  satisfying (\ref{intnumb}),  and write  
 $$
  \nu(s) = - \left[  \sum_{k=1}^N \,\Gamma_k   \star d_s G(s, s_k) \right] \oplus  \eta,\,\,\,\,\,\, \eta = \sum_{\ell=1}^g \,\left(  A_{\ell} {\alpha_{\ell}} + B_{\ell} {\beta_{\ell}} \right) .$$
 \vspace{1mm}

\begin{teo}(\cite{Gustafsson}) \label{gus} The ODEs for the motion of $N$ vortices  $ s_j, \, 1 \leq j \leq N$ of strengths
$\Gamma_j$ on a closed surface (without boundary),  coupled to the $2g$ harmonic fields,  are 
{\small
\begin{equation} \label{Guseq1}
  \dot{s}_j   =  -\frac{1}{2}\, \Gamma_j \,  (\star dR)^\sharp(s_j) +  \sum_{k\neq j} \, \Gamma_k \,  v_{s_k}^{vort}(s_j)
+
\sum_{\ell=1}^g \,\left(  A_{\ell} {\alpha^{\sharp}_{\ell}}_{|s_j} + B_{\ell} {\beta^{\sharp}_{\ell}}_{|s_j} \right)
\end{equation}
\begin{equation} \label{Guseq2}
\begin{split} 
  \dot{A}_k & =   (\Gamma/V) \left( \sum_{\ell=1}^g \, A_{\ell} \oint_{a_k} \star \alpha_{\ell} +  \sum_{\ell=1}^g \, B_{\ell} \oint_{a_k} \star \beta_{\ell} \right)   + \, \sum_{j=1}^N \, \Gamma_j \,  \beta_k(\dot{s_j})  \\
  \dot{B}_k & =   (\Gamma/V) \left( \sum_{\ell=1}^g \, A_{\ell} \oint_{b_k} \star \alpha_{\ell} +  \sum_{\ell=1}^g \, B_{\ell} \oint_{b_k} \star \beta_{\ell} \right)   - \, \sum_{j=1}^N \, \Gamma_j \,   \alpha_k(\dot{s}_j) \,\, .
 \end{split}
 \end{equation}
 }
\noindent where  $\Gamma = \sum_{\ell=1}^N \, \Gamma_{\ell}$ is the total vorticity, $V$ is the surface area and  
the line integrals are given by Riemann's relations (\ref{staralphabeta}). We insert   the velocities of the vortices (\ref{Guseq1})  in (\ref{Guseq2}).
\end{teo}

\proofn The first set of equations (\ref{Guseq1}) is quite intuitive.  The harmonic components act on the vortices  as they were steady {\it outside agents}. 
 It corresponds to the first equation in  Theorem  \ref{coupledeq}  in the singular limit of concentrated vorticity in the $s_j$.

It remains to get the equations (\ref{Guseq2})  for the harmonic piece. Using (\ref{Guseq1}) as an \textit{Ansatz}, 
the equations for  $\dot{A}, \dot{B}$  follow by
applying the two lemmas  of  section \ref{mainlemmas}   to the Euler flow (\ref{Hodgedecompflow}). 
  One gets, ``turning the algebra crancks":
\begin{equation} 
 \begin{split} 
  \dot{A}_k & =   (\Gamma/V) \left( \sum_{\ell=1}^g \, A_{\ell} \oint_{a_k} \star \alpha_{\ell} +  \sum_{\ell=1}^g \, B_{\ell} \oint_{a_k} \star \beta_{\ell} \right)   + \, \sum_{j=1}^N \, \Gamma_j \,  \frac{d}{dt}  \oint_{a_k} \star dG_{s_j} \\
  \dot{B}_k & =   (\Gamma/V) \left( \sum_{\ell=1}^g \, A_{\ell} \oint_{b_k} \star \alpha_{\ell} +  \sum_{\ell=1}^g \, B_{\ell} \oint_{b_k} \star \beta_{\ell} \right)   + \, \sum_{j=1}^N \, \Gamma_j \,   \frac{d}{dt} \oint_{b_k} \star dG_{s_j}\,\,, 
 \end{split}
 \end{equation}
and  the ODEs  (\ref{Guseq2}) result from   the ``hat trick" applied to the last terms.   The line integrals   are given in terms of the Riemann coefficients  $P$'s, $Q$'s, $R$'s (\ref{staralphabeta}, \ref{staralphabeta1}).  See \cite{Gustafsson} for details. \qed

\begin{rem}    Note the second terms.  $A$ and $B$ change due to the presence of vortices, even in the special case when   the total vorticity vanishes.
    \end{rem}

 \subsection{Hamiltonian structure.}
\vspace{1.5mm}

In the paper  \cite{Gustafsson},  the coordinates taken for the harmonic component  are the circulations
\begin{equation}
 \tilde{A}_k = \oint_{a_k} \nu \,,  \, \tilde{B}_k = \oint_{b_k} \nu
 \end{equation}
of the Euler flow $ \nu(t) =   
     \eta(t) -  \sum_k\, \Gamma_k  \star  dG(\,\cdot\,  ; s_k(t) ) .$   The  symplectic form nicely splits as
     \begin{equation}
  \Omega= \left( \sum_{j=1}^N \Gamma_j\mu_{s_j} \right) -\frac{V}{\Gamma} \left( \sum_{i=1}^gd\tilde A_j\wedge d\tilde B_j .\right). \label{form3}
  \end{equation}
In  these variables the Hamiltonian has mixed terms in the $s_k$ and the $\tilde{A}, \tilde{B}$.  Alternatively, 
in  the companion paper \cite{Ragazzo} the Hamiltonian structure was given using instead
\begin{equation}  A _k = \oint_{a_k} \eta \,  ,\,    B_k = \oint_{b_k} \eta \,\, ,\,\,\,\, 1 \leq k \leq g.
\end{equation}
The Hamiltonian splits on  a pure vortical part
 $H_{vort}$ and the harmonic part $H_{harm}$, but in turn,  the symplectic form will  contain
 mixed terms that  resemble magnetic 2-forms, as follows:
 \vspace{1mm}

 \begin{teo} \label{Hamstru}  
 \label{rag} The symplectic structure  $\Omega(X_H , \cdot ) =    dH $ 
   for the $N$-vortex system is given by  
    $$H =  H_{vort}(s_1, \cdots, s_N) + H_{harm}(A,B), $$
\begin{equation}  \label{Hamsymp}
\begin{split}
& H_{vort} =   \, \sum_{j=1}^N\frac{\Gm_j^2}{2}\, R(s_j)  + \sum_{j=1}^N\sum_{k\ne j}^{N}
                 \frac{\Gm_j\Gm_k}{2} \, G(s_j,s_k) \\
  &  H_{harm} =  \frac{1}{2}   \, \mathlarger{\int}_{\mathsmaller{\Sigma}} \eta \wedge \star \eta = \frac{1}{2}   \, (A^\dagger, B^\dagger)
 \left( \begin{array}{ll} P  & R \\ R^\dagger & Q    \end{array}  \right) \left( \begin{array}{l}  A\\ B     \end{array}     \right)
 \end{split}
\end{equation}
\begin{equation}
  \Omega=\sum_{j=1}^N \Gamma_j\mu_{s_j}-\frac{V}{\Gamma}
  \sum_{j=1}^g\big(dA_j-\sum_{k=1}^N\Gm_k\beta_{js_k}\big)\wedge\big(dB_j+\sum_{\ell=1}^N\Gm_{\ell}
  \alpha_{js_{\ell}}\big)\,.
  \label{form2}
  \end{equation}
  \end{teo}
Nota bene: The subscripts $s_k, s_\ell$ added to the  harmonic forms mean solely that the $\alpha$'s and $\beta$'s will be applied on the (corresponding) tangent vectors in the $k$-th and $\ell$-slots of $\Sigma \times \cdots \times \Sigma$.  \\The sign choice  $\Omega (X_H,\cdot)=+dH$ gives  the correct result for the unit  vortex in a simply connected domain,  namely $ 
\dot{x}=\frac{1}{2}\frac{\partial R}{\partial y}, \quad \dot{y}=-\frac{1}{2}\frac{\partial R}{\partial x}. 
$ 
 
    \proofn
  Let us  just  show   that $\Omega$ is symplectic.
  The 1-forms $\alpha_{j}$ and $\beta_{j}$ are closed, hence in any contractible
  open ball $\mathcal U_{s_o}$ centered in a given $s_o \in \Sigma$ there exist functions $\xi_{j,s_o}(s) (= -U_{b_j})$
  and $\zeta_{j,s_o}(s) (= U_{a_j})$ such that
  $\alpha_{j}=d\xi_{j,s_o}$ and $\beta_{j}=d\zeta_{j,s_o}$. Consider the map
  $$\big(s_1,\ldots s_N,(A,B)\big) \in \mathcal U_1\times\ldots\mathcal U_N\times {\rm Harm}^1(\Sigma)\to  \big(s_1,\ldots s_N,(\tilde A,\tilde B)\big)  \in \mathcal U_1\times\ldots\mathcal U_N
  \times {\rm Harm}^1(\Sigma)   $$
  \[
    \tilde A_j=A_j-\sum_{k=1}^N\Gm_k\zeta_j(s_k)\,, \ \tilde B_j=B_j+\sum_{k=1}^N\Gm_k\xi_j(s_k)\,, \
j=1,\ldots,g
\]
defining new coordinates on  $\mathcal U_1\times\ldots\mathcal U_N\times {\rm Harm}^1(\Sigma)$. In these coordinates,
which were used in \cite{Gustafsson} (see equations (5.25. 5.26) there)
\begin{equation} \label{gusform}
  \Omega=\sum_{j=1}^N \Gamma_j\mu_{s_j} -\frac{V}{\Gamma} \sum_{i=1}^gd\tilde A_j\wedge d\tilde B_j\,  
  \end{equation}

    The total vorticity $\Gamma$ in the denominator  could raise eyebrows, but curiously
the equation  of motion for the harmonic part 
(\ref{Guseq2}) actually becomes simpler for $\Gamma = 0$.  Note that in this case the harmonic part still changes, due to the vortex velocities\footnote{We hope that  experts on symplectic/Poisson  structures will be intrigued by this situation. Such singular  structures are now a fashionable theme (log-symplectic, b-Poisson, \cite{Miranda0},
    \cite{Zambon}). We believe that the symplectic form  (\ref{gusform}) as  $\Gamma \rightarrow 0$ could be cast in these frameworks.}.

If all the vortex strengths are set to zero, equations (\ref{Guseq1}) is interpreted as motion of markers on a potential flow. In this case, equation (\ref{gusform}) will make sense  formally, since $dA = dB = 0$. 

\vspace*{-2mm}

\subsection{A sample of consequences}  \label{sample}

\vspace{1mm}

  \begin{prp} \label{propeq}
 The equilibria of the incomplete system (the equation (\ref{Guseq1}) with
  $A_{\ell} = B_{\ell} = 0$)  extend to equilibria for the full system. A local minimum of the Hamiltonian of the 
  incomplete system  extend to stable equilibria of the full system.
   \end{prp}
    \begin{prp}  \label{harmonic torus} (proof in the next  section \ref{flatnonflattori}).
    In the case of a torus (genus 1)  with  the  flat metric    the vorticity and harmonic components do not couple. The harmonic part stays stationary. 
    \end{prp}

 \begin{prp} (Kimura's conjecture \cite{Kimura}).   A vortex dipole follows a geodesic.
\end{prp}
 \proofn In \cite{Gustafsson} it is  shown that the effect
 of the harmonic part on a  the motion of a close pair of opposite vortices  is of higher order, rescuing the insights in \cite{KoillerBoatto} with  the incomplete equations.  \qed
\begin{rem} In a way  a vortex dipole  is a ground state. The interaction energy between two opposite vortices $\pm \Gamma$ at close distance behaves  as   $-\Gamma^2 (- \log \ell(s_1,s_2)) \sim - \infty$.  
\end{rem}

\subsection{Steady vortex metrics and their probabilistic interpretation} \label{steadyvm}  
\vspace{1.5mm}

One may think that  when there is no  initial harmonic field,  a single vortex will not move on a surface of constant curvature.   Not so!  A  vortex pinned at $s_o$ will produce the flow with stream function $G(s;s_o)$. The ``rest of the world"  may  react, making the vortex to drift according to the regularized  Hamiltonian $R$.  It was shown in  \cite{Ragazzo0}
that the  Robin function  $R(s)$ is not constant in  Bolza's surface,  the  Riemann surface of genus two with  the maximal number of discrete symmetries. It  has an interesting 
structure of singularities and separatrices,  see Fig. 5   in section \ref{Bolzasurf}. 

This prompts the concept of  a \textit{steady vortex metric} (SVM),  defined by the requirement that the Robin function is
a  constant, so a  single vortex  will stay still.   
Being SVM means a hydrodynamical balance condition in the sense that there is no  vortex drift \textit{anywhere} in the surface. 

Metrics
of constant curvature are SVM only in surfaces of genus 0 and 1 \cite{RagVig}.  This gives rise to a question for geometric analysts.
For genus $\geq 2$ could the steady vortex metrics rival with the customary  canonical metrics?  Say,  the hyperbolic, Arakelov, Hahn and the capacity metrics?

Surprisingly, a  SVM is  related  to a certain uniformity in  Brownian motion, a fact that  may have some biological interest. 
 
 A closed compact manifold  $M$  is stochastically complete.  While the probability is zero, starting at $r$,   to reach  any particular point $ s $, the probability of reaching  an $\epsilon$-ball  $B_{\epsilon}(s)$  is  1,  for any $\epsilon > 0$ arbitrarily small.   Let  the random variable $\tau_{\epsilon}^{r \to s}$ be the \textit{first hitting time, or first passage} for a Brownian motion beginning at $r$ to reach that ball around $s$  of radius $\epsilon$.  If $\ell(r,s) < \epsilon$, we take by convention that  $\tau_{\epsilon}^{r \to s} = 0$,  hence  the integrations below can be done in the whole surface. 
   The expected value  $E_{s,\epsilon}(r) = \mathbb{E}[\tau_\epsilon^{x\to y}] $  is usually   called  the \textit{narrow escape time}  (NET)  in  cellular biology  \cite{HolmanSchuss0, Schuss, HolmanSchuss}. It   can also be written $ E_{\epsilon}(r; s, \epsilon)$  but it is not symmetric in $r,s$.
$E_{s,\epsilon}(r)$  satisfies the exterior  Dirichlet problem
\begin{equation} \Delta_r E_{s,\epsilon}(r) = 1,  \,\,  r \in M - B_{\epsilon}(s), \,\,\,  E_{s,\epsilon}(r) = 0,\,\, r \in \partial(M - B_{\epsilon}(s)).
\end{equation}

Steiner and Doyle   \cite{SteinerDoyle2009} consider the `hide-and-seek' games on a surface $M = \Sigma$.
Normalizing  the total area  of $\Sigma$ to 1,   the area form $\mu$  becomes a
 probability measure\footnote{
 Historical note.  In the 20th century the physical and biological sciences have been revolutionized
 in probabilistic terms. Mathematicians began to look at differential equations, number theory and combinatorics in that light too.  The applications  expanded, among other areas,   to financial mathematics
 and artificial intelligence.
The most
important person at   the  origin of 
probabilistic potential theory was   
Norbert Wiener\footnote{Wiener died in 1964 from a heart attack at the entrance of the math department of  the KTH Royal Institute of Technology in Stockholm.
  But we hope our readers will not be discouraged to visit KTH.}.
The  seminal work  is his   paper from 1923 (one hundred years back from now), the first rigorous construction of a Brownian motion process
  \cite{Wiener}.
Curiously, Brown made his experiment in 1827, one year before Green's paper.
The probabilistic analogies in   electrostatics and ideal fluids are natural consequences of the fundamental
role of the Laplacian   in all these subjects. For the intuition in  electrical networks in the discrete
context (graphs and Markov chains),  see \cite{doylesnell}, and the review in \cite{stolar}.  
}. Their notation  for Robin function $R$ is  changed to $m$,  called the Robin  {\it mass}:
$ m(s) - G(r,s) = \frac{1}{2\pi} \log \ell(r,s) + O(\ell(r,s)) .\,\,
$   They show\footnote{The Robin mass 
 relates with a spectral invariant, 
 the Zeta function 
$Z(p) =: {\rm Trace (\Delta}^{-p}) = \sum_{j=1}^{\infty} \lambda_j^{-p} ,  \,\,\re p >1,
$
that can be  continued to a meromorphic function  with a simple pole at $p=1. $
$ \tilde{Z} = \lim_{p \to 1} \left( Z(p) - \frac{1}{p-1}  \right)  
$ is the regularized trace.
Morpurgo \cite{Morpurgo-a} proved that 
$\tilde{Z} = \int_{\Sigma} R(s) \mu(s) + \frac{1}{2\pi}(\gamma - \log2),\,\,\,\,\,\,
\gamma \sim .5772\,\,  (\rm{Euler}  \, {\rm constant).} $ See also  Jean Steiner \cite{Steiner-a}).
}:
\vspace{-2mm}
\begin{equation}
E_{\epsilon,s}(r) =  -\frac{1}{2\pi} \log  \epsilon -  G(r,s) +  m(s) +  u_{\epsilon}(r),  \,\,\,  ||u_{\epsilon}||_{C^0} = O(\epsilon) .
\end{equation}
where $u_{\epsilon}$ harmonic in the complement of the ball.
The divergence of
$-\log \epsilon$  as  $\epsilon \to 0$  reflects the fact that the probability of a stochastic path to  hit  $s$ precisely is zero, so the expected time needed to  reach that  specific point  $s$ is infinity. 
It follows that  the expected  time for  the first passage, letting $s$ span the whole surface, with a fixed seeker starting at $r$ 
has the leading term given by  
\begin{equation}  \label{game1a}
\int_{\Sigma} E(r; s , \epsilon) \mu(s) = -\frac{1}{2\pi} \log \epsilon + \int_{\Sigma}  m(s) \mu(s) + O(\epsilon) .
\end{equation}
Neglecting the $O(\epsilon)$ this  is independent of where the seeker $r$ starts.
On the other hand, if  the hider $s$ is fixed, and the seeker $r$  is chosen at random,  the expected time is
\begin{equation} \label{game2a}
 \int_{\Sigma} E_{s, \epsilon}(r) \mu(r) = -\frac{1}{2\pi} \log \epsilon +  m(s)   + O(\epsilon) .
\end{equation}

This expression shows that steady vortex metrics have a probabilistic interpretation, namely:  the spatial average of the narrow scape times   $E_{s, \epsilon}(r), $  with respect
to a uniform distribution of initial points $r$ in $\Sigma$, does not depend on the
position of the escaping window $B(y, \epsilon)$.

In a forthcoming article by one of the coauthors  \cite{ragsub-a}, the notion is generalized to higher dimensions. The acronym  SVM  will  then change  to UDM,  ``uniform drainage metrics",
where the spatial average of NET with respect
to a uniform distribution of initial points  $r$ does not depend on the
position of the escaping window $B_{\epsilon}(s)$.\\

 \subsection{How Green and Robin functions respond  under a conformal change of metric} \label{GreenRobinconformal}

  In this section (only) we will use the letter  $g$ to denote a Riemannian metric in the conformal class of $\Sigma$.
Consider two conformally related metrics,
$g$ and  $g_{\rho} = \rho g,   \, \rho > 0$.     Recall that if  $\mu$ is the area form of metric $g$, then  $\rho \mu$ is the area form of metric $g_{\rho}$, and that
$  \Delta_{\rho} =  \rho^{-1} \Delta .
$
 Denote
\begin{equation}  |\Sigma| = \int_{\Sigma}\, \mu,  \,\,\,  |\Sigma|_{\rho} = \int_{\Sigma} \rho \mu,  \, \,\,\,  \bar{\rho} = \frac{1}{|\Sigma| } \int_{\Sigma}  \rho \mu    =    \frac{|\Sigma|_{\rho}}{|\Sigma|  }. 
\end{equation}

Let $G, G_{\rho}$ be the Green functions for the respective Laplacians $\Delta,\Delta_{\rho}$. The last term  in the following formula is a constant that does not matter for vortex motions, but is relevant for geometric function theory aspects, such as comparisons of  Zeta functions.  

\begin{lem} (Baernstein formulas,  \cite{Morpurgo-a})  \label{Baernsteinlemma}
\begin{equation}
\begin{split}
& G_{\rho}(s_1,s_2) =   G(s_1,s_2) - \frac{1}{ |\Sigma|_{\rho} } \left[ \Delta^{-1}  (\rho - \bar{\rho})(s_1) + \Delta^{-1}  (\rho - \bar{\rho})(s_2)
\right] + \frac{1}{ (|\Sigma| _{\rho})^2 } \int_{\Sigma} (\rho - \bar{\rho})
\Delta^{-1}  (\rho - \bar{\rho}) \mu \\
& R_{\rho} (s) =   R(s) + \frac{\log \rho(s)}{4\pi} -  \frac{2}{|\Sigma|_{\rho} } \Delta^{-1}(\rho - \bar{\rho})(s) +  \frac{1}{(|\Sigma|_{\rho})^2 } \int_{\Sigma}( \rho - \bar{\rho}) \Delta^{-1}(\rho - \bar{\rho}) \mu .
\end{split}
\end{equation}
\end{lem}

\proofn
For the first, fix $s_2$ and apply  $\Delta_{\rho} =  \Delta_{|s_1} /\rho(s_1)$ on the $s_1$ slot of the right hand side.  We get \begin{equation*}
\begin{split}     \frac{1}{\rho(s_1)} \Delta_{|s_1} G( s_1, s_2) -   \frac{1}{ \rho(s_1)  |\Sigma|_{\rho} } [ \rho(s_1)  -  |\Sigma|_{\rho}/   |\Sigma|  ] & = \\  =  \frac{\delta_g(s_1)}{w(s_1)}  - \cancel{\frac{1}{|\Sigma| w(s_1)}}
- \frac{1}{ |\Sigma|_{\rho} } +  \cancel{\frac{1}{|\Sigma| {\rho}(s_1)} } & 
 = \frac{\delta_g(s_1)}{w(s_1)}  - \frac{1}{ |\Sigma|_{\rho}} =   \delta_{\rho g}(s_1)  - \frac{1}{ |\Sigma|_{\rho}} .
\end{split}
\end{equation*}
Thus
$$ G^{\rho}(s_1,s_2) =   G(s_1,s_2) - \frac{1}{ |\Sigma|_{\rho} } \left[ \Delta^{-1}  (\rho - \bar{\rho})(s_1) \right] + c_{\rho}(s_2), \,\, {\rm for} \,\,\, s_2 \neq s_1.
$$
To find $c_{\rho}(s_2)$,  multiply both sides by $\rho(s_1)$  and integrate in $s_1$ in the area form $\mu(s_1)$, 
 using that $ G^{\rho}$  has
mean zero in the metric $ g_{\rho}$.  The proof of the second formula is immediate. \qed
\smallskip
 
\begin{teo} \label{conformalteo}  Relative to the conformal metric $ g_{\rho}$,  the symplectic form   in Theorem \ref{Hamstru} is modified by changing   the area forms  to $\rho \mu$ and the total volume in the magnetic term.   The Hamiltonian corresponding to metric $\rho g$  is the Hamiltonian (\ref{Hamsymp}) of  metric $g$ amended with the terms
\begin{equation} \label{extra}
 \sum_{j=1}^N  \frac{1}{8\pi} \Gamma_j^2 \log \rho(s_j) - \frac{\Gamma}{|\Sigma|_{\rho} }
 \sum_{j=1}^N   \Gamma_j  \Delta^{-1}  (\rho - \bar{\rho})(s_j) .
\end{equation}
\end{teo}
 Interestingly,  if the total vorticity $\Gamma = \sum_{k=1}^N \, \Gamma_k $ vanishes, the nonlocal terms disappear. 
 \proofn We revisit the derivation in \cite{KoillerBoatto}.  
Neglecting the constants and taking the 1/2 factors in (\ref{Hamsymp}) into account,  the Hamiltonian will receive extra terms. For
the Robin part: 
$$  \sum_{j=1}^N  \frac{1}{8\pi} \Gamma_j^2 \log \rho(s_j) -  \frac{1}{|\Sigma|_{\rho} } \left( \sum_{j=1}^N   \Gamma_j^2 \Delta^{-1}(\rho - \bar{\rho})(s_j) \right) .
$$
For the mutual interactions (the Green functions terms)
$$  - \frac{1}{2 |\Sigma|_{\rho} } \left[ \sum_{j\neq k=1}^N  \Gamma_j \Gamma_k \left( \Delta^{-1}  (\rho - \bar{\rho})(s_j) + \Delta^{-1}  (\rho - \bar{\rho})(s_k) \right) \right]= - \frac{1}{|\Sigma|_{\rho} }
 \sum_{j=1}^N   \Gamma_j  \Delta^{-1}  (\rho - \bar{\rho})(s_j)  (\sum_{k\neq j} \Gamma_k).
$$
An algebra juggling leads to  (\ref{extra}).  Remarkably, the total vorticity pops  up from the summation.    \qed

\subsection{Recovering the pressure} \label{recoverpress}
\vspace{1.5mm}

 An Engineering oriented reader may complain that in (\ref{EulerArnold}) we are ``throwing  the baby out with the bathwater"  by eliminating  the pressure differential $dp$  (together with  all terms  that can be  collected as  an exact form).  
However, as  Sir James Lighthill argued,   vorticity is   more reliable than   pressure as the driving entity for   fluid motion \cite{Lighthill, Howe}.  While pressure often has enormous peaks,   vorticity varies smoothly. Moreover,
 for incompressible fluids the speed of sound  is
infinite (morally, this is the reason for the appearance of the Laplacian all over). 

We now write the Hodge projections more explicitly
in order to recover  the pressure.
Recall that
using that $L_{\nu^\sharp}\nu=di_{\nu^\sharp}\nu+ i_{\nu^\sharp}d\nu$,
our equation (\ref{Euler}) becomes
\begin{equation}
  \partial_t \nu +i_v d\nu=\partial_t \nu +i_v \omega=
  -d\left( p+ \frac{|v|^2}{2}\right):=-d h\,,\label{euler0}
\end{equation}
where: $v=\nu^\sharp$, $\omega=d\nu$,  and  $h$ is  the total enthalpy.
Taking the external derivative of equation (\ref{euler0})
we obtain the first equation
in our system (\ref{omegaeta}), namely
 $ \partial_t \omega +di_v\omega=0\,.  \label{euler2} $

 Let  $\Sigma$  a closed surface without boundary.  The   $L_2(\Sigma)$
inner product of two one-forms $\nu$ and $\sg$  on $\Sigma$ is defined as (Appendix B).
         \begin{equation}
      \langle \nu,\sg\rangle= \int_{\Sigma}  \nu\wedge \star \sg .
        \label{Un}
         \end{equation}

For concreteness, we now particularize for two dimensions.       Let   $a_1, b_1,\ldots,a_g, b_g $   a canonical  basis of the homology of $\Sigma$ as in section \ref{Riemrel},   and the dual basis  $\{\alpha_{1s},\ldots,\alpha_{gs},\beta_{1s},\ldots,\beta_{gs}\}$  of harmonic forms, where we added a dummy letter $s$.
      Using two dummy letters $r, s$   varying in $\Sigma \times \Sigma$, we  define the formal expression
         \begin{equation}
           B_{sr}=\sum_{j=1}^g\alpha_{jr}\beta_{js}-\beta_{jr}\alpha_{js}\,\,\,\, \text{ (Bergman type kernel)}
         \end{equation}

\begin{lem}
         The projection of any one-form $\nu$ according to the $L_2(\Sigma)$ inner
         product to the subspace of harmonic forms is
\begin{equation}
 \eta(r)  = (\pi_{har} \nu)(r) = \int_{\Sigma_s}\nu_s\wedge \star B_{sr} =\langle \nu_s,B_{sr}\rangle
                      \label{B2}
         \end{equation}
         \end{lem}
     If we apply this projection to both sides of  equation (\ref{euler0})
         and use that any exact form is in the kernel of the projection, then
         we obtain our second equation in the system (\ref{omegaeta}), namely
         \begin{equation}
  \partial_t \eta + \pi_{har} i_v \omega=0\,,\label{euler3}
\end{equation}
Equation  (\ref{euler3}) requires  the velocity $v$ (or equivalently $\nu=v^\sharp$).
Hence, in order  to close the system we need the   \textit{ third }
equation
\begin{equation}
  \nu=\eta+\delta \Psi\qquad\text{where}\qquad d\delta \Psi=\Delta\Psi=
  \omega\,,\label{euler4}
\end{equation}

  Taking the divergence of  equation (\ref{euler0})
we obtain:
\begin{prp}  The enthalpy (so the pressure) can be recovered via a Poisson equation,
\begin{equation}
\Delta h (= \delta d h) =  - \delta i_v \omega \, , \label{euler5}
\end{equation}
whose right hand side is  a  function of $\nu$ and
its derivatives.
\end{prp}
\begin{rem}
If we solve  this Poisson equation using the Green's function of the Laplacian, and  then substitute in
the original Euler equation, we obtain  an
integro-differential equation that must be equivalent to  the above splitting into three equations.
Note that this splitting
is determined by the Riemannian metric  on $\Sigma$ and the natural  inner product it
induces on the space of one-forms.
The vorticity 2-form is uncontroversial by  definition, but note that different projections  $\pi_{har} $ on the harmonic space, different from our  Theorem \ref{coupledeq}
 could  be more relevant on a specific problem.
 \end{rem}
 
 $$ $$

 \section{TORI}  \label{flatnonflattori}

\vspace{2mm}

We give  first a  proof  for  Proposition \ref{harmonic torus}:  {\it for flat tori the harmonic component is constant.}
There is nothing to to add to  studies on vortex streets and lattices such as 
\cite{Tkachenko, Neil, StremlerAref,Stremler, crowdyrect, KilinArtemova}. 

However, for non-flat metrics  the situation  will change drastically!  
For a single vortex the phase space has dimension four. Two coordinates correspond to the vortex position and the other two represent the harmonic flows on the torus. Works like \cite{Green, Sakajo, Sakajo-a, guenther} could be supplemented,  considering the interplay with the  harmonic components.

In section \ref{non-flatchosen} we present a simple but  generic  example, showing that in this case the exchange vortices-harmonic cannot be ignored.  This example serves as a a preview  for  a more detailed  study on vortex motions on tori with non-flat metrics that vorticists\footnote{\textit{Vorticists} is the name coined by H. Aref  \cite{Arefob} for the community.  An interesting artistic movement  in the early XX century used the same name \url{https://www.tate.org.uk/art/art-terms/v/vorticism} .}  could be interested in pursuing.   

For instance, looking at metrics with $ S^1$ symmetries such as those listed in the end of the section. For which a single vortex these systems are integrable, but some of  these symmetries are not  immediate to visualize.

\subsection{Uniformization of tori. A convenient change of basis of 1-forms}
\vspace{1mm}

The required data for the problem are  the modulus $\tau$ for  the  conformal structure on the torus and the conformal factor $\lambda$ to the flat metric.  We use  complex variable notation.
 For  flat tori  the Green function depends only on the relative positions
between $z$ and  $w$     (see e.g.  \cite{Stremler}, p. 3),
$$
G(z;w)=G_{\rm flat}(z-w) . $$
As an immediate consequence, the Robin function is constant.
The expression for $G_{\rm flat}$ in terms of elliptic functions can be found in several sources, see e.g.  \cite{chineses}, where the authors  describe the equilibrium curves for  $G_{\rm flat}(z-w)$ as a function of the modulus $\tau$.

\newpage

 Every Riemannian two torus is conformally equivalent to a flat
         torus ${\mathbb T}_{a\,b}:=\R^2 /(\vec a\,\Z+\vec b\,\Z)$, where the generators of the torus lattice are
        \begin{equation}
         \vec a=a_x\partial_x+a_y\partial_y \,\,\, {\rm and} \,\,\,
         \vec b=b_x\partial_x+b_y\partial_y.
         \end{equation} 
        
        One  usually takes $a=1$, and $b = \tau \in \C$ is the \textit{modulus}, but
       we prefer  to have the flexibility to fix the volume to be equal to 1.
         The volume form    is given by
         \begin{equation} \mu= \rho \, dx\wedge dy ,\,\, \rho = \lambda^2 > 0
         \end{equation} where
         $ \lambda $ is the conformal factor to the flat metric, 
         \begin{equation} ds^2 = \lambda^2 (dx^2 + dy^2) . \end{equation}
         Thus we  assume  without loss of generality that the volume of both metrics on ${\mathbb T}_{a\,b}$ are equal to 1, 
         \begin{equation}  \label{normtorus}
          V=\int_{{\mathbb T}_{a\,b}} \mu=1\quad\text{and}\quad   a_x b_y-a_y b_x=1\,.
           \end{equation}
           The generators of the homology of the torus are the curves
           $$a:=\{t\, \vec a: t\in[0,1)\} \,\,\, {\rm and} \,\,\, b:=\{t\,\vec b: t\in[0,1)\} . $$
           The basis of harmonic one forms $\{\alpha,\beta\}$
           such that $$\int_a\alpha=\int_b\beta=1 \,\,\, {\rm  and}\,\,\,
           \int_a\beta=\int_b\alpha=0$$ is given by
           \begin{equation}
             \alpha=b_y dx-b_x dy\,,\qquad \beta=-a_y dx+a_x dy\,,\label{abexp}
           \end{equation}
           which implies
            \begin{equation}  \label{alphabetadxdy}
            \begin{split}
            &  \star \alpha=b_x dx +b_y dy\,,\qquad \star \beta=-a_x dx-a_y dy\, \\ & \, \\
           & \,\, \alpha \wedge \beta = \star \alpha \wedge \star \beta =  dx \wedge dy
           \end{split}
           \end{equation}

  It is convenient  to use the coordinates  $(\eta_x,\eta_y)$  for the harmonic forms given by
 \begin{equation} \label{etatorus}
\eta= \eta_x dx+\eta_y dy \,\,\,(= A\alpha+B\beta).
  \end{equation}
  \vspace{-1.2mm}
\begin{lem} In these coordinates
 \begin{equation}
 H_{harm}   = \frac{1}{2} \big(\eta^2_x+\eta_y^2\big) .
  \end{equation}
  \begin{equation} \label{changeeta}
 \begin{split}
    A&=a_x \eta_x+a_y \eta_y\,,\quad B=b_x\eta_x +b_y\eta_y\,, \\ \, \\
    \eta_x&  =A b_y - B a_y\,,\quad  \eta_y = - A b_x  + B a_x  
    \end{split}
    \end{equation}
    \end{lem}

The  Hamiltonian system in Theorem \ref{Hamstru} particularizes to
\begin{prp}
\begin{equation}
 H=\sum_{j=1}^N\frac{\Gm_j^2}{2}R(s_j)+\sum_{j=1}^N \sum_{k\ne j}^{N}
      \frac{\Gm_j\Gm_k}{2} G(s_j,s_k)
 +\frac{1}{2} \big(\eta^2_x+\eta_y^2\big)\,,\label{H2}
\end{equation}
with symplectic form
\begin{equation}
\Omega  =  \sum_{j=1}^N \, \Gamma_j \mu_{s_j} - \frac{1}{\Gamma} \left(dA - \sum_{k=1}^N \Gamma_k \beta_{s_k}  \right) \wedge  \left(dB + \sum_{\ell=1}^N \Gamma_{\ell} \alpha_{s_{\ell}}  \right)
\end{equation}
where:
\begin{equation}
\begin{split}
& \mu_{s_j} = \rho(x_j,y_j) dx_j \wedge dy_j,\,
   dA =   \, a_x d\eta_x + a_y d\eta_y, \, dB = b_x d\eta_x + b_y d\eta_y,  \\ \, \\
&  \alpha_{s_j} =    b_y dx_j - b_x dy_j,  \,\,\,
 \beta_{s_j} = - a_y dx_j + a_x dy_j .
 \end{split}
 \end{equation}
\end{prp}

After some manipulations, one verifies that the equations of motion are given by 
    \begin{equation} \label{etaN}
    \begin{split}
           &  \,\,\,\,\,\,\,\,\,\,\,  \dot \eta_x-\sum_{j=1}^N\Gamma_j\dot y_j=
               -\Gamma\eta_y\,\,\,\,,\,\,\, 
   \dot \eta_y+\sum_{j=1}^N\Gamma_j\dot x_j=
               \Gamma\eta_x  \\
 &  \rho(x_j,y_j)\, \dot x_j= \ \, \frac{\partial }{\partial y_j}\left\{\frac{\Gamma_j}{2} R(x_j,y_j)+
\sum_{k\ne j}^{N}\Gm_k G(x_j,y_j;x_k,y_k)\right\}
                +\eta_x 
                \\
 &   \rho(x_j,y_j)\, \dot y_j= -\ \, \frac{\partial }{\partial x_j}\left\{\frac{\Gamma_j}{2} R(x_j,y_j)+
\sum_{k\ne j}^{n}\Gm_k G(x_j,y_j;x_k,y_k)\right\}
                +\eta_y  
                \end{split}
                \end{equation}

Note that when  $\Gamma=\sum_{j=1}^N \Gm_j=0$  
one gets   the conservation laws
$$\eta_x-\sum_{j=1}^N\Gamma_j y_j= \text{const.}\,\,\,\,,\,\,\,  \eta_y+\sum_{j=1}^N\Gamma_j x_j= \text{const.}
$$

\subsection{Flat tori: the Robin function and the harmonic field  are constants} 
\vspace{2mm}

As we mentioned, the Robin function is constant because the Green function depends only  on the difference of the coordinates.
The equations become:
     \begin{equation}
     \begin{split}
           &    \dot \eta_x-\sum_{j=1}^N\Gamma_j\dot y_j=
               -\Gamma\eta_y \,\,\,,\,\,\,
   \dot \eta_y+\sum_{j=1}^N\Gamma_j\dot x_j=
               \Gamma\eta_x\\
 &    \dot x_j= \ \, \frac{\partial }{\partial y_j}\left\{
\sum_{k\ne j}^{N}\Gm_k G(x_j,y_j;x_k,y_k)\right\}
                +\eta_x \\
&  \dot y_j= -\ \, \frac{\partial }{\partial x_j}\left\{
\sum_{k\ne j}^{N}\Gm_k G(x_j,y_j;x_k,y_k)\right\}
                +\eta_y
                \end{split}
    \end{equation}

\vspace{3mm}
\begin{lem}
    If the metric is flat the following identity holds:
        \begin{equation} \label{equali} \sum_{j=1}^N \frac{\partial }{\partial y_j}\left\{
            \sum_{k\ne j}^{n}\Gm_j\Gm_k G_{\rm{flat}}(x_j,y_j;x_k,y_k)\right\}=
        \sum_{j=1}^n \frac{\partial }{\partial x_j}\left\{
\sum_{k\ne j}^{N}\Gm_j\Gm_k G_{\rm{flat}}(x_j,y_j;x_k,y_k)\right\}=0\end{equation}
\end{lem}
   \proofn   An easy consequence of the
translation invariance in the period lattice, $$
G(z,w)=G_{\rm{flat}}(z-w) .$$
This  implies
    \begin{equation*}
    \begin{split}
        & \dot \eta_x=\sum_{j=1}^N\Gamma_j\dot y_j
               -\Gamma\eta_y\,\,\,,\,\,
   \dot \eta_y=-\sum_{j=1}^N\Gamma_j\dot x_j+
               \Gamma\eta_x\,\,\,,\,\, \\
   &
   \sum_{j=1}^N\Gamma_j\dot x_j=
                \Gamma \eta_x\,\,, \,\,
                  \sum_{j=1}^N\Gamma_j\dot y_j= \Gamma\eta_y .
  \end{split}   \end{equation*}
 
Therefore, as we wanted to show, 
    \begin{equation*}
      \dot\eta_x=\dot\eta_y=0 \quad\Longrightarrow \quad \eta={\rm constant\,.}
      \end{equation*}
\qed

\subsection{A typical non-flat torus metric: a single vortex system is non-integrable} \label{non-flatchosen} 
\vspace{0.5mm}

We will now show, by means of an example, that if the metric is non-flat, there will be interplay  between the vortices and harmonic fields. 
\begin{exa}
The square torus $\R^2/(\Z\times\Z)$ with this very simple conformal factor 
     \begin{equation} \label{exaomega}
      \rho(x,y)= \lambda^2= 1+\frac{1}{4}\Big(\cos(2\pi x)+\sin(2\pi y)\Big)\,.
     \end{equation}
     \end{exa}
  
We use the notation of section \ref{GreenRobinconformal}:  $|\Sigma|_\rho=1$ and $\overline \rho=1$. 
     The solution to $\Delta\phi=\rho-\overline \rho$ is immediate
  \begin{equation}
       \phi=\Delta^{-1}(\rho-1)= \frac{1}{16\pi^2} \Big(\cos(2\pi x)+\sin(2\pi y)\Big) = \frac{1}{4\pi^2}(\rho-1).
       \end{equation} 
     Since for the flat torus the Robin function is constant, we obtain from Baernstein Lemma \ref{Baernsteinlemma} 
\begin{equation}
    R_{\rho} =   \frac{1}{4\pi}\log\rho - \frac{1}{2\pi^2}(\rho-1) + {\rm const}\,\,\, \text{($\rho$   given by (\ref{exaomega}))}.
\end{equation}

 We now anticipate   the numerical results for
     the equations of motion  (\ref{motion1v}) in the next section. 
 
     Using the convenient coordinates $(\eta_x, \eta_y) $ for the cohomology, 
we found that there are four equilibria, all of them with $\eta_x=\eta_y=0$ and
$(x,y)$ equal to critical values for $\rho$, actually:\\

$(0,1/4)$ (maximum of $R$),  

$(0,3/4)$ and  $(1/2,1/4)$
(saddles of $R$), 

$(1/2,3/4)$ (minimum of $R$).\\

For values of $H$ greater than
$R(0,1/4)\approx 0.007$ the energy level is a 3-torus.
The equations  were numerically integrated for several initial conditions the level set $H=0.12754$
satisfying: 

\centerline{$x(0)=0$, $y(0)=0.5$, $\sqrt{\eta_x^2(0)+\eta_y^2(0)}=0.5$, and
$\eta_x>0$.}

\begin{figure}[h]
    \includegraphics[width=0.65\textwidth]{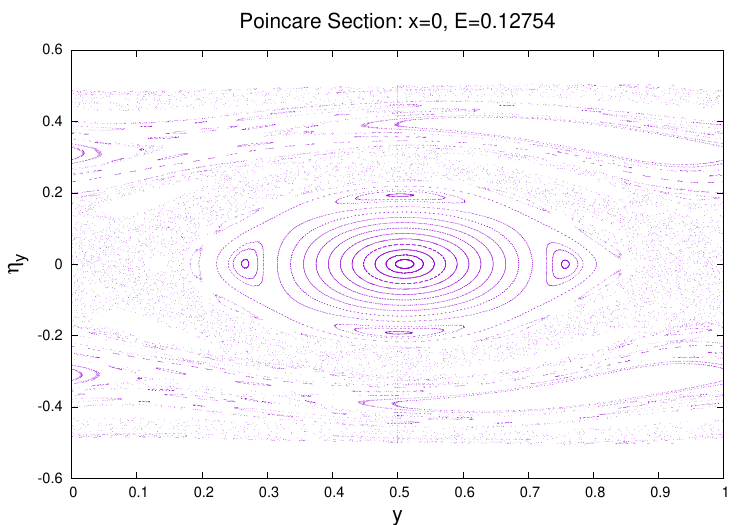}
\caption{Poincar\'e  section 
defined by
$x=0\mod 1$, $\eta_x>0$, and $H=0.127. $ The surface is parameterized by the
variables $(y,\eta_y)$. Some initial conditions
are contained in invariant
surfaces while others wander in a portion of the phase space with positive
volume, so the only integral is the Hamiltonian.}
\label{figtorus}
\end{figure}

In Figure \ref{figtorus} we show the points where the solutions generated by
these initial conditions intersect the  Poincar\'e section  defined by
$x=0\mod 1$, $\eta_x>0$. 
The surface is parameterized by the
variables $y,\eta_y$. We clearly see that some initial conditions
are contained in invariant
surfaces while others wander in a portion of the phase space with positive
volume, and  so the only first integral is the Hamiltonian
function.

          \subsection{Robin function  for an arbitrary  torus metric } \label{moreBaerst} 
          \vspace{1.5mm}

    The general expansion for a function $\rho$ in  $\mathbb T_{a\,b}$ with average $\bar{\rho} = 1$ is
           \begin{equation}
           \rho =  \lambda^2(x,y)=1+\sum_{(m,n)\ne (0,0)} c_{mn}\exp\{i\,2\pi[m \alpha(\vec x)+
            n\beta(\vec
            x)]\}\,,\,\,\,\,  \text{where}\ \vec x=(x,y)   \label{lb2} 
          \end{equation}
          where we  recall  that    we assumed (\ref{normtorus}) with $   \alpha (\vec a)=1, \,\beta(\vec b)=1\,,\, \alpha(\vec b)=
            \beta(\vec a)=0\, $.

        From Baernstein's  lemma we get   the    Robin function of the  non-flat torus  
                   \begin{equation}
               R=\frac{1}{4\pi}\log \rho + 2 \phi+ const\,,\label{R}
             \end{equation}
             where $\phi$ is the solution to the equation on
             $ \mathbb T_{a\,b}$
             \begin{equation}
               \frac{\partial^2 \phi}{\partial x^2}+
               \frac{\partial^2 \phi}{\partial y^2}=\rho -1\,\quad\text{with}
               \quad\int_{ \mathbb T_{a\,b}}\phi \,dx\wedge dy=0 .
              \label{Laplamb}
          \end{equation}

\begin{prp}
          \begin{equation}
  \phi(x,y)=\sum_{(m,n)\ne (0,0)} \phi_{mn}\exp\{i\,2\pi[m \alpha(\vec x)+
            n\beta(\vec
            x)]\}\,,\label{phif}
          \end{equation}
          \vspace{-1.5mm}
          \begin{equation}
            \phi_{mn}=-\frac{c_{mn}}{4\pi^2}\left\{
            \big(m,n\big)\underbrace{\left(
\begin{array}{cc}
 b_x^2+b_y^2 & -a_xb_x-a_yb_y \\
 -a_xb_x-a_yb_y & a_x^2+a_y^2 \\
\end{array}
\right)}_{:=M}\left(\begin{array}{c} m \\ n\end{array}\right)\right\}^{-1}\,.
\label{phimn}
\end{equation}
\end{prp}
\proofn   The Poisson equation (\ref{Laplamb})  separates. Remarkably, $M$ expresses the  Riemann relations.  \qed

In conclusion, given the conformal factor, the Robin function in equation
(\ref{R}) can be explicitly computed using the Fourier expansion  of $\phi$ given by (\ref{phif}, \ref{phimn}).

\subsection{Equations of motion for one vortex  of unit intensity} 
\vspace{1.5mm}
  
           Let $(x, y)$ denote the position of the vortex   and
 $ \eta=A\alpha+B\beta=\eta_x dx+\eta_y dy$  (see (\ref{etatorus}, \ref{changeeta}))
 denote the harmonic form that is dual to the harmonic part of the
 velocity field of the fluid.
Applying Theorem \ref{Hamstru} we have for the phase space 
$\mathbb T_{a\, b}\times \R^2$ that\\ $\,$ \\  
i) The Hamiltonian function  is
 \begin{equation}
 H=\frac{R(x,y)}{2}+\frac{1}{2} \big(A,B\big)\left(
\begin{array}{cc}
 b_x^2+b_y^2 & -a_xb_x-a_yb_y \\
 -a_xb_x-a_yb_y & a_x^2+a_y^2 \\
\end{array}
\right)\left(\begin{array}{c} A \\ B\end{array}\right)
\end{equation}
ii) The symplectic form is
\begin{equation} 
   \Omega=\rho dx\wedge dy -(dA-\beta)\wedge(dB+\alpha)\,.\label{Omega0}
  \end{equation}
 From (\ref{abexp}, \ref{etatorus}) we get
   \vspace{2mm}

   \begin{teo}
    In the variables $(x,y,\eta_x,\eta_y)$:
  \begin{equation}
  \begin{split}
 H & =\frac{1}{2} [R(x,y)  + \big(\eta^2_x+\eta_y^2\big)]  \,\,, \,\,\,\\
  \Omega & = (\rho-1) dx\wedge dy -d\eta_x\wedge d\eta_y
  +dx\wedge d\eta_x+dy\wedge d\eta_y \label{Omega1} 
    \end{split}
  \end{equation}
It is readily seen that equations of motion become (a special case of (\ref{etaN}))
           \begin{equation}      \label{motion1v}
                    \begin{split}
                \rho\, \dot x&= \ \, \frac{1}{2}\frac{\partial R}{\partial y}
                +\eta_x \,\,\,,\,\,\,
                \rho\, \dot y = -\frac{1}{2}\frac{\partial R}{\partial x}
                +\eta_y\\
 \rho \dot \eta_x&= -\frac{1}{2}\frac{\partial R}{\partial x}
                -(\rho-1)\eta_y \,\,\,,\,\,\,
 \rho \dot \eta_y = -\frac{1}{2}\frac{\partial R}{\partial y}
 +(\rho-1)\eta_x
\end{split} 
\end{equation}
\end{teo}

\begin{cor} If the metric is flat, namely $\rho=1$, then the Robin function is constant
and 
          \begin{equation}
                \dot x=\eta_x\,,\quad
\dot y= \eta_y\,,\quad
 \dot \eta_x=0\,,\quad \dot \eta_y=0
\label{t3.5}\end{equation}
In this case the velocity of the vortex is equal to the harmonic
part of the fluid velocity. The vortex moves along
the winding geodesics of the flat metric, with constant velocity $(\eta_x, \eta_y)$. 
\end{cor}
    \vspace{1.5mm}
 \begin{prp}  The symplectic form $\Omega$ given by  (\ref{Omega0}) is exact.  
  \end{prp}
\proofn   Observe that $d(A\alpha + B\beta + A dB) = (dA-\beta)\wedge(dB+\alpha) + \alpha \wedge \beta$, and we know from (\ref{alphabetadxdy})  that  $\alpha \wedge \beta = dx \wedge dy$.  It remains to show that $(\rho-1) dx \wedge dy$ is exact. It integrates to zero, so it is a differential.  Equivalently,  $\rho-1$ is a sum of products of sines and cosines in $x$ and/or $y$. 
\vspace{1mm}
\begin{rem}
Something special should be happening in genus 1.  For the sphere, no matter the  metric, the vortex symplectic form (there is no harmonic)
cannot be exact. For genus $\geq 2$? 
\end{rem}

\subsection{Some  torus families} 

The tori in the classes listed below have   $S^1$ symmetry, and  therefore, the 1-vortex system will be  automatically integrable. However it could be not  so immediate to describe the $S^1$-action in the symplectic manifold $\T \times \R^2$ with the symplectic 1-form (\ref{Omega0}).    
Finding the action-angle variables  for them could be an interesting exercise\footnote{Various tori families are depicted in  \url{https://www.math.uni-tuebingen.de/user/nick/gallery/} .}. One could start with  the  tori of revolution  in $\R^3$ and their images under inversion about  an external sphere, the  \textit{cyclides}. \\

 Even more interesting are:\\

\noindent  \textit{Willmore tori}. Willmore surfaces are the critical points of the functional $\int_{\Sigma} \, (H,H) \,  \mu$, where $\mu$ is the area form of the induced metric and $H$ is the mean curvature vector. The celebrated Willmore conjecture \cite{Willmore} for tori in $S^3$ was solved in 2012  by Marques and Neves \cite{Coda}.  The literature is vast, see eg. \cite{Pinkall}, \cite{Magdalena}).  
M. Barros  and coworkers  \cite{Barros2001, Barros2014} have a methodology to get $S^1$ -invariant Willmore tori,  using Kaluza-Klein   conformal structures on  principal $S^1$ bundles. \\

 \noindent  \textit{Constant mean curvature tori}. In 1986 Wente (\cite{Wente}, 1986;  see also \cite{Abresch})  constructed the first constant mean curvature immersed torus in $\R^3$, re-opening a rich field of research stemming from the times of Hopf, Blaschke and Alexandrov\footnote{The only constant mean curvature surface of genus zero is a round sphere.}.  An embedded  CMC torus in $S^3$  is axially symmetric \cite{Andrews, Hauswirth}. \\
 
 \noindent \textit{Clifford-Lawson tori} \cite{lawson, Penskoi}.  Those are the minimal surfaces with flat metrics  parametrized by
 $  (x, y) \mapsto  (\exp(imx) \cos y,  \exp(inx) \sin y) \in S^3,$ $m,n$ relative prime. The usual Clifford's torus  is for $ m=n=1.$   Choosing a pole in $S^3$ and projecting to $\R^3$, one gets conformal metrics  
 in the images\footnote{See animations in \url{https://virtualmathmuseum.org/Surface/clifford_torus/clifford_torus.html .}}.\\
 
 \noindent \textit{Hopf tori.}  Let $S^3 \to S^2$  be the Hopf fibration. A Hopf torus, as defined by Pinkall \cite{Pinkall1}  is  the inverse image of any closed curve on $S^2$. \\
 
  \noindent \textit{Lagrangian tori.}  Those are immersions in a symplectic manifold such that the symplectic form vanishes on them. The minimal Lagrangian tori have been classified in $\C P^2$ and $\C P^3$ with the Fubini-Study 2-form \cite{Mironov}.\\
  
  \noindent \textit{Okikiolu tori.}  Steady vortex metrics constructed by K. Okikiolu \cite{Okikiolu} and explored in \cite{ragsub-a}.

\newpage

$$ $$

$$ $$

$$ $$

  \begin{figure}[h]
        \includegraphics[width=1\textwidth]{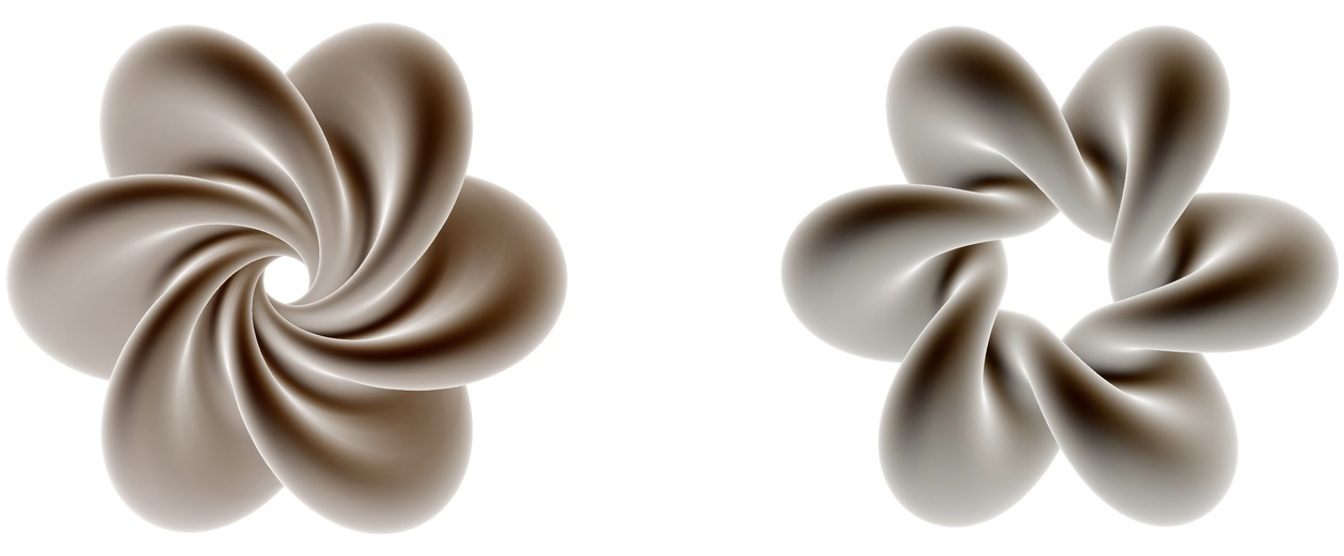}
\vspace{2mm}
\caption{Left:  6-lobed Hopf torus ;  Right: 6-lobed equivariant Willmore torus. The $S^1$ invariance is not easily recognized. Adapted from  GeometrieWerkstatt/Tuebingen.}
\end{figure}

\newpage

\section{VORTEX CRYSTALS IN THE HYPERBOLIC PLANE}  \label{crystals}

\vspace{1.5mm}

\begin{quote}
{\footnotesize
``As the theory is extended to
vortices on manifolds, we may hope that profound connections will arise between
mathematical entities not commonly thought to be related."
\cite{Arefcrystals} }
\end{quote}
 
\vspace{-2mm}

 The name \textit{vortex crystals} has been mainly used for  patterns in the plane or in the sphere that remain in equilibrium or move rigidly.  The collective crystal  motion is generated by an  infinitesimal translation or rotation. The latter  forming  rings of vortices. 
Vortex crystals have also been studied  on the universal cover of the cylinder
(periodic strip) and torus (periodic parallelogram), see for example the already mentioned  works \cite{Tkachenko, Neil, StremlerAref, Stremler, crowdyrect, KilinArtemova}.
Such configurations have  physical applications (for example in superfluids)  and can also model natural phenomena, for instance   Karman's street behind an obstacle (see the discussion in \cite{KMontaldi}).

In the final section of the above quoted paper \cite{Arefcrystals}
the authors suggest looking for  vortex crystals in Poincar\'e disk
$D$, the open unit disk with the hyperbolic metric  $2/(1-|z|^2)|dz|$ with constant negative curvature -1.  
They observed that,  as  in the planar case, due to the $S^1$ symmetry, a regular polygon of identical vortices is a relative equilibrium.
An  vortex may be placed in the center, and a special choice of an opposite strength will stop the rotation.
Yet, they did not develop further.

Let us put the problem in context. Upon uniformization, a compact Riemann surface can be realized geometrically as a quotient of  a tesselation of the Poincar\'e disk.
Let $\Gamma$ be a co-compact Fuchsian subgroup  of  $\C \cup \infty$ leaving the unit circle invariant\footnote{We found this M.Sc. thesis quite interesting and readable
\url{https://macsphere.mcmaster.ca/handle/11375/9044}. We apologize for  using the symbol $\Gamma$ for these discrete groups, since it  is traditional. We are sure that no  confusion will arise in this section with the same symbol being used for vorticities.}. 

A  compact fundamental domain  $\cal{F}$ for $\Gamma$  inside $D$ with $4g,\, g\geq 2$ sides  can be constructed, identifying the circular arcs $a_1b_1 a_1^{-1}b_1^{-1}  \cdots a_gb_g a_g^{-1}b_g^{-1}$.   The metric  of curvature -1  in the Poincar\'e disk descends to the closed Riemann surface $\Sigma = D/\Gamma$, which has genus $g \geq  2$.

The patterns  we consider in  the Poincar\'e disk consist of an infinite  periodic distribution  of vortices   invariant under 
  $\Gamma$. 
  The simplest ``vortex jelly" is formed by a single vortex on a fundamental domain,  together will all its replicas.  The motion is governed by the Robin function.  
  We will  give  examples of crystals  on the Poincar\'e disk  with the metric of constant curvature -1,  corresponding to systems of $N\geq2$ vortices on an underlying Riemann surface with discrete symmetries, as suggested in the final section of \cite{Ragazzo0}. The main  results are  given in   section \ref{extension}. 

\vspace{1.5mm}
 
 Suppose that on a fundamental domain $\cal{F}$ one has $N$ vortices.
The  motion of the infinite pattern in the Poincar\'e disk is  just the lift of the  Hamiltonian system of Theorem \ref{Hamstru} for $N$ vortices in the underlying  Riemann surface $\Sigma = D/\Gamma$,  coupled with the $2g$ dimensional harmonic component.  

Puzzling as it may seem at first sight, the harmonic part cannot be neglected in the dynamics,  although the universal cover $D$ is simply connected!
 Periodicity forces the appearance of harmonic forms,   in the Poincar\'e disk differentials of harmonic functions. 
 At any rate, it is readily seen from the ODEs (\ref{Guseq1}, \ref{Guseq2})  that at an equilibrium   the harmonic component  vanishes. 
For this reason,  for economy sake we  can,  and will,  work  in this section with the incomplete equations of \cite{KoillerBoatto}, ignoring  the harmonic contribution from  the complete system.  
 
  \begin{prp} \label{simplified}   Steady vortex crystals   are  the critical points of the   incomplete Hamiltonian
   $$\Omega_{\rm vort} = \sum_{j=1}^N  \Gamma_j \mu(s_j) \,\,,\,\, H_{\rm vort} = \sum_{j=1}^N\frac{\Gm_j^2}{2}R(s_j)+\sum_{j=1}^N \sum_{k\ne j}^{N}
      \frac{\Gm_j\Gm_k}{2} G(s_j,s_k) ,  \,\,\,\, (s_1, \cdots, s_N) \in  \Sigma \times \myeq \Sigma  . $$  
\end{prp}
\textit{Nota bene.}
To analyze the linear stability one needs to use the complete equations.
\vspace{1mm}

\begin{rem}
As far as we know   the existence of  \textit{ steady translating patterns}  on the Poincar\'e disk is an open problem\footnote{Translations  move points along  a geodesic at  a constant speed. However,   because of the failure of Euclid's fifth  postulate, most likely an  the extra requirement is needed for a  precise definition of  a steady translating pattern.}.
 For 1-vortex in the flat torus the corresponding  lattice in the cover moves along geodesics (parallel straight lines)  and the harmonic part is constant. For a steady vortex metric (section \ref{steadyvm}),
the Robin function is constant but even so Equations (\ref{Guseq1}, \ref{Guseq2})  for a 1-vortex still depend on the metric. If the surface has genus $g \geq 1$  how is this motion?  It seems that for the Okikiolu's torus \cite{Okikiolu}, that has an SVM, the motion of a single vortex
does not follow a geodesic.
\end{rem}

\subsection{Bolza's surface}  \label{Bolzasurf} 

\vspace{2mm} 

  Introduced by Oskar Bolza in 1887 (\cite{Bolzapaper}),  it is the   compact Riemann surface 
 with the highest possible order,   48 (with reflections 96),  for a conformal automorphism group in  genus 2. 
For the algebraic geometer viewpoint it is the Riemann surface of the equation $y^2 = x(x^4-1)$.

The Fuchsian group defining the Bolza surface is a subgroup of the group generated by reflections in the sides of a hyperbolic triangle with angles $\pi/2, \pi/3, \pi/8;$ i.e, the $(2,3,8)$ Schwarz triangle surface  \cite{Schwarztriangle}. 
The full symmetry group is generated by (see Figs. 3 and 4):

\vspace{1mm}
 
\indent i) rotation of order 8 about the centre of the octagon;\\
\indent ii) reflection in the real line;\\
\indent iii) reflection in the side of one of the sixteen  (4,4,4) triangles that partition the octagon;\\
\indent iv) rotation of order 3 about the centre of a (4,4,4) triangle.
\vspace{1mm}

  Fig. 3  shows the standard Dirichlet fundamental domain $\cal{F}$, a regular polygon with angles $\pi/4$ on the vertices.  In the left, the symmetry generators are shown. The geodesic triangle shown in the right   (sides $a, b$, and half of a side of the octagon) is a sub-fundamental domain in ${\cal F}$.
The angles in the figure are $p=\pi/8$ and $q=\pi/4$.
The Euclidean lengths of the segments $a, b$ are 
 $$E(a) = (2^{1/2} - 1)^{1/2}, \,\,\, 
  E(b)=2^{-1/4} . $$    The curvilinear sides belong to circles centered externally that  meet the unit disk at right angles  having radius 
    $E(c)=\sqrt{(\sqrt{2}-1)/2}. $ The hyperbolic lengths of the segments $a$ and $b$ are
    $$\ell(a)=2\arctanh E(a), \,\,\,\, \ell(b)=2 \arctanh E(b). $$

\vspace{1.5mm}
 \subsection*{Robin function of Bolza's surface} 
   
\vspace{1.5mm} 
  
    The level lines of Robin function $R$ on Bolza's  surface were computed  
in \cite{Ragazzo0}.  Only after  the 46 equilibria were located numerically,  it was  perceived  that they were fixed points for  involutions from the Bolza's surface automorphism group (both orientation preserving and reversing).     Six among the equilibria are the  Weierstrass points of Bolza's surface.

 Using only symmetry considerations, it was argued  that one $+1$  vortex  in the center of the octagon, surrounded by eight opposite ($-1$) vortices, together with all the replicas in the disk, should be in equilibrium.
Although one may get  some ``feeling" of imbalance, this configuration is, indeed, the lift to the universal cover of a system of two opposite vortices placed at symmetric points (as if they were north-south poles)
on the Bolza surface\footnote{It is similar to the lift to the universal cover of
an equilibrium position of a pair of opposite vortices on the Schottky
double of a planar domain.}.
\vspace{1.5mm}

  In the next section we recall the definition of discrete symmetries of a Riemann surface
 and summarize the  reasonings in \cite{Ragazzo0} about their role  in finding equilibria.

 \bigskip \bigskip

     \begin{figure}[h]   \label{fundamdomainbolza}
         \includegraphics[scale=0.5]{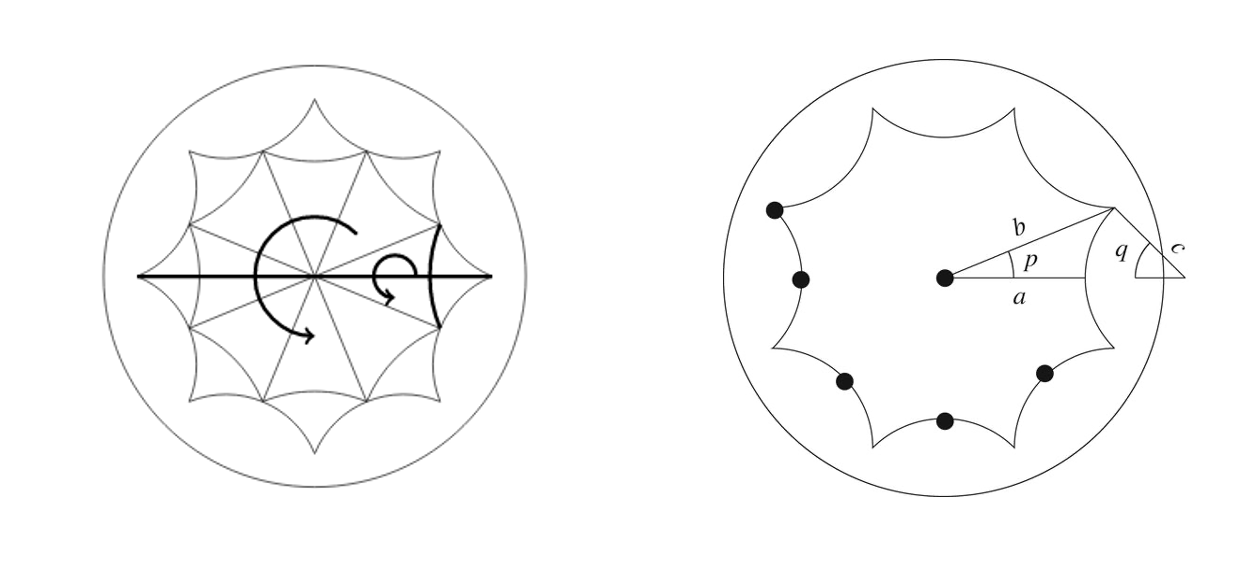}
\caption{Left. Symmetry generators of Bolza's surface (\url{https://en.wikipedia.org/wiki/Bolza_surface}).
Right.  The eight vertices of the octagon   represent the same point in $\Sigma$ and opposite sides represent   the same geodesic arc.
$z \mapsto -z$ is the hyperelliptic involution  fixing  the six marked Weierstrass points.}
\end{figure}

   \begin{figure}[!ht] \label{bolzasurf}
 \centering
\includegraphics[scale=0.36]{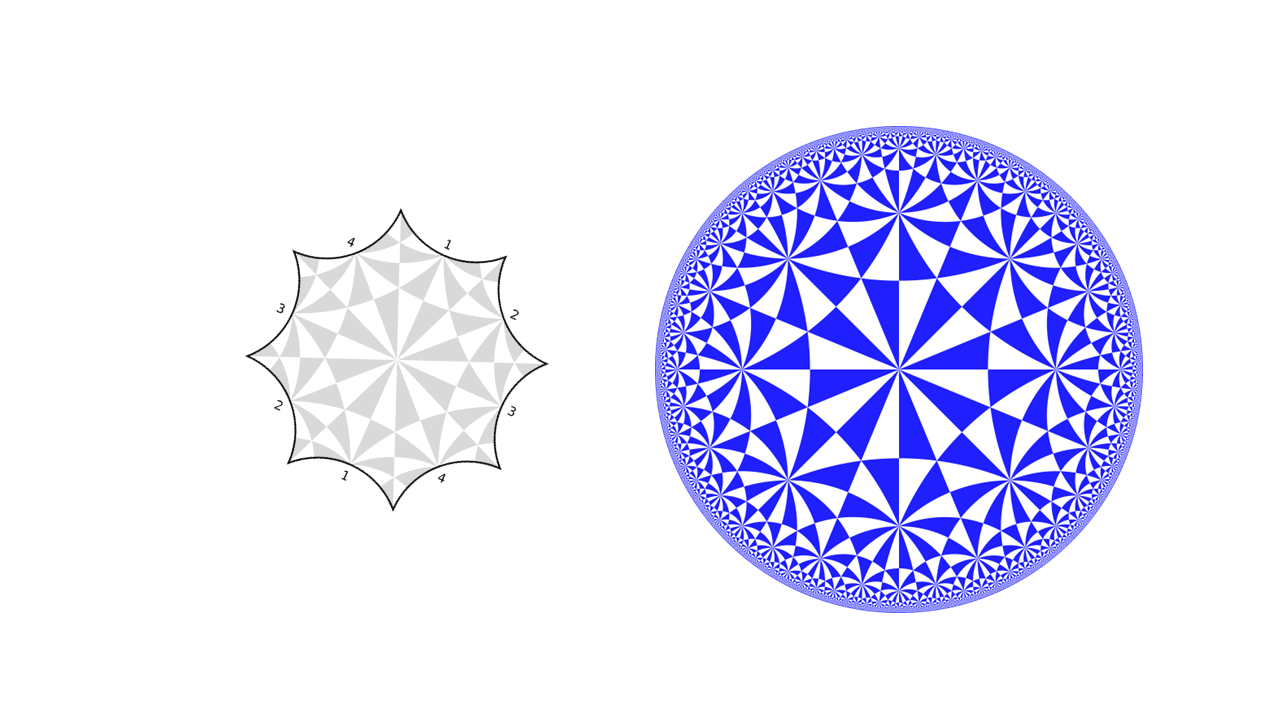}
\vspace*{-1cm}
\caption{The tesselation of the Poincar\'e disk.   In Fig. 4, the reader can count the number of Schwarz triangles in the fundamental domain.  See  \cite{Balazs}, Fig. 16, for more details.
} 
  \end{figure}

\newpage

$$ $$  $$ $$
  
\begin{figure}[!ht] \label{fig2}

\includegraphics[scale=0.4]{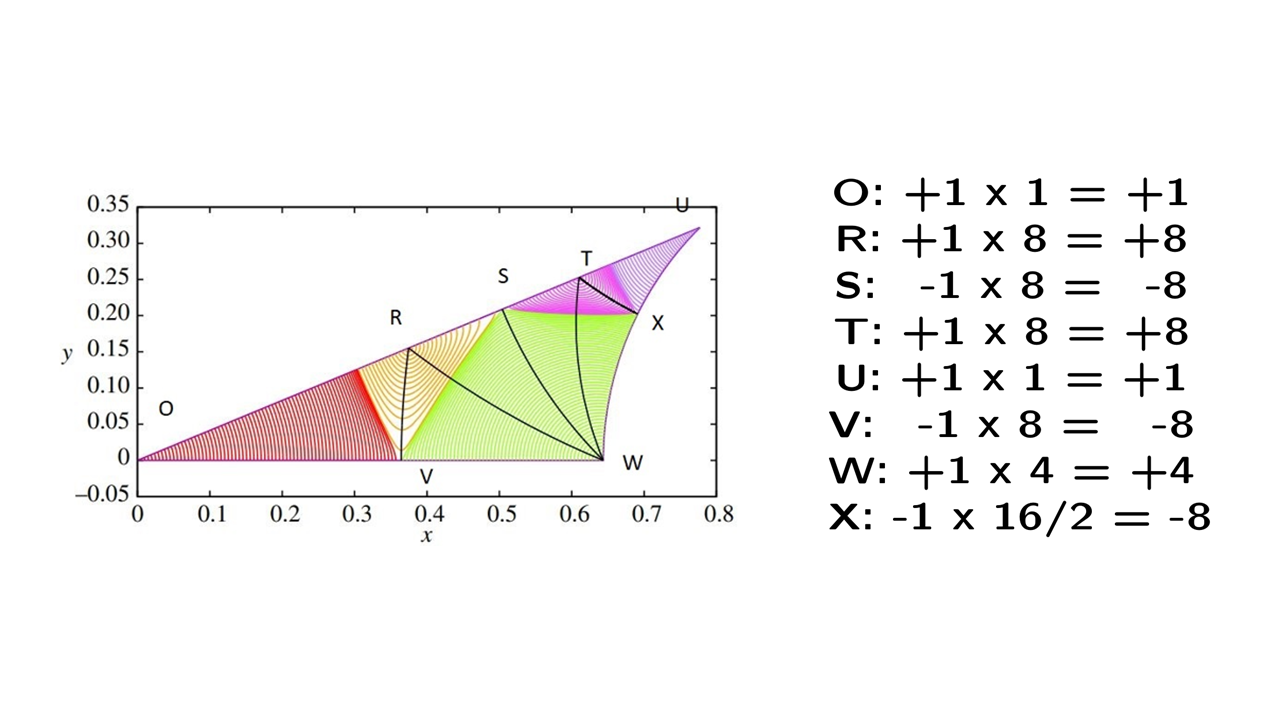}

\includegraphics[scale=0.65]{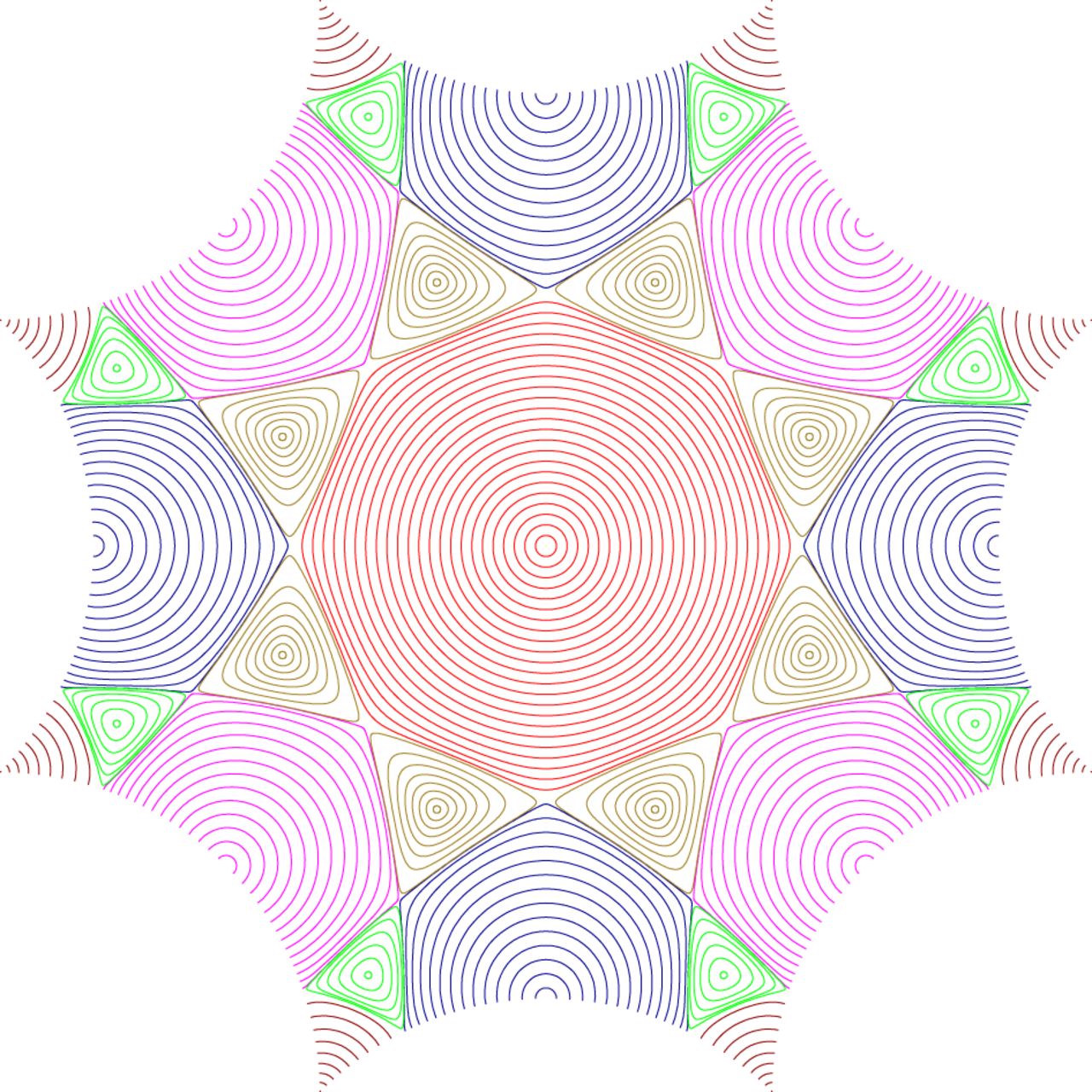}

\caption{The total number of critical points of Robin function is 46. The sum of indices is  $-2$, as predicted by   Euler's formula   $  2 - 2 \, g$.    Weierstrass points are marked by the letters  $O$, $U$, and  $W$.}

\end{figure}

\newpage

\subsection{ Symmetries and involutions, Weierstrass points, hyperelliptic surfaces,}

\vspace{1mm}

\begin{definition} A M\"obius transformation (also orientation reversing ones)  $\sigma$ is a {\it symmetry} of a Riemann surface  $ \Sigma =  D/\Gamma$  if
\begin{equation}    \sigma^{-1} \gamma \sigma \in \Gamma \,\,\,, \,\forall \, \gamma \in \Gamma .
\end{equation} 
\end{definition}
\vspace{-1.5mm}  
\begin{lem}
A symmetry $\sigma$ 
 of  $ \Sigma = D/\Gamma $  preserves  both the Green and Robin functions.
 \end{lem}

Since the area form is preserved when $\sigma$ is orientation preserving and changes sign when it is reversing, we have for any symmetry $\sigma$ an immediate consequence:  
\begin{cor} \label{invHamil} (\cite{Ragazzo})
 For the incomplete 1-vortex  
vector field 
 $ X = X_R\,\,, 
i_X \mu  = \frac{1}{2} dR,  $  
\begin{equation}
 \sigma^* X_R = \left\{  \begin{array}{l}  +X_R  \,\,\,  \text{(when   $\sigma$  is orientation preserving)} \\
  - X_R  \,\,\,   \text{(when $ \sigma $ is orientation reversing). }\end{array}
   \right.
\end{equation}

This also holds for the incomplete systems $X_H$  with  any number of vortices.
\begin{equation}
 \sigma^* X_H = \left\{  \begin{array}{l}  +X_H  \,\,\,  \text{(when   $\sigma$  is orientation preserving)} \\
  - X_H  \,\,\,   \text{(when $ \sigma $ is orientation reversing). }\end{array}
   \right.
\end{equation}
\end{cor} 
\proofn  In view of Proposition  \ref{simplified}  we do not have to worry about the harmonic part.  The fact that the lemma works also 
for $N \geq 2$   is due  to the invariance of both $R$ and $G$ under symmetries.   
The symplectic form is a combination of area forms, and the Hamiltonian vector field $X_H$ comes from $R$'s and $G$'s.  An orientation reversing symmetry of the surface  $\sigma$ is  anti-symplectic, it
reverses the sign of the symplectic form. 
Hence it  generates a time-reversible transformation in
phase space.   \qed 

\noindent There is just a  notational subtlety: 
the parity of $ \sigma^* X_H$ is at the individual component $\Sigma$, not in the product space (for which an  orientation  reversing for $\Sigma$ would become also volume preserving  in the product for an even number of vortices). 

\begin{rem} If one wants to extend the lemma to the complete system   there is a point to be careful about. A symmetry transforms harmonic forms into harmonic forms. However, the  symmetry could mess up the intersection numbers.  For example, in the square torus one of the symmetries is the rotation by 90 degrees. It changes $dx$ into $ dy $  and $dy$  into $-dx$.  This issue could certainly be addressed, but we do not have the need  to do it, at least for now.
\end{rem}

\vspace{2mm}

Recall the definition and main facts about Weierstrass points (see e.g. \cite{FarkasKra} for the proofs).   
\begin{definition} A point $s$ on a closed Riemann surface of genus $g \geq 2$ is a   \textit{Weierstrass  point}  if there is a non-constant meromorphic function that has just one pole, at $s$, of  order less or equal $g$ (no Weierstrass points exist for $g=0$ and $1$).
\end{definition} 
\vspace{-2.5mm} 
For $g \geq 2$ there are  at least $2g+2$ Weierstrass points, and no more than $g(g^2-1)$. 

\begin{definition}
   Hyperelliptic surfaces are Riemann surfaces that are two-sheeted coverings of the
Riemann sphere.
\end{definition} 
\vspace{-2.5mm} 
 
 They admit an automorphism $J$, called the hyperelliptic involution,
that interchanges the two sheets (and   is central in the automorphism group).

Many  authors restrict the term hyperelliptic for genus $g \geq 2$.
By the Riemann-Hurwitz formula there are $2g + 2$  branch points on the
surface which are fixed by $J$. These are the precisely the  Weierstrass points. 
Every surface of genus $2$ is hyperelliptic  with 6 Weierstrass points (\cite{FarkasKra}, III.7.2).

\begin{exa}
 The reader could confirm by examining  Fig. 5  that  the total number of critical points  of Robin function of Bolza surface is $46$, with total sum of indices equal to  $-2$, as predicted by   Euler's formula   $  2 - 2 \, g$.
 Weierstrass points  (total 6) are marked by  $O$, $U$, and  $W$, fixed by $J: z \mapsto -z$.
\end{exa}

The next  lemma comes from direct computations in the group of M\"obius transformations.

\begin{lem} \label{symD} (\cite{Ragazzo0})
Let   $\sigma$   be  an involution ($\sigma = \sigma^{-1}$) of $\Sigma = D/\Gamma$   with   a fixed point.\\ \vspace{1mm}
(a) Orientation preserving:  $\sigma$ is either the identity or has the form 
$$    \sigma:\,  z \to  \frac{az+b}{\bar{b} z + \bar{a}},\,\,\, |a|^2 - |b|^2 = 1 \,\,{\rm and} \,\, a = \pm i|a|.
$$
Moreover, if the fixed point is at $z = 0$ and $\sigma$  is not the identity, then $\sigma(z) = - z$.\\ \vspace{1mm}
(b) Orientation reversing:  $\sigma$ is of form 
$$   \sigma:\,  z \to  \frac{a \bar{z}+b}{\bar{b} \bar{z} + \bar{a}},\,\,\, |a|^2 - |b|^2 = 1 \, \,,\,\, b = \pm i|b| \,.
$$
Moreover, if the fixed point is at $z = 0$, then  $ \sigma(z) =  e^{i \theta} \bar{z}$.  
\end{lem}

The next results hold  for a \textit{single}  vortex
and the constant curvature hyperbolic metric. 
For a single vortex only the Robin function appears. 
Equilibria are critical points for  the Robin function with the components of the harmonic vector field equal to zero.

 \begin{teo} (Theorem 3.2 in \cite{Ragazzo0}, case a)) \label{symprop}\\ \vspace{1mm}
 Let $ \sigma $ be an involutive orientation preserving symmetry of
$ \Sigma = D/\Gamma $ that is not the identity. Any fixed point  $ s_* $ of $ \sigma$ is also a singularity of the single vortex system.  In particular, Weierstrass points of hyperelliptic surfaces are fixed points of the 1-vortex system.
\end{teo}
 \begin{teo} (Theorem 3.2 in \cite{Ragazzo0}, case b)) \label{symprop1}\\ \vspace{1mm}
  Let $ \sigma_1 $  and $ \sigma_2 $  denote two different orientation reversing symmetries of $ \Sigma = D/\Gamma$. Suppose that $ s_*  $ is a fixed point of both $ \sigma_1 $  and $ \sigma_2 $. Then $ s_* $ is also singularity of the 1-vortex velocity field.
 \end{teo}

\noindent \proofnsymprop  

\noindent  An important preliminary  (*): Without loss of generality, we can assume $ s_*  = 0.$
 The reason goes as follows:  let $ h :  D  \rightarrow   D $  be  a M\"obius transformation such that $ h(s_*) = 0. $  Then
 $$ \Sigma =  D /\Gamma  \equiv \tilde{\Sigma} =  D /h \Gamma h^{-1}   $$
and $\tilde{\sigma} = h \circ \sigma \circ h^{-1}  $  is
an involutive symmetry of $ \tilde{S}$.  If $\sigma(s_*) = s_*$ then   $ \tilde{\sigma}(0) = 0. $  \qed 

We now  remove the tildas. 
\vspace{1mm}

\noindent  Case (a).   As $ \sigma $ is not the identity, Lemma \ref{symD}  implies that $  \sigma(z) = -z $. The Robin function is invariant under isometries of the constant curvature hyperbolic metric;  therefore, it is even:
 $$   R(z) = R(-z)   \,\,\,\  \Rightarrow    \,\,\,   dR(0) = 0    \,\,\,  \Rightarrow    \,\,\,    X(0) = 0 \, . $$
\noindent Case (b).   If  $ \sigma $ 
 is orientation reversing, then 
 $$ \sigma_*X(0) = -X(0).   $$
So
  $ 
   \,  (\sigma_1)_*(0) X(0)  = (\sigma_2)_*(0) X(0)  = -X(0) .\,
    $  Again we  know by  Lemma \ref{symD} that
 $$   \sigma_1(z) = e^{i\theta_1} \bar{z}\,  ,  \, \sigma_2(z) = e^{i\theta_2} \bar{z}\,\,, \,\, \theta_1 \neq \theta_2  . $$

  The 1-dimensional eigenspaces of $  (\sigma_1)_*(0) $  and $ (\sigma_2)_*(0) $ associated with the eigenvalue -1 are transversal. But they have the same eigenvector. Hence  $ X(0) = 0.  $ \qed

Note that case (b) cannot be reduced to (a) by composing $\sigma_1 \sigma_2$ because we cannot guarantee it remains an involution.  For the relation of fixed points of orientation preserving involutions and their relation with period matrices, see \cite{Gilman}. For bounds on the number of fixed points by an automorphism and its relation with simple closed geodesics,
see \cite{Schaller, Haas, Parlier}  and references therein.

\subsection{All the equilibria of   Bolza's
 surface Robin function come from  cases (a) or (b)}

We know that each equilibrium, together with all of its images,   forms   a steady  vortex crystal in $D$.
We claim  that the 46  equilibria  in the fundamental domain  found numerically  in \cite{Ragazzo0}, all of them,    correspond to the situations (a) or (b).  Let us check this out.  

Referring to Figs. 2-4  and  6-8 in \cite{Ragazzo0}, or Figs. 3-5 
 here, the orientation preserving symmetries of Bolza are generated by the eight  $z \mapsto e^{\ell i \pi/4} z, 0 \leq  \ell \leq 7$ rotations and the M\"obius transformation $\sigma_1$ that maps the triangle $R O V$  to $R W S$.
   $R$ is a fixed point of $\sigma_1$  (but it is not an involution).

As we  mentioned,   the six Weierstrass points are:  $O$, the four $W$ and $U$,    fixed by $J: z \mapsto -z$.

\begin{prp}  Points type $R, S, T, V, X$ are fixed points of two orientation reversing symmetries.
\end{prp}

\proofn All contiguous
triangles in our Fig. 4 and Fig.5  
(top) are mapped by reflections over the common side.  For instance, consider the anti-M\"obius transformations $\beta_1$, $\beta_2$   described in Figs. 2,3 of \cite{Ragazzo0} (explicit formulas in (4.1) there). For our purposes is enough to register that  for  $\beta_1$ all the points the arc SW are fixed, and $\beta_1$ interchanges $R \leftrightarrow T$ and $O \leftrightarrow W$. For $\beta_2$  the arc RV is fixed.

More generally, one can get the full 96 symmetries (either area preserving or reversing) by composing  $\beta_1, \beta_2$  at will, together  with the  eight  rotations and  with the reflection $z \mapsto \bar{z}$.

It is immediate to see that each one of the points $R, S, T, V$ is fixed by  two such reflections.
 Points type $X$ as well.  In Fig. 5 (top),  $X$is fixed by the reflection over the arc $XT$ such that $U \leftrightarrow W$.
There is another reflection that fixes  $X\,$, mapping the triangle $WTX$ outside the fundamental domain (fixing the arc $WX$).   \qed

\begin{exa} Two fixed points of Robin function that are mapped to each other by an orientation reversing involution (such as $\beta_1:\,  R \leftrightarrow T$  or  $ \beta_1:\,O \leftrightarrow W$) produce +1, +1 steady vortex crystals. The reason is again the anti-symmetry of the symplectic form under that involution. 
\end{exa}

\subsection{Generalization for $N$ vortex systems, $N \geq 2$} \label{extension}
\vspace{1mm}

 We begin with a question addressed to algebraic geometers. We are taking uniformizing coordinates  $z= x+i y$. Fix  arbitrarily $z_j^*, 1 \leq j \leq N$ on a fundamental domain. 
 Express all partial derivatives
of $R$'s and $G$'s  
  in terms of the $2N$  forms $dx_j , dy_j, 1 \leq j \leq N$. 
Forcing $dH_{|(z_1^*, \cdots , z_N^*)} = 0 $   will give $2N$ quadratic equations for the $N $ vorticities  $ \Gamma$'s taken as unknowns. Is it possible to prescribe  conditions for the existence of real valued solutions for them? What can singularity theory say - at least generically -  about   branchings when the positions   $s_j^*$ are varied?

For now  on the values  of circulations $\Gamma_j$ will be arbitrary.  In  the next results we explore  involutions of  the Riemann surface $\Sigma$,  that are orientation preserving or reversing isometries of the hyperbolic  metric. 
We look for configurations where not only the $dR$'s, but also the $dG$'s vanish pairwise. Since vortex Hamiltonians are combinations of $R$'s and $G$'s,  by finding common zeros of $dR$ and $dG$  we will get a plethora of steady vortex crystals.
In Bolza's surface there are  46 critical points of $R$.
A natural  question is:  for which  pairs among them  the $dG$'s  also vanishes?

\vspace{-1.5mm}

\subsection*{Extending Theorem \ref{symprop} (Theorem 3.2 of \cite{Ragazzo0}, case (a))} 

\vspace{1mm}

\begin{teo}\label{caseaext}  Assume $s_1^*, s_2^*$ are fixed points of the {\it same} area preserving involution $\sigma$ which is non trivial (different from
the identity). Then not only $dR(s_1^*) = dR(s_2^*)=0$ but as well $d G (s_1^*, s_2^*) = 0$.
  \end{teo}
   \proofn We can take $h_1$ and $h_2$ as in the preliminary reasoning  (*) so that  $h_1(s_1^*) = 0$ and  $h_2(s_2^*) = 0$.  In this way the Green function $G(s_1,s_2)$ around  $(s_1^*, s_2^*)$  will be represented by a function $\bar{G}(z_1,z_2)$ around $(0,0)$.
 In this format  the symmetry $z_1 \leftrightarrow z_2$ is lost, but this is not important.  What we need remains true, namely the invariance of $G$ under $\sigma$ entails  $\bar{G}(z_1,z_2) = \bar{G}(-z_1,-z_2)$ on a neighborhood $U_1 \times U_2$ of $(0,0)$.
 Thus $d \bar{G}(0,0) = 0$, and this amounts to say that  $d G (s_1^*, s_2^*) = 0$,   as desired.
 Actually this  reasoning proves more. If we have a non trivial orientation preserving involution  $\sigma$  fixing $m$   points $s_1^*, \cdots, s_m^*$,  any $\sigma$ invariant function of $(s_1, \cdots, s_m)$  will have  them and their permutations as  critical points. \qed

\begin{cor} The $2g+2$  Weierstrass points of an hyperelliptic Riemann surface yield steady states for any system with equal or less number of vortices, independently of the assigned intensities. 
(For Bolza's surface, we insist once more: those are points $O, U,$ and the four $W$.)
\end{cor}

\subsection*{Extending Theorem \ref{symprop1} (Theorem 3.2 of \cite{Ragazzo0}, case (b))} \label{anticonf}

\vspace{1mm}

 Assume that there exists {\it two}  orientation reversing involutions $\sigma_1, \sigma_2$ of $\Sigma$ {\it both fixing} the two points $s_1^*, s_2^*$.
Consider   the (artificial) Hamiltonian $ H = G(s_1, s_2)$  in $\Sigma \times \Sigma$,  with symplectic form being an arbitrary combination sum of the  two individual area forms. Denote by $ X_H = (X_1, X_2)$.

 Each $\sigma_i$ acts diagonally on $\Sigma \times \Sigma$. It follows that at   $(s_1^*, s_2^*)$ (each of them mapped to the origin by
$h_1$ and $h_2$),  then -1 is an eigenvalue with eigenvector $(X_1,X_2)$ of   $$(\sigma_i)_* (U, V) =
((\sigma_i)_* U   , (\sigma_i)_* V),\, i=1,2 . $$  
 The same reasoning of Theorem  \ref{symprop1} would apply to each one of the two slots, implying: 
 
\vspace{1mm}

 \begin{teo} 
  Consider two  points $s_1^*, s_2^*$ that are fixed by two different orientation reversing involutions $\sigma_1, \sigma_2$.
 Then not only $dR(s_1^*) = dR(s_2^*) = 0 $  but  also $dG(s_1^*, s_2^*) = 0$. 
\end{teo}

Putting these facts together, we get:

\begin{teo} Suppose a compact Riemann surface has $N$ points such that every pair is  either of type (a: Theorem 8)
or (b: Theorem 9). 
 Consider a system of $m  \leq N$ vortices, the
vorticity strengths being arbitrary assigned. 
Then there are at least $N!/(N-m)!$
different  equilibrium configurations. 
 \end{teo}
 \proofn Both
  the $dR$ and the $dG$ vanish at those points and pairwise. \qed

\vspace{1mm}

\begin{exa}    Case (b) in Bolza's surface.  We saw that case (a) (Theorem 8) takes care of all Weierstrass points.  We can  make an `overkill' for  three of them. Both orientation
reversing involutions $z \mapsto \bar{z}$ and  $z \mapsto -\bar{z}$ fix  the Weierstrass points
at the center of the octagon,  at the horizontal
axis, and  at the vertical axis  (see Fig. 3,  right). 
\end{exa}

\begin{exa}  The Schottky double (formally defined in the next section) of the planar domain in  Fig.6  is a closed Riemann surface of   genus 2.   The three  axis inversions
 $I_x : x \mapsto -x,\,  I_y:  y \mapsto -y,\,$  and $  I_z: z \mapsto -z$  (this by abuse of notation means exchanging faces) are orientation reversion isometries of the Euclidian  metric. As we will see in the next section, the curvature is concentrated on the boundaries. Upon uniformization of the Schottky double, these maps conjugate to hyperbolic involutions. So we can pretend that we are `seeing'  Fig. 6 in the uniformization. $I_x$ and $I_z$ fix the pink points. All pairs of  points in the horizontal axis are fixed by $I_y$ and $I_z$.  A question for experts:  which  are the Weierstrass points?  
\end{exa} 

\vspace*{-0.4cm}

\begin{figure}[!ht] \label{fig3}
\includegraphics[scale=0.12]{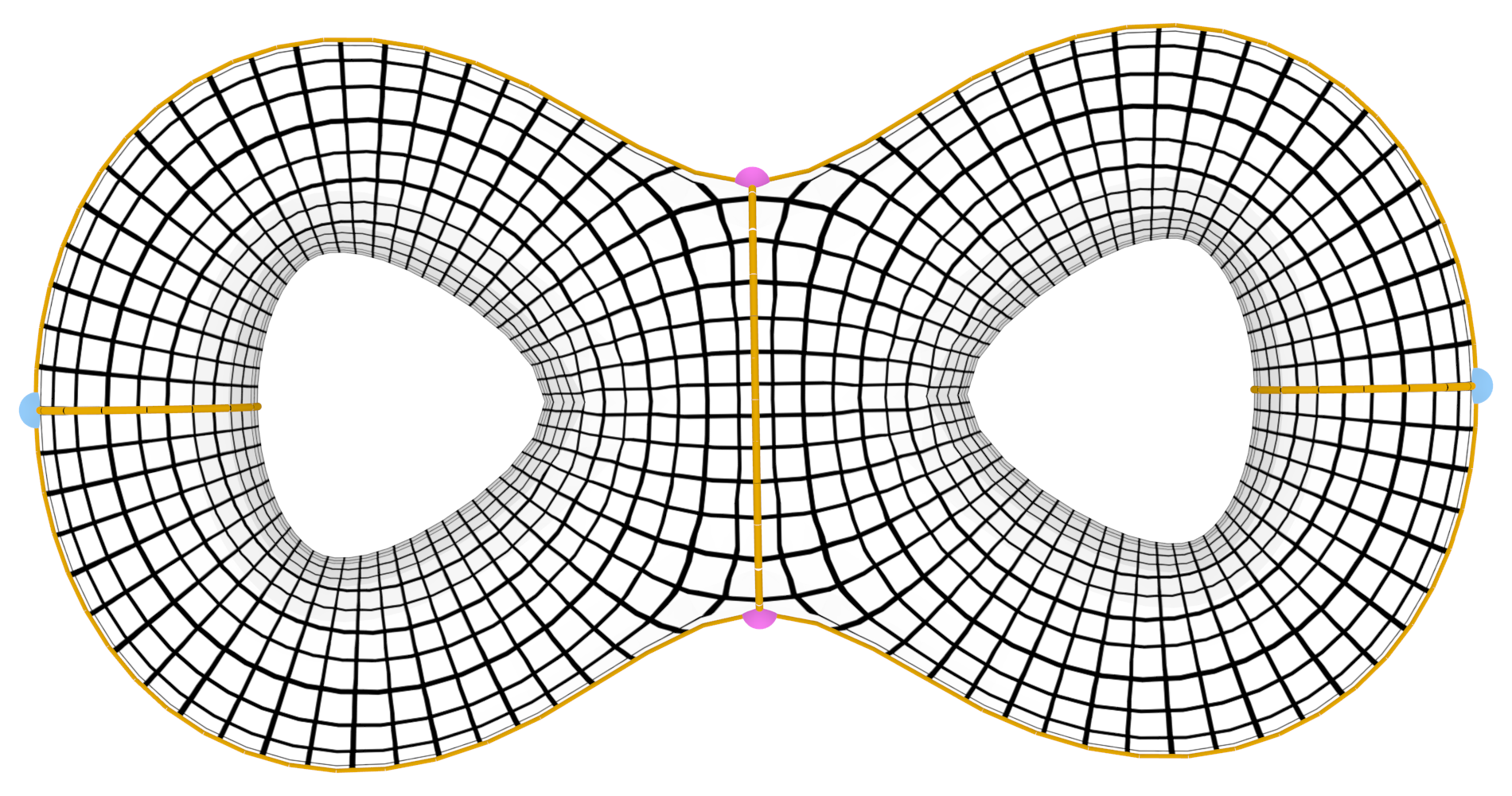}

\caption{From \url{https://sechel.de/}.  Let $x$ be  the horizontal axis, $y$ the vertical,
axis $z$ pointing outside of the paper.
The two pink points are fixed both for
  $I_x : x \mapsto -x$      and    $I_z: z \mapsto -z$. The two blue points (and 4 others not marked along the $x$ axis) are fixed by $I_y: y \mapsto -y$      and    $I_z: z \mapsto -z$. }
\end{figure}
\begin{rem} \label{insight}
Why are  all these results   true independently of  the strengths  of the vortices?  A physical  insight  could be as follows. 
 Each of these points  (denote $s_*$) is an equilibrium of the corresponding single vortex system.   The fluid
velocity (in the covector representation) due to the unit vortex {\it bound} at $s_*$  is $ -\star d_s G(s, s_*)$. 
In both cases (a) or (b)  we have  $d_{s=s_j} G(s, s_*) = 0,  \, s_j \neq s_*$.  

This means that  the velocity field of
the fluid has  all the other points  as \textit{stagnation points}.  We can give any vorticity value at
to these stagnation points.  Neither they will move nor, by symmetry of reasoning,  they will displace
the vortex at $s_*$. The simplest situation is the round sphere:  vortices at antipodal points remain at rest
no matter what are their   intensities.
\end{rem}

\vspace{0.3mm}

\section{SCHOTTKY DOUBLE OF A  BOUNDED PLANAR DOMAIN }  \label{scho}

\vspace{1mm}

\vspace{1mm}
  In this and subsequent sections we will consider bounded multiply-connected
planar domains $\Sigma\subset \C$. 
  We assume    all boundaries analytic\footnote{It would be interesting to extend the discussion of this and the next two sections to curved surfaces (i.e. non flat metrics)  with boundaries, and  the extension to  dimensions  $\geq 3$
 \cite{Friedrichs-1955-a, Schwarz-1995-a, Morrey-a}.}.
We will denote by $b_o$  the external boundary 
 and $b_k, 1 \leq k \leq g$  the inner boundaries of $\Sigma$, all run according to the conventions of section \ref{signconv}. 
  
  Our aim is to relate  C. C. Lin's  formulation \cite{Lin-a} with our approach on compact boundaryless  Riemann surfaces  by considering the Schottky double $\hat{\Sigma}$.    
 
\begin{definition} Double of a manifold $\Sigma$ with boundary. Take a copy
$\Tilde{\Sigma^o}$ of the interior $\Sigma^o$ of the manifold and glue the two
along their common boundary $\partial \Sigma$.
The result is a closed manifold, set theoretically written as $\Hat{\Sigma}= \Sigma^o\cup \partial \Sigma \cup \Tilde {\Sigma^o} . $
\end{definition}

\vspace{-2mm}

The complex structure can be extended analytically: $\Hat{\Sigma}$ is
 a closed genus $g$ Riemann surface,  having a natural  involution $I$ that exchanges the faces\footnote{The front and back faces are  `morally' indistinguishable, being anti-conformally equivalent. The construction goes back to F.~Schottky \cite{Schottky-1877}. See \cite{Schiffer-Spencer-1954-a}, \cite{Hawley-Schiffer-1967} for  general discussions.}.
 From the metric point of view  $\Hat{\Sigma}$ is a Lipschitz manifold, whose curvature is concentrated on the boundaries.
 \vspace{1.5mm}

\indent \textit{The three Green functions}. $  \,\, G_{\rm {electro}} $ is the more traditional Green function on $\Sigma$   with zero-valued Dirichlet boundary conditions  (sometimes called ``of the first kind").  $ \, G_{\rm {hydro}}$  has prescribed inner boundary circulations $p_j (1\leq j \leq g) $ and  constant (unknown) values on the boundaries.
$G_{\rm {double}} $ is the Green function of the Laplace-Beltrami operator  of $\Hat{\Sigma}$ with the Lipschitz metric. 
\vspace{1mm}

 The  difference  $G_{\rm hydro} - G_{\rm electro}$ is harmonic in each slot.  We will  be able to express it in a quite interesting way
by means of the  \textit{hydrodynamical capacity matrix $Q$},  revisiting \cite{Gustafsson-1979-b,  Flucher-Gustafsson-1997-a, Flucher-1999-a}.  
We do not resist to tell  a spoiler: $Q$ is one of the blocks in Riemann's relations.
 \vspace{1mm}
 
 $ G_{\rm {double} }$ yields $G_{\rm {electro}}$ in a trivial  way,  but so far there is no
general procedure to get $G_{\rm {double}}$ from $G_{\rm {electro}}$. For simply connected domains this can be done on a roundabout way.

\vspace{1mm}

  Consider a $N$-vortex system in $\Sigma$,  the boundaries being impenetrable. The `image  method' proposed   by  Green himself, and popularized by Thomson, consists to extend the system to the Schottky  double by taking  $N$ virtual
`image vortices' by the involution in the back side,    with opposite strengths.
These configurations form a special case of the setting in Theorem \ref{Hamstru}:   under the dynamics in $\Hat{\Sigma}$ these configurations  will form an \textit{invariant submanifold}.

     Moreover, the curves in $\Hat{\Sigma}$ corresponding to the boundaries will be marker streamlines for the Euler flow, no matter how the $N$ true  vortices move in the front side.
Since no flow crosses $\partial \Sigma$, the non-smoothness of the metric structure over $\partial \Sigma$ never causes problems.

By Helmholtz' theorem, there will be $g$ constants of motion, the circulations of the flow along the curves $b_k, 1\leq k \leq g$.
We will discuss a \textit{gedanken} experiment:   let the flow start impulsively by suddenly creating vortices from the fluid at rest.  What could be the  boundary 
 circulations?

  Marsden-Weinstein theorem allows reducing to the vortex space,  of dimension $2N$. Assuming the vortex motion being solved, reconstruction  will allow identifying the harmonic $B$-components of the flow, that  at each instant of time are $L^2$-orthogonal to the `pure' vortical component with respect to the Lipschitz metric in the double.  The $A$-components are zero by symmetry.    

\newpage

There will be three ways to derive the Hamiltonian  (yielding the same result): \\
\indent  i)  the Marsden-Weinstein reduction; \\ \indent ii) the
Kirchhoff-Routh hamiltonian of C. C. Lin, and \\\indent  iii) using the full hydrodynamical Green function.\
\vspace{1mm}
  
With additional  work   one can orthogonally separate the vortical from {\it two } other components of  the \textit{Hodge decomposition  for domains with boundaries}  (some informations  in Appendix B).      As we mentioned in the introduction, 
  this finer decomposition  may be  important in practice, hopefully for physiological flows, specially vascular models \cite{Raza-a, Saqr-a, Razaf-a}.  We referred to  \cite{Poelke-a, Desbrun-a} for computational geometry algorithms.  We strongly recommend the recent paper \cite{SanDiego}, with a video  in the supplementary materials  that gives life  to ``fluid cohomology".

\subsection{Schwarz function:  conformal structure remains smooth in $\Hat{\Sigma}$ \\ but not the metric}
\vspace{1mm}

This material is well known, we present it only for completeness. When $\Sigma$ is a planar domain, any point $s\in\Sigma$ is  a complex number $z\in\C$. The obvious holomorphic coordinate $\phi_1$
is \begin{equation} \phi_1(s)=z. \end{equation}
The backside $\Tilde{\Sigma}$ is to have the opposite conformal structure, and therefore a suitable coordinate is
\vspace{-0.7mm}
 \begin{equation} \phi_2(\tilde{s})=\bar{z},
\end{equation}  where $\tilde{s}\in\Tilde{\Sigma}$ denotes the point on $\Tilde{\Sigma}$  which is opposite to $s$.
The map $I:s\mapsto \tilde{s} \,\,\, {\rm and}\,\, \tilde{s}\mapsto s$ defines the anti-conformal involution on $\Hat{\Sigma}$.   We will show now that there is a Lipschitz metric structure in $\Hat{\Sigma}$, and, no surprise,  denoting $\hat{\mu}$ the area form in $\Hat{\Sigma}$, we have
\begin{equation} \label{nosurp}
I^*(\hat{\mu}) = - \hat{\mu} .
\end{equation}

When $\partial \Sigma$ is analytic there is a unique analytic function $S(z)$, known as the {\it Schwarz function} for $\partial\Sigma$,
defined in a neighborhood of $\partial \Sigma$ in the complex plane and satisfying
\begin{equation}
S(z)=\bar{z}, \quad z\in \partial \Sigma.
\end{equation}

See  for instance \cite{Davis-1974}, Chapter 5. The meaning is that $z\mapsto \overline{S(z)}$ is the (local) anti-conformal reflection in $\partial \Sigma$
within the complex plane.  The Schwarz function is the unique analytic function which at each point $z$ along $\partial \Sigma$ takes the value $\bar{z}$.

\begin{exa}  \label{disk1}
The double $\Hat{D}$ of the  unit disk  $D$ 
 is  identified with the extended plane $\C\cup\{\infty\}$ with involution
$I:z\mapsto 1/\bar{z}$.
The back-side  is `folded up' to the exterior of $D$.
\end{exa}
\vspace{-1.5mm}

One can make more general assumptions for  the  boundary
curves. 
All is needed is that the domain is  finitely connected and no boundary component
consists of just a single point. Any such domain can be mapped conformally onto a
domain bounded by analytic curves.
Nonetheless, most authors assume a little more. To make sense to physical quantities,  piecewise
smooth, or just Lipschitz, is usually enough. 
With analytic boundaries, using $S(z)$ one can ``fold up'' $\Tilde{\Sigma}$ locally near $\partial \Sigma$ so that it becomes represented as part of the exterior of
$\Sigma\cup \partial \Sigma$ in $\C$.
Thus we see that the conformal structure in $\hat{\Sigma}$ is smooth. On the other hand, $\Sigma$ has a Riemannian structure,
but  the symmetrically extended metric structure is usually not smooth across $\partial\Sigma$.

In order to  detail  this  we return more carefully to the two holomorphic coordinates $\phi_1$ and $\phi_2$, introduced above.
The domains of definition of these do not fully cover $\Hat{\Sigma}$: the boundary $\partial\Sigma$ is missing.
Nonetheless,  when $\partial\Sigma$ consists of analytic curves then the Schwarz function can be used to construct analytic
extensions of each of $\phi_1$ and $\phi_2$ across $\partial \Sigma$.
\begin{lem} The Schwarz function is the transition function between the front and the back side.
\begin{equation}
S=\phi_2\circ \phi_1^{-1}.
\end{equation}
\end{lem}
\proofn This is verified  just by noting that $\phi_2\circ \phi_1^{-1}$ is holomorphic in its domain of definition (a neighborhood of $\partial\Sigma$)
and that it on $\partial\Sigma$ takes $z$ to $\bar{z}$, these being exactly the defining properties of the Schwarz function $S$.

\begin{lem} Let $\Sigma$ be  given a Riemannian metric $\lambda(z)^2|dz|^2$.  The density of the metric, still expressed in
the   coordinate $\phi_1$  but across the boundary $\partial\Sigma$
will  acquire  the factor
\begin{equation}
|\frac{d\phi_2}{d\phi_1}|^2=|(\phi_2\circ \phi_1^{-1})^\prime|^2=|S^\prime|^2.
\end{equation}
\end{lem}
Thus we see that it is only Lipschitz continuous in general.
\proofn   The metric is given more precisely by
$\lambda(s)^2|d\phi_1(s)|^2$,   with $\lambda$ considered as a function on $\Sigma$. Copying symmetrically to $\Tilde{\Sigma}$ gives there
$\lambda(\tilde{s})^2|d\phi_2(\tilde{s})|^2$, $\tilde{s}\in\Tilde{\Sigma}$, and on letting $\tilde{s}\to s\in \partial\Sigma$ we have
$$
\lambda(\tilde{s})^2|d\phi_2(\tilde{s})|^2\to \lambda({s})^2|d\phi_2({s})|^2
=\lambda({s})^2|\frac{d\phi_2(s)}{d\phi_1(s)}|^2|d\phi_1({s})|^2.
$$   \qed

\noindent{\bf  Continuing the Example \ref{disk1}.}
If for example $\lambda=1$ in $\Sigma$, which is the usual choice for planar domains, and if we take $\Sigma=D$
then  in the same coordinates
\begin{equation}  \label{mm}
\lambda(z)=|S^\prime(z)|=\frac{1}{|z|^2} \quad  (|z|>1).
\end{equation}
One may notice that the behavior at infinity becomes the same as for the spherical metric, namely $|dz|/(1+|z|^2)$,
but  in the case of $\hat{D}$  all the Gaussian curvature becomes concentrated to $\partial \Sigma$.
The Gaussian curvature is a singular
distribution on $\partial\Sigma$, and its density with respect to Euclidean arc length on $\partial\Sigma$ can be found to be $\pm |S''|$
(or\, $\I S''/(S')^{3/2}$, to also get the sign).
This is twice the curvature of  $\partial\Sigma$ as a plane curve,
as is also consistent with  the fact  that the total Gaussian curvature of $\Hat{D}$ (topologically a sphere)
is $4\pi$ while the total arc length of $\partial D$ is $2\pi$.

\vspace{1.5mm}
\subsection{Getting $G_{\rm electro}$ from $G_{\rm double}$ is immediate}

\vspace{1mm}
 In Physics,  the existence of  the electrostatic Green function 
 $\,\, G_{\rm electro}(z,z_o) =  -\frac{1}{2\pi} \log|z-z_o| +  {\rm harmonic}, $
vanishing for $z$ (or $z_o$)  in all the $g+1$  boundary curves  is taken for granted:    if a unit charge is placed at a point $s_o$  in the interior of $\Sigma$  and if all the $g+1$ boundaries are grounded by infinitely thin wires,   they all will have zero potential.
 The literature has a large number of analytical and numerical methods for obtaining  the  Green function $G_{\rm electro}$, a particular case of Dirichlet boundary conditions (see e.g. \cite{Cohn-1980-a}).

 Clearly, if one already knew the Green function   $G_{\rm {double}}$ of the Laplace-Beltrami operator of the double, then for $s,s_o$ in the front side,

\begin{prp}
\begin{equation}  \label{electro} 
\begin{split}
& G_{\rm electro}(s, s_o)=G_{\rm {double}}(s , s_o)-G_{\rm {double}}(s ,I(s_o)) \\
& R_{\rm electro}(s) = R_{\rm double}(s) -   G_{\rm double}(s, I(s)) \\
\end{split}
\end{equation}
\end{prp}
In particular 
\begin{equation} 
\begin{split}
&G_{\rm {double}}(s;s_o)=G_{\rm {double}}(I(s),I(s_o)) \\
& R_{\rm {double}}(s)=R_{\rm {double}}(I(s)) \\
&G_{\rm {double}}(s , I(s_o)) = G_{\rm {double}}(I(s), s_o) .
\end{split}
\end{equation}

\begin{rem}  \label{Gelectroback}  
On the other hand,  changing  $G_{\rm electro}$ into $G_{\rm double}$ seems not an easy task.  A conceptual reason is that  the former is conformally invariant, while the latter depends crucially on the boundary shape. 
At any rate, one would need first to  extend to the double the electrostatic Green function
$G_{\rm electro}(s, ;s_o), $  which is defined  originally  only for $s, s_o \in \Sigma$.
One does it by making it  an odd function in each variable, hence as an even function if both variables are extended simultaneously. 
If we take $\Sigma$ to be the upper half-plane, for example, then 
$$
G_{\rm electro}(z,a)=-\frac{1}{2\pi}\log |\frac{z-a}{z-\bar{a}}|, \quad z,a\in \Sigma.
$$
This formula extends to all the Riemann sphere, and we get, with $I(z)=\bar{z}$,
$$
G_{\rm electro}(I(z),I(a))=-\frac{1}{2\pi}\log |\frac{\bar{z}-\bar{a}}{\bar{z}-{a}}|= + G_{\rm electro}(z,a).
$$ 
Note that in spite of this,  the  vortex  with positive strength +1 at $a$ corresponds to the  vortex of negative strength -1  at $I(a)$ because the involution inverts the sign of  the  area form. \\

We now show one case where we can get $G_{\rm double}$ from $G_{\rm electro}.$ 
\end{rem}

\subsection{ $G_{\rm {double}}(\hat{D})$  of the unit disk} \label{diskcake}

The Green function $G_{\rm {double}}$ of the  Schottky double  (`pancake)'   
is frequently considered in geometric function theory \cite{Schottky-1877, Schiffer-Spencer-1954-a, Hawley-Schiffer-1967}. In spite of the non-smoothness of the metric at the boundary, its existence can be verified with an atlas of  Lipschitz continuous transition functions. One may visualize  by a limiting process, eg., making  the smaller axis  of a triaxial ellipsoid shrink.

As we mentioned, it is not known a general  recipe to derive $G_{\rm {double}}(\hat{\Sigma})$ from $G_{\rm  electro}(\Sigma)$, because the latter has a positive vortex in the front side and an image  negative vortex in the back side. 
Informally speaking, one needs to spread the negative concentrated vorticity to the whole surface.

This is a partial balayage \cite{Gustafsson-Roos-2018}). The result is in this particular case  is simply the normalized area measure $\frac{1}{V}\mu$. In practice, this procedure
 amounts to constructing $G_{\rm double}(\Hat{\Sigma})$ directly.  
In some cases 
it is possible to elaborate explicit formulas for $G_{\rm double}(\Hat{\Sigma})$ by matching well-known formulas.

For instance, one can guess the formula for the Green function of the disk pancake $\Hat{D}$.
It is well known that the ordinary Dirichlet Green's function for $D$ is, in complex variable notation,
\begin{equation}\label{green1}
\begin{split}
G_{\rm electro}(z,w)& =  -\frac{1}{2\pi}\log \big| \frac{z-w}{1-z\bar{w}}\big| \\
& =\frac{1}{2\pi}\Big(\log\frac{1}{|z-w|}-\log\frac{1}{|z-1/\bar{w}|}+\log |w|\Big)\,,\,\,\, |z|, |w| < 1.
\end{split}
\end{equation}
We have  the metric in $\Hat{D}$,  which is $ds=\lambda(z) |dz|$  given by $\lambda=1$ for   $|z| < 1$ and  $1/|z|^2$ 
for $|z| > 1$
(recall \ref{mm})).
Observe that the total area of $\Hat{D}$ then is $2\pi$ (=twice the area of $D$).

The Green's function of the Laplace-Beltrami operator in $\Hat{D}$  is given in \cite{Gustafsson-Roos-2018}, end of section~12 there. In our notations, the Schottky double (monopole) Green function is 
\vspace{2mm}
\begin{prp}  \label{Greendoubledisk} Green function for the Schottky double of the unit disk with the flat metric:
\begin{equation}
\begin{split}
8 \pi G_{\rm double}(z,w)=& \,4 \log\frac{1}{|z-w|}+ |z|^2+|w|^2 + c \,\,\,\,\,\,\, (|z|<1, |w|<1),\\
=& \,4 \log\frac{1}{|z-w|}+ |z|^{-2}+|w|^2 + 4 \log |z| + c \,\,\,\, (|z|>1, |w|<1), \label{Gdiskpancake}\\
=& \,4 \log\frac{1}{|z-w|}+ |z|^{2}+|w|^{-2} + 4 \log |w| + c \quad (|z|<1, |w|>1),\\
=& \,4 \log\frac{1}{|z-w|}+ |z|^{-2}+|w|^{-2} + 4 (\log |z|+\log|w|) + c \,\,\,\, (|z|>1, |w|>1).
\end{split}
\end{equation}
\end{prp}
\vspace{-3.5mm}
\proofn  Applying the usual Laplacian  with respect to the slot $z$ in the first expression,   for $w\neq z$  we get $(2+2)/8\pi = 1/2\pi$, which is the inverse of the area of the pancake.  The  other expressions give the same result.
One easily checks that the matching across $|z|=1$ is $C^1$, so there is no distributional contribution to the Laplacian
on the unit circle. The above Green's function have all the required properties. The symmetry, for instance is evident. 
 \qed

\vspace{1mm}
The constant $c$ is to be chosen so that  
 $ \int_{\Hat{D}}  G_{\Hat{D}}( \cdot ,w) \mu = 0. $  and 
 we can choose $w = 0$.  The computation can be done in polar coordinates since there is no angular dependence. Let $r = |z|$.\begin{equation*}
 \begin{split}  & \int_0^1 \, (4 \log \frac{1}{r} + r^2 + c) \cdot 1 \cdot r dr + \int_0^1 \, (4 \log \frac{1}{r} + r^{-2} + 4 \log r + c) \cdot r^{-4} \cdot r dr   =  \frac{3}{2} + c.
\end{split}
\end{equation*}
Thus we shall have   $c = -  3/2,$
and the contribution to the Green's function itself becomes $-3/16\pi$.

From the above expressions   (\ref{green1}) and (\ref{Gdiskpancake}), it is instructive for the reader  to  confirm, in the case of the disk  pancake,
 the general identities  (\ref{electro}), by taking   $I(w)=1/\bar{w}$.

\vspace{1.5mm}
\noindent \textit{Green function of $S^2$.  } For the 
 round metric,  in stereographic coordinates, projecting  to the equator, $z=0$ corresponding to the south pole,
  $\lambda(z)=2(1+|z|^2)^{-2}$.   Analogous computations yield
\begin{equation} \label{roundsphere}
\begin{split}
G_{S^2}(z,w) &=-\frac{1}{4\pi}\Big(\log\frac{|z-w|^2}{(1+|z|^2)(1+|w|^2)}+1\Big) \\ &= \Big(-\frac{1}{4\pi} \log ||x-y||, \,\, ||x||=||y||=1 \Big)
\end{split}
\end{equation}
where $||\,||$ is the Euclidian norm in $\R^3$.
 
\vspace{1mm}

\subsection{A method for 
 $G_{\rm {double}}(\hat{\Sigma})$ of   a simply connected  planar domain $\Sigma$} \label{simplyconnSho}
Baernstein's lemma \ref{Baernsteinlemma}  can be used to get $G_{\rm {double}}(\hat{\Sigma})$ from the  Green function $G_{\rm {double}}(\hat{D})$  for the disk pancake  (Proposition \ref{Greendoubledisk}).  For this, one  just  need to get  
 the Riemann mapping of $\Sigma$  to the unit disk $D$ and the Schwarz mapping of the analytic boundary $\partial \Sigma$. Computationally, it may be useful  to   inflate  the disk pancake  to the round sphere by  means of the two stereographic projections. Then we will have 
to  solve a Poisson equation on $S^2$   (which is easy  by means of spherical harmonics), the source term being  the product of the conformal factors.

 Actually,  for vortex problems, via Theorem \ref{conformalteo},  the need of  computing  the Green function for the pancake $\Hat{\Sigma}$ can be avoided.
  Moreover, since the total vorticity vanishes due to the image vortices, there will be no need to include the nonlocal terms.
At the expense of including those,  one can study general  vortex systems in $\Hat{\Sigma}$ (i.e, not restricted to $N$ opposite pairs). One transfers the equations of motion equations to the  representing  sphere.
  See \cite{RKCRCD} for an example.

 $$ $$
 
\subsection*{A detour: anti-conformal involutions; vortices on non-orientable surfaces}  \label{nonorientable}

\vspace{1.5mm}

In general the fixed point set  of an involution $\sigma$ on a Riemann surface  consists of a finite number
 of analytic curves. There are two cases: either they separate the surface into two halves, in which case the full surface can be
considered as the Schottky double of each of them, or else they do not separate the surface. In that case the full surface can be considered
as the double of the non-orientable surface obtained by identifying the (at most) two points on each orbit of $\sigma$. Other types of doubles exist, for example the  \textit{orienting doubles}     \cite{Alling}.
\begin{exa}
 The two involutions $\sigma_1(z)=1/\bar{z}$ and $\sigma_2(z)=-1/\bar{z}$  on  the Riemann sphere illustrate these two possibilities. With $\sigma_2$
the Riemann sphere becomes the double of the real projective plane (a ``cross-cap'').
On the torus there are several anti-conformal involutions, corresponding to the torus being the double of an annulus,
a M\"obius strip or a Klein's bottle.
Consider   the period lattice $\C/(2\Z\times 2\I\Z)$,
with fundamental domain $[-1,+1]\times \I [-1,+1]$. We have: 
\vspace*{-2mm}
\begin{enumerate}
\item Double of an annulus (two symmetry lines, defined by $y=0$ and $y=1$ plus translations of these with period $2$)
$
I_1:z\mapsto \bar{z}.$
\vspace*{-3mm}
\item  Double of an annulus (two symmetry lines, $x=0$ and $x=1$):
$
I_2:z\mapsto -\bar{z}.
$
\vspace*{-3mm}
\item  Double of a Klein's bottle (no symmetry line)
$
I_3:z\mapsto \bar{z}+1.
$
\vspace*{-3mm}
\item  Double of a Klein's bottle (no symmetry line)
$
I_4:z\mapsto -\bar{z}+\I.
$
\vspace*{-3mm}
\item Double of a M\"obius strip (one symmetry line, $x=\frac{1}{2}$)
$
I_5:z\mapsto -\bar{z}+1.
$
\vspace*{-3mm}
\item  Double of a M\"obius strip (one symmetry line, $y=\frac{1}{2}$)
$
I_6:z\mapsto \bar{z}+\I.
$
\end{enumerate}
\end{exa}

 We refer to   recent work by Vanneste \cite{Vanneste} and Balabanova and Montaldi \cite{Balabanova} about vortices on non-orientable surfaces,  that we think should attract further interest.

\begin{rem}
 Schiffer-Spencer \cite{Schiffer-Spencer-1954-a} consider also doubles of closed surfaces, the double then being two
copies of the same surface, but having {\textit{opposite} } conformal structures. 
One might think that this is  of  little interest, but these situations may  appear  naturally, see e.g. Proposition~3 in  \cite{Gustafsson-Tkachev-2011}.  There   it is considered, for instance, the Riemann surface associated to the level lines of the real polynomial $Q(z,\bar{z}) = \alpha$, where
$
Q(z,w)=z^2w^2-z^2-w^2-2r^2zw \,\,\, \text{(with $r>1$ as a parameter).}
$
The involution is $(z,w)\mapsto (\bar{w},\bar{z})$.
When $\alpha=-(r^2+1)$ it consists two spheres, the involution being the identity map between them,
which is anti-conformal since the spheres have opposite conformal structures.
 For   $\alpha=-(r^2-1)^2$
the Riemann surface is again two spheres, but this time the involution consists of an ordinary involution within each sphere, so that it the double of two disjoint disks.
 \end{rem}
 
 \newpage

\section{HARMONIC MEASURES} \label{electrosection}

\subsection{Choice of homology basis  $\{ a_k , b_k \}, \, 1 \leq k \leq g $ for the Schottky double $\tilde{\Sigma}$} \label{conventions}

 We assumed   $\Sigma\subset \C$   bounded and denoted
  by $b_o$  the external boundary  -  which is not in the homology basis.  The following  choice abides with the conventions in section \ref{signconv}. It  gives a simple relationship between two different bases for the Abelian differentials of the first kind (the holomorphic ones) on the Schottky double, 
  see   Lemma 2.3 in \cite{GuSebar}; it is the same used in \cite{Krichever}, \cite{Yamada}.
  
  \begin{enumerate}
\item Each  \textit{inner} boundary component $b_k,\, 1 \leq k \leq g$   goes  \textit{clockwise}. 
\vspace{-1.5mm}
\item  Each curve $a_k, \, 1 \leq k \leq g$, goes from the outer boundary $b_o$ (the boundary of the unbounded component of $\C\setminus \Sigma$)  to $b_k$ and
 then back along the same track on the back-side.
\end{enumerate}

Then $a_1 b_1 a_1^{-1} b_1^{-1} \cdots       a_g b_g a_g^{-1} b_g$ is  a {\textit canonical homology basis} (see \cite{FarkasKra}, 0-3 and I-2-5),  that is, the intersection numbers  (\ref{intnumb}) are
$a_i . b_j = \delta_{ij}, \, a_i . a_j = b_i . b_j = 0.$    
\vspace{1mm}

\begin{prp}  Let   $\{\alpha, \beta \} $ be the dual  cohomology basis (see section \ref{Riemrel}). Then
\begin{equation}  
\begin{split}  \label{signs}   I^* \alpha_j & = - \alpha_j\,\,\,, \,\,\, I^* \beta_j =  \beta_j  \,\,\,\,\,,\,\,\,
   I^*( \star \alpha_j ) =  \star \alpha_j\,\,\,, \,\,\, I^*(\star \beta_j )=  - \star \beta_j  
   \end{split}
\end{equation}
\end{prp}
\proofn  $\,$ Note that  $I_*a_j= -a_j$ (with the obvious meaning,  $I$ reverses directions) and $I_*b_j = + b_j$ (the boundary curves
are invariant under $I$). The sign exchange in the last two equations comes from the fact that $I$ inverts the area form. \qed

\vspace{1mm}

  \begin{prp}  \label{vanishes} 
For the Riemann relations (Proposition \ref{RR}) in a Schottky double we have
\vspace{1mm}

\centerline{  i) $R=0\,\,\,\, $ and ii) $QP=PQ=I\,\,\,\,\,\,$ (both positive definite).}
\end{prp}
\proofn $\,$
Once i) is proved ii) follows immediately from  Corollary \ref{naomanjado}, section \ref{Riemrel}.   
Now, for  the proof of  i)  see  \cite{Gustafsson}, section 9.   Here is a  short argument. 
 We have by (\ref{signs}): 
$$\oint_{a_j}\star\alpha_k=-\oint_{I_*a_j}\star\alpha_k=-\oint_{a_j}I^*(\star \alpha_k)=-\oint_{a_j}\star\alpha_k .$$
so we conclude that the integral  has to be zero.  Now, 
$$R_{jk} = \int_{\Hat{\Sigma}} \beta_k \wedge \star \alpha_j = \sum_i \left( \cancel{\int_{a_i} \beta_k} \int_{bi} \alpha_j - \int_{bi} \beta_k  \cancel{\int_{a_i} \star \alpha_j } 
\right) = 0.    \text{\qed}
$$ 

\begin{figure}[h]
\includegraphics[scale=0.1]{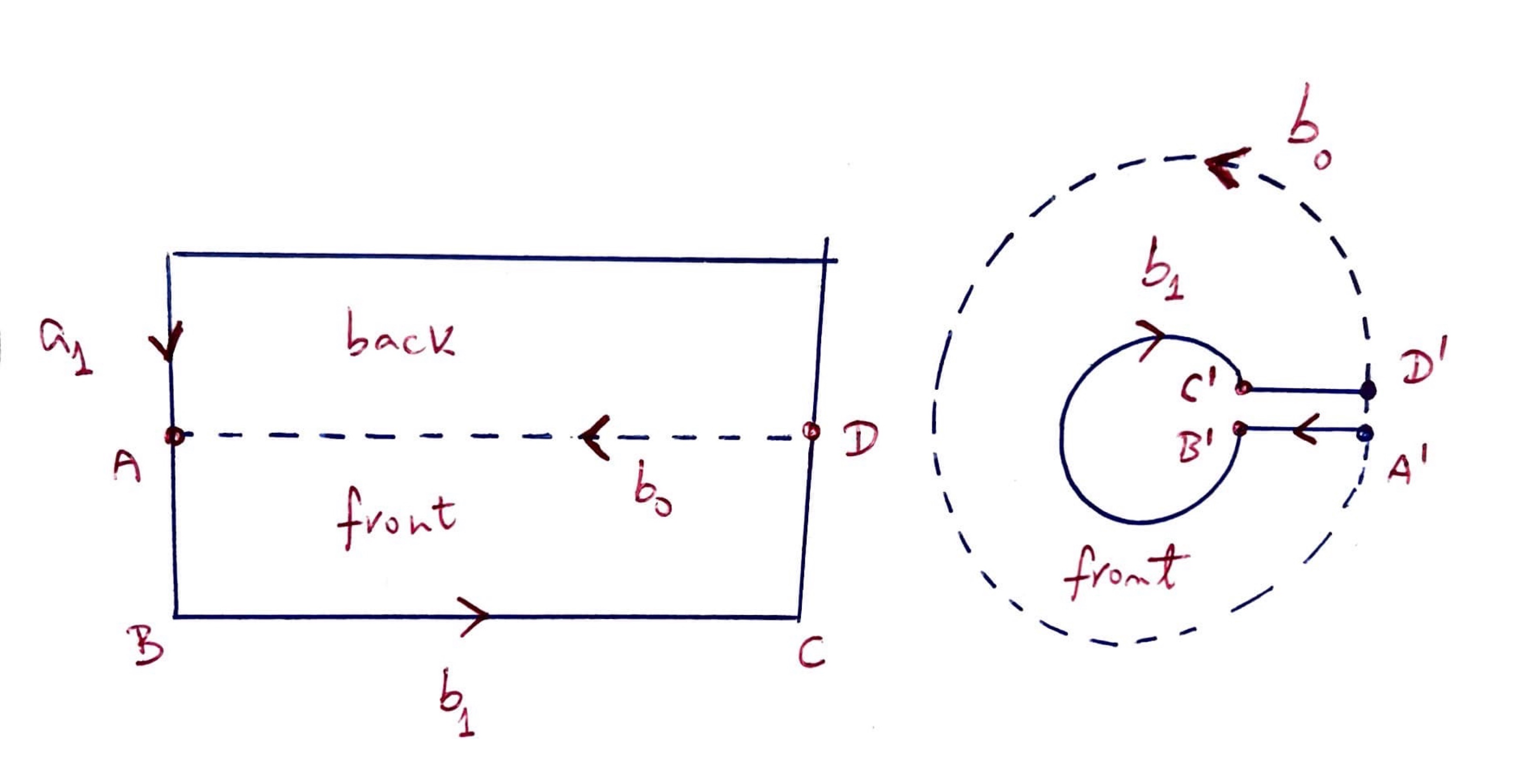}
\caption{Homology basis for the double of an annulus}
\end{figure}

\newpage

\subsection{Harmonic measures for the electrostatic Green function  }

\vspace{1.5mm}

Recall  the most remarkable property of  $G_{\rm electro}$: it gives the solution of Dirichlet's boundary value problem on $\Sigma$. This result actually goes back to Green's  1828 \textit{Essay}  \cite{GGreen}. 
We write in  complex variables notation ($z, \xi \in \C$),  so the formulas  will be familiar.
\begin{teo}  \label{Dirichsol}
Given the values   $f(\xi)$ on the (possibly multiple) boundary, it can be extended harmonically to the interior,
\begin{equation}  \label{Dirichsol1}
F(z)=-\oint_{\partial\Sigma} f(\xi) \,\partial_n G_{\rm electro}(z,\xi)|d\xi| 
\end{equation}
where  $\partial_n $  is the normal derivative at the boundaries with respect to the second variable $\xi$, the normal vector directed \textit{outward }  the region.
\end{teo} 
The kernel, $-\partial_n G_{\rm electro}(z,\xi), $   
for  the unit disk  is the well-known  Poisson kernel, 
$$
-\partial_nG_{\rm electro}(z,\xi)|d\xi|=\frac{1}{2\pi}\frac{1-r^2}{1-2r\cos (\theta -t)+r^2}dt\,\,\, \,\text{with}\,\, \,\,z=re^{i\theta}, \, r < 1, \, \xi=e^{it}. 
$$

 \begin{definition}  The harmonic measures $u_k, \,  0 \leq k \leq g$ are the harmonic functions   in $\Sigma$  satisfying 
 \vspace{1mm}
\begin{equation} u_k = 1  \,\text{ on the curve} \,\, \,  b_k  \, \,\,\, \text{and} \,\,  u_k=0  \, \text{on all the other boundaries}.
\end{equation}  Thus by (\ref{Dirichsol1})
 \begin{equation} \label{uformula}
 u_{\ell}(z) = - \oint_{b_{\ell} }  \partial_n G_{\rm electro}(z, \xi) |d\xi| ,   \,\, (1 \leq \ell \leq g)  
  \end{equation}
  \end{definition}

\noindent   {\bf In shorthand}:
\begin{equation} \label{rho}
u_{\ell}(s)  :=   - \oint_{b_{\ell}} \star_r d_rG_{{\rm electro}}(s,r),\,  s \in \Sigma^o,\,\,\, (1\leq \ell \leq g)\footnote{This formula will appear frequently in what follows. It can be viewed as a  link between analysis and differential geometry. The same quantity is viewed in two different ways: 
i) integrating 
 a function over  a measure, $|d\xi|$, defined   \textit{on}  the (non-oriented) unit circle. 
ii) a differential form 
with respect to $z$, evaluated \textit{along} the oriented unit circle.}. 
 \end{equation}

 \noindent We just  need to  translate to our more intrinsic formulation. 
We assert that
$$\partial_n G_{\rm electro}(z, \xi) |d\xi|  =  \star dG_{\rm electro}(z, \xi) .$$

\proofn 
 Write for short $G = G_{\rm electro}$.  We have
  $ \star dG = = - G_y dx + G_x dy .$ 
Let  $\xi(t) = (x(t), y(t)), $   then the \textit{external} normal to any of the boundary curves (according to our direction conventions)  is
$\hat{n} = - i (\dot{x} + i \dot{y})/\sqrt{\dot{x}^2 + \dot{y}^2} = ( \dot{y} , - \dot{x})/ \sqrt{\dot{x}^2 + \dot{y}^2} 
$ 
(mixing the vector and complex variables notation). Therefore
$$  \partial_n G =  \langle (G_x ,  G_y) , ( \dot{y} , - \dot{x}) \rangle / \sqrt{\dot{x}^2 + \dot{y}^2} 
\,\,\,  \Rightarrow \,\,\, \partial_n G  |d\xi(t)| =  G_x  \dot{y } -  G_y \dot{x} .  \,\,\,\,\,\,\,\,\,\,\,\,  \text{\qed} 
$$
\vspace{1.5mm}

\begin{prp}  \label{harm-meas} In the front face $\Sigma$: 
  \begin{equation} \label{alphau}
  du_k =  2  {\alpha_k}_{|\Sigma}  \,
,\,\, 1 \leq k \leq  g \,\,,\,\, \sum_{k=0}^g u_k \equiv 1,  \,\,   \, 0 \leq u_k \leq 1 .
     \end{equation}
 \end{prp}
\proofn By construction, each  of the real differentials $du_k$ vanishes identically along  all boundary curves $b_{\ell}$, 
hence  all the $du_k$  must be combinations of the base forms $\alpha_{\ell}$.   We can say more. The complex line integral  $ \oint_{a_j}  \alpha_k  = \delta_{jk}$ on the closed loop $a_k$,
can be divided into two pieces (front and back), yielding by symmetry the same result.  On the other hand,   the following calculation can be done in the front face, denoting $a_k^{{\rm front}} $ the front half of the loop  $a_k, \, 1 \leq k \leq g$:
$$ \int_{a_k^{{\rm front}} } du_j =  (u_j)_{|b_k} -  (u_j)_{|b_o}  =   (u_j)_{|b_k} = \delta_{jk} \, .
$$
  The inequality  follows from the maximum principle  (compare \cite{Krichever}).  \qed

\section{THE IMAGE METHOD \\AND   MARSDEN-WEINSTEIN REDUCTION}  \label{mixedblessing}

We will consider special vortex motions on $\Tilde{\Sigma}$: given
$N$ point vortices $s_1,\dots, s_N$  in the  front face $\Sigma^o$,  we  associate $N$ counter-vortices at 
$\tilde{s}_1,\dots, \tilde{s}_N\in\Tilde{\Sigma}^o$,
of opposite strengths, images under the anti-conformal involution $I$.
In particular  the total vorticity $\Gamma = 0$, i.e,
there will be no uniform compensating counter-vorticity.  The non-smoothness of the metric at  $\partial \Sigma$ never causes problems:

\textit{Even if the solid boundaries are removed, the flow $\nu(t)$  stays in $\Sigma^o$,   symmetrically copied to $\Tilde{\Sigma}^o$}.

\vspace{1mm}
We now specialize theorems \ref{gus} and \ref{Hamstru}  for this setting:

\begin{lem}
During the time evolution  $A(t) \equiv 0$. The energy of the harmonic  component  energy  ({\text{see}}  \,(\ref{Hamsymp})) reduces to
\vspace{-2.5mm}
\begin{equation} \label{Harmener}
H_{harm}=\frac{1}{2}B^\dagger QB  
\end{equation}
\end{lem}
\proofn  The reasoning is similar to Proposition \ref{vanishes}. The pull-back of the flow $\nu$ must be even with respect to $I$,
i.e  $I^*\nu=\nu$. We  know that  $\star dG_{\rm double}$  and the $\beta_k$ are  even (the latter because of (\ref{signs})). 
But the $\alpha_k$ are odd, so their coefficients should vanish. 
   \qed

   A marker $s \in \Hat{\Sigma}$ which is not a vortex point will move according to
   \begin{equation*} 
    \dot{s}^{\flat} =  - \sum_{k=1}^N \Gamma_k \star dG_{\rm double}(s, s_k(t)) + \sum_{k=1}^N \Gamma_k \star dG_{\rm double}(s, I(s_k(t))) + \sum_{\ell=1}^g B_{\ell}(t)\beta_{\ell}(s) .
  \end{equation*}
  In view of (\ref{electro}) we get, combining  the image vortices, that the Euler flow  governing the motion of a marker $s$ in  the front side is given by
  \begin{equation} \label{absorbed1}
  \nu(s,t) =  \dot{s}^{\flat} = -\sum_{j=1}^N \Gamma_j\,  \star d_s  G_{{\rm electro}}(s;s_j) + \sum_{\ell=1}^g B_{\ell} \beta_{\ell}(s) .
\end{equation}

\bigskip
 
Regularizing   in the usual way at the vortices,  according to Theorem \ref{gus} we get      \bigskip

\begin{teo} \label{caseSchottky} $\,$\\  $\,$ \\ 
i)  The equations of motion  for  the $N$  vortices in the front side: 
\begin{equation}  \label{sflat}
(\dot{s}_j)^\flat=
-\frac{\Gamma_j}{2}\star d R_{\rm electro} (s_j)
- \sum_{k\ne j} \Gamma_k \star d|_{s=s_j}G_{\rm electro} (s, s_k)
+\sum_{\ell=1}^gB_\ell\beta_\ell (s_j).
\end{equation}
ii) The  coupled  equations for the harmonic components  ($1\leq \ell \leq g$):
\begin{equation} \label{Pell}
\dot{B}_{\ell} = \sum_{j=1}^N \Gamma_j ( \alpha_{\ell}(\dot{\tilde{s}}_j )- \alpha_{\ell}(\dot{s}_j ) )=
 - 2 \sum_{j=1}^N \Gamma_j \alpha_{\ell}(\dot{s}_j ) =  - \sum_{j=1}^N \Gamma_j  du_{\ell}(\dot{s}_j ) .
\end{equation}
\end{teo}
\begin{rem}
 We gladly observe that the ODEs  for the $A$'s in Theorem \ref{gus}  are coherent: due to  (\ref{signs}) the right hand side of  $\dot{A}$ vanishes.  Notice also the elegant expression  (\ref{Pell}) for $\dot{B}$. 
 \end{rem}

\begin{cor} \label{motionconst} There are $g$ constants of motion,  the circulations of the Euler flow  $\nu(t)$ around the inner boundaries. They are given by
\begin{equation} \label{constantmotion}
p_{\ell} =  \oint_{b_{\ell}} \nu =   B_{\ell} (t) + \sum_{j=1}^N  \Gamma_j \,u_{\ell}(s_j(t)) = {\rm const.}\,\,\,  (1 \leq \ell \leq g) .
\end{equation}
\end{cor}
\proofn   
Follows immediately from  
and (\ref{rho}), (\ref{alphau})  and   (\ref{Pell}).  \hfill \qed
 
 \vspace{3mm}
  Clearly, this  result must be true.   In the double  the curves $b_{\ell}$  are ``made of fluid particles", but they
   behave as if they were solid boundaries - which they effectively are.  Helmholtz' theorem implies that the circulations of the Euler flow around  the  boundaries (or stationary flow curves on the double) $b_{\ell}$ are constant.  

\newpage

\subsection{Marsden-Weinstein reduction leads to C.C. Lin's theorem}

The system reduces to 
 the  $N$  vortices that live in the front face.  
One inserts   in  the harmonic energy  part
$H_{\rm harm}$  given by  (\ref{Harmener}) of   the total energy
$ H = H_{\rm vort} + H_{\rm harm}$ the  expression 
  \vspace{-1mm}
\begin{equation} \label{Bofs}
 B(s_1, \cdots s_N)  = p -  \sum_{k=1}^N   \Gamma_k \, u(s_k) \in \R^g \,\,\,,\,\,\, p = (p_1, \cdots, p_g) \,,\,\, u = (u_1, \cdots, u_g ). 
  \end{equation}
where $u_{\ell}$ are the harmonic measures, and  the inner circulations  $p_{\ell}$ are interpreted  as parameters.

 \vspace{1.5mm}
 
  As for the reduced symplectic form there is  a pleasant surprise.  
  Look at  (\ref{form2}).  Although there is a    $\Gamma = 0 $ in the denominator of the magnetic term, this  causes no harm. Again due  to (\ref{Pell}):\\

 \centerline{ \textit{For the Schottky double system, the second factor in the  magnetic term of  (\ref{form2}) vanishes.} }

  \vspace{1.5mm}

 \begin{lem}  Consider the embedding
 $$ (s_1, \cdots , s_N) \in \Sigma \times \overN \times \Sigma  \,\, \longrightarrow \,\,  (s_1, \cdots , s_N,  I(s_1), \cdots, I(s_N) ) 
 \in \Hat{\Sigma} \times \overtwoN \times \Hat{\Sigma}  .
 $$
The pull back of the $2N$ vortices  part of the symplectic form  in $\Hat{\Sigma}^{2N}  \times \R^g$    is  
$ 2 \sum_{k=1}^N  \Gamma_k \mu(s_k) .
 $
 \end{lem}
 \vspace{1mm}
\proofn   
Applying Theorem \ref{Hamstru} to the double, the  opposite signs of the image vortices  are compensated by  the area  form inversion (\ref{nosurp}).    \qed

\begin{teo}  
 \label{MWred} The reduced Hamiltonian in
${\Sigma} \times \overN \times   {\Sigma}$  with symplectic form  $\Omega = \sum_{k=1}^N \Gamma_k \mu(s_k)$ is
\begin{equation} \label{reducedHamilt}
H_{\rm red}
=\frac{1}{2}\sum_{j=1}^N \Gamma_j^2 R_{\rm electro} (s_j)
+\frac{1}{2}\sum_{j=1}^N \sum_{1\leq k\ne j}^N \Gamma_j\Gamma_k G_{\rm electro} (s_j, s_k)
+\frac{1}{4} B Q B^\dagger . 
 \end{equation}
with $B(s_1, \cdots , s_N) = p -  \sum_{k=1}^N   \Gamma_k \, u(s_k) $ (written as row vector, see (\ref{Bofs})).
  \end{teo}
  \proofn   Collect terms carefully in the Hamiltonian of Theorem \ref{Hamstru} for this special situation. Divide by the common factor 2  both in the symplectic form and in the  resulting pulled back Hamiltonian. \qed

\begin{definition}
\begin{equation} \label{GLindef}
\begin{split} 
  & G_{\rm Lin}(s,r)   : = G_{\rm electro}(s,r)  +  \frac{1}{2}  u(s) Q u(r)^\dagger \,\,,\,\,
 R_{\rm Lin}(s)   : = R_{\rm electro} + \frac{1}{2}   u(s) Q u(s)^\dagger \\ 
 & \psi(s ; \mathbf{p})    :=  - \frac{1}{2} \,\mathbf{p} \, Q\,  u(s)^\dagger   \,\,\,\, \text{(``outside agency")}
\end{split}
\end{equation} 
\end{definition}
\begin{teo} (C.C. Lin's Theorem, 1941)  \label{Linteo}  The $N$-vortex system is governed by   the Kirchhoff-Routh Hamiltonian  
\begin{equation}  \label{KRHamilt}
H_{\rm red} =   H_{\rm Lin} =   \frac{1}{2}\sum_{j=1}^N \mathbf{\Gamma_j}^2 R_{\rm Lin} (s_j)
+\frac{1}{2}\sum_{j=1}^N \sum_{1\leq k\ne j}^N \mathbf{\Gamma_j\Gamma_k} G_{\rm Lin} (s_j, s_k) +
   \sum_{i=1}^N \mathbf{ \Gamma_i} \,  \psi(s_i, \mathbf{p}) 
\end{equation}
that was obtained by C. C. Lin in \cite{Lin-a}.
\end{teo}
\proofn Expanding  $BQB^\dagger $ in    $H_{\rm red}$    one gets $H_{\rm Lin}$  (up to a constant).  \qed   
\begin{rem} \label{coherence}
The Green and Robin  functions, the harmonic measures $u$ and the matrix $Q$  are dimensionless.  To emphasize the dimensional coherence of the Kirchhoff-Routh hamiltonian in the right hand side, we wrote vorticities and circulations in bold.    Physically,  vortex strengths and circulations have dimension 
$L^2/T$, so the Hamiltonian is  $(L^2/T)^2 = L^2 (L/T)^2$.   It will have the dimension of energy if we multiply by  the 
 density (mass/area), that one assumes constant. 
\end{rem}

\section{HYDRODYNAMICAL GREEN FUNCTION}     \label{hydrogreen0} 
\vspace{1.5mm}

To complete the proof of Theorem \ref{Linteo} it remains to show that $G_{\rm Lin}$ given by (\ref{GLindef}) satisfies the prescriptions to be the {\it modified} Green function as defined by C.C. Lin, namely,  having zero circulation on the inner boundaries.
  It will be a particular case  of  the formula (\ref{G2})  for the full hydrodynamical Green function introduced by Flucher and Gustafsson 
  \cite{Gustafsson-1979-b, Flucher-Gustafsson-1997-a, Flucher-1999-a}. 

 Recall (once more) the conventions in section \ref{signconv}.  The flow   corresponding to  a stream function $\psi$ is $\nu = -\star d\psi $.  Its  {\it circulation} $p_k$ around a boundary component $b_k$  is
\begin{equation}
p_k=  \oint_{b_k}  \nu\,\,\, ,\,\,\, \nu = -\star d\psi \,\,\, \text{ (see (\ref{correctchoice}))}.
\end{equation} 
(according to  our conventions  a boundary curve $b_k$ is parametrized seeing the surface on the left). 
\vspace{0.5mm}

  As before  we assume  for simplicity that $\Sigma$  is bounded by an external curve $b_o$. We now formally define the hydrodynamical Green function: it     
  is  characterized by having prescribed  circulations $p_k, \,\, 1 \leq k \leq g $ on inner boundaries 
  (parametrized  clockwise)\footnote{ Crowdy and Marshall use the name  \textit{modified}   Green functions those for which 
 the circulation is nonzero    in one (only) of the  $g+1$ curves. Their work is highly recommended \cite{Crowdy, Crowdybook, Crowdy0,Crowdy1,Crowdy-prime}.}. In view of (\ref{byresidue}) one has for the outer boundary (which runs counter-clockwise) the value
 $
 p_o = - \sum_{i=1}^g\, p_i + 1 . $
 If there is no external boundary,  $p_o$  can be  thought as circulation at the point at infinity.  

We have a  `gedanken experiment'  to create  prescribed circulations: 
suppose a flow originally on a simply connected domain bounded by a closed curve $b_o$  has
$g+N$ vortices in it, with  vortex strengths  $\Gamma_i, \, 1\leq i \leq g+N$.   The circulation around $b_o$ (run counterclockwise) is 
$p_o = \sum_{i=1}^{g+N} \, \Gamma_i .
$
Suppose that at a  \textit{snapshot } one inserts instantaneously $g$ solid boundaries $b_i \,(i=1, \cdots,  g)$ coinciding with closed streamlines around
 $g$ of the vortices, say, $s_{N+i}, \,1\leq  i \leq g$.   Then these curves, run clockwise,  will have circulations $p_i = - \Gamma_{N+i}, \,  i=1, \cdots,  g \,\,\,\, \text{with relation (\ref{byresidue}) being satisfied.} $ 
  
   \vspace{1mm}

The existence  and uniqueness proof for these $G_{\rm hydro}$ is attributed to Paul Koebe  \cite{Koebe, Koebe1}\footnote{Koebe constructs canonical conformal mappings by means of orthogonal series of analytic functions. 
From this he gets the hydrodynamic Green function with zero  inner boundaries periods, 
which  was used by C.C. Lin.}.

\begin{definition}
\textit{ The hydrodynamic Green function} $G_{\rm hydro}(s,r; \, p),\,\, p = (p_1, \cdots, p_g)$  is characterized: 
\vspace{-1mm}
\begin{enumerate}
\item  by being symmetric in $s$ and $r$;
\item  by being harmonic on each slot $r, s$ except for  the logarithmic singularity $-(1/2\pi) \log \ell(s,r)$;
\item for each  $r \in \Sigma^o$ fixed,  the function  $G_{\rm hydro}(s,r)$ on the first slot $s$
   has   constant values  on the boundary components. These values in general  depend on $r$ and are not specified.
\item by having
prescribed
  periods in the inner boundaries  
 \begin{equation} \label{anyso}  p_k = - \oint_{b_k} \star_r d_rG_{\rm hydro}(s,r)   
\end{equation}
($r$ parametrizes the inner boundary $b_k$ and runs clockwise).
\item
$
\oint_{\partial\Sigma} G_{\rm hydro}(\cdot;r) \star dG_{\rm hydro}(\cdot; s)=0 \quad \text{for all } r,s\in\Sigma \quad\text{(normalization)}.
$
\vspace{3mm}
\end{enumerate}
 \end{definition}

\newpage

The  integral (\ref{anyso})  gives  the same value $p_k$ no matter where the interior point $s$ is situated. 
Requirement iii) implies that  for each fixed vortex position  in $r$  we have a stream function. The difference of the values of a stream functions between two streamlines is the flux through a line segment joining them. The stream function changes when the vortex $r$ is displaced.

\begin{rem}
About Koebe, we quote from Ahlfors'  \cite{Ahlfors-1973}:
\begin{quote}
``We shall prove the famous \textit {uniformization theorem} of Koebe. This is perhaps the most important theorem in the whole theory
of analytic functions in one variable (\dots)  The only constructive element is contained in the Perron method for solving the Dirichlet problem (\dots) The proof rests on repeated applications of the maximum principle." 
\end{quote}
\end{rem}

\subsection{Motion of $N$ vortices on a planar domain via  $G_{\rm hydro}$}

Consider the stream function in $\Sigma$ given by
\begin{equation}
\psi(s) = \sum_{k=1}^N \, \Gamma_j  G_{\rm hydro}(s ,  s_j; p')\,\,,\,\,\,\, p' = (p'_1 , \cdots , p'_g ) 
\end{equation}
where $p'$  is a vector with dimensionless parameters.  Let us dimensionalize  the vorticities ($L^2/T$).

It is immediately seen  that the inner boundary circulations are  given by
\begin{equation}   \mathbf{p }= \mathbf{\Gamma} p' \, ,\,\,\, \Gamma = \sum_{i=1}^N \Gamma_i .
\end{equation}

\begin{teo}  Let $\mathbf{p}$ the dimensionalized vector of  inner circulations.   When  
$\Gamma \neq 0$ one can write the dynamics for the vortices $s_j$   via the Hamiltonian system 
\begin{equation} \begin{split}
\Omega  &= \sum_{i=1}^N \, \Gamma_i \mu(s_i) \,\,,\,\,
H_{\rm hydro} = \frac{1}{2}\sum_{j=1}^N \Gamma_j^2 R_{\rm hydro} (s_j; p')
+\frac{1}{2}\sum_{j=1}^N \sum_{1\leq k\ne j}^N \Gamma_j\Gamma_k G_{\rm hydro} (s_j, s_k; p') .
\end{split} \end{equation}
where  $p'= p/\Gamma$ is nondimensional. 
\end{teo}
\begin{rem}
If  $\Gamma = 0$  then $H_{\rm hydro} = H_{\rm Lin}$ and one must add the  terms with the outside agency. 
One could try different $p'_{s_i}$,  each representing  a vortex carrying its own harmonic complement,  to represent the same stream function
$$ \psi (s) = \sum_{i=1}^N \, \Gamma_i G_{\rm hydro}(s , s_i ; p'_{s_i}) \,\,\,  \text{where}\,\,\,p = \sum_{i=1}^N \m \Gamma_i  p'_{s_i},
$$
 Although valid, such expressions
do not seem to yield (at least in a simple way) a  Hamiltonian representation for the $N$-vortex motion.
\end{rem}

\subsection{Obtaining $G_{{\rm hydro}}$  from  $G_{{\rm electro}}$} \label{capmatrix}

Staring at the expression  $B Q B^\dagger $ with row vector  $B = p - \Gamma u, $  that appeared naturally in the 
Marsden-Weinstein reduction, one is   lead to  have an educated guess\footnote{There are actually a lot of indices and sums   when expanding $B Q B^\dagger $.} for the following result, that  to our  surprise simplifies  substantially earlier 
ones \cite{Gustafsson-1979-b, Flucher-Gustafsson-1997-a, Flucher-1999-a}. 

\vspace{1mm} 

 Let  $p = (p_1, \dots, p_g)$ be a prescribed list of periods (nondimensional).

\begin{teo} \label{capacteo}  
\begin{equation}\label{G2}
G_{\rm hydro}(s , r ; p )=G_{\rm electro}(s;r)+\frac{1}{2}\sum_{i,j=1}^g Q_{ij}U_i(s)U_j(r)\,\,\,,\,\,\, U_i=u_i-p_i.
\end{equation}
where the $u_i$ are the harmonic measures given by (\ref{uformula}). In particular  $G_{\rm Lin}$ corresponds to $p = 0$.
\end{teo}

We may  call $Q$  the {\it hydrostatic capacity matrix}, due to  analogies with electrostatics\footnote{We know from Proposition \ref{vanishes} that  $Q = P^{-1}$, and $P$ can be also called by the  name of  
electrostatic capacity. We plan to  discuss the various capacities notions  in a review paper.}. 

\proofn  
Items i)-iii) from the definition are obvious from construction. 
To compute iv), the period around $b_k$ ($1\leq k\leq g$),  we need a lemma:
\begin{lem}
The matrix $P$  in  the Riemann relations  for $\Hat{\Sigma}$ (Proposition \ref{RR}) can be written as  
\begin{equation}
 P_{ij}=\int_{\Hat{\Sigma}} \alpha_i\wedge \star\alpha_j=2\int_{{\Sigma}} \alpha_i\wedge \star\alpha_j
=\frac{1}{2}\int_\Sigma du_i\wedge\star du_j . 
\end{equation}
\end{lem}
This result we already know,  Lemma \ref{harm-meas} about harmonic measures.   So we continue  the derivation. The  second slot  is $r$ is fixed in this computation, so the operations of $\star, d$ and integration refer to the first slot.
We use  (\ref{rho}) and Theorem \ref{Dirichsol}. 
\begin{equation*}
\begin{split}
&
\oint_{b_k}\star dG_{\rm electro}(\cdot;r)+\frac{1}{2}\sum_{i,j=1}^g Q_{ij}U_j(r)\oint_{b_k}\star dU_i = \\
&
\,\,\,\,\, \,\,\,\,\,\,\,\,\,\,\, =\oint_{\partial \Sigma}u_k\star dG_{\rm electro}(\cdot;r)+\frac{1}{2}\sum_{i,j=1}^g Q_{ij}U_j(r)\oint_{\partial \Sigma}u_k \star dU_i \\
& \,\,\,\,\, \,\,\,\,\,\,\,\,\,\,\,
=-u_k(r)+\frac{1}{2}\sum_{i,j=1}^g Q_{ij}(u_j(r)-p_j)\int_{ \Sigma}du_k\wedge \star du_i \\
& \,\,\,\,\, \,\,\,\,\,\,\,\,\,\,\,
=-u_k(r)+\frac{1}{2}\sum_{i,j=1}^g Q_{ij}(u_j(r)-p_j)2P_{ki} \,\,\,\,\,\,\, \text{(by the lemma)} \\
& \,\,\,\,\, \,\,\,\,\,\,\,\,\,\,\,
=-u_k(r)+\sum_{i,j=1}^g Q_{ij}P_{ki}u_j(r)-\sum_{i,j=1}^g Q_{ij}P_{ki}p_j =\\
&  \,\,\,\,\, \,\,\,\,\,\,\,\,\,\,\, \,\,\,\,\, =
-u_k(r)+u_k(r)-p_k=-p_k .
\end{split}
\end{equation*}

\newpage

As for the normalization  (item v)) we have, using that the circulation around $b_o$ equals $$p_0= -\sum_{k=1}^g p_k + 1 :$$
\begin{equation*}
\begin{split}
&
\int_{\partial\Sigma} G_{\rm hydro}(\cdot;r) \star dG_{\rm hydro}(\cdot; s)
=\sum_{k=1}^g p_k G_{\rm hydro}(\cdot;r)|_{b_k} +(-1  + \sum_{k=1}^g p_k)G_{\rm hydro} (\cdot; r)|_{b_o} \\
& \,\,\,\,\,\,\,
=\sum_{k=1}^g p_k\Big(\frac{1}{2}\sum_{i,j=1}^g Q_{ij}U_i|_{b_k}U_j(r)\Big)
+(-1  +  \sum_{k=1}^g p_k)\Big(\frac{1}{2}\sum_{i,j=1}^g Q_{ij}U_i|_{b_o}U_j(r)\Big) \\
& \,\,\,\,\,\,\,
=\sum_{k=1}^g p_k\Big(\frac{1}{2}\sum_{i,j=1}^g Q_{ij}(\delta_{ik} - p_i)U_j(r)\Big)
+(-1 + \sum_{k=1}^g p_k)\Big(\frac{1}{2}\sum_{i,j=1}^g Q_{ij}p_iU_j(r)\Big) \\
&  \,\,\,\,\,\,\,
=\frac{1}{2}\Big(\sum_{k,j=
1}^g Q_{kj}p_kU_j(r)
 -   \sum_{k,i,j=1}^g Q_{ij}p_kp_iU_j(r)
-\sum_{i,j=1}^g Q_{ij}p_iU_j(r)
\,  +  \sum_{k,i,j=1}^g Q_{ij}p_kp_iU_j(r)\Big) = 0
\end{split}
\end{equation*} 
\qed 

\begin{rem}
 In hindsight, this should now be crystal clear.  Note that
$$
 G_{\rm hydro}(s,r) = G_{\rm Lin}(s,r)  - \frac{1}{2} \,p \,Q\, u(s)^{\dagger}  - \frac{1}{2}\, p \,Q u(r)^{\dagger} +  \frac{1}{2}\, p \,Q p^{\dagger}.
$$
The stream function $G_{\rm Lin}(s,r)$ produces zero periods, hence  all the circulations must come  from
 $$\psi(s) = -   \frac{1}{2} \,p \, Q \, u(s)^{\dagger} . $$
 
Take  the case where only  $p_1= 1$  is nonzero:  we have  $$\psi =  -   \frac{1}{2} ( Q_{11} u_1 + \cdots  + Q_{g1} u_g ). $$ 
\begin{equation*}
\begin{split}  &\,\,\,\, -  \star d\psi =    \frac{1}{2}  \star ( Q_{11} du_1 + \cdots Q_{g1} du_g)
 =  Q_{11}  \star  \alpha_1  + \cdots Q_{g1}  \star  \alpha_g \,\,\,\, \, \text{(by (\ref{alphau})) }
\end{split}
 \end{equation*}
This means that $ -  \star d\psi  =   Q_1  \star \alpha$
where $Q_1$ is the first row of $Q$ and $\alpha$  is viewed as a column vector of 1-forms.
Now,  by the Riemann relations  we have $\star \alpha = P \beta$.  Therefore
$$ -  \star d\psi  = Q_1 P \beta = (1,  0 , \cdots , 0) \beta = \beta_1
$$
which produces (unit) circulation only on boundary curve $b_1$, and similarly for the others. 
\end{rem}

\newpage

\subsection{Redundancy between normalization   and symmetry}
  \vskip 1truemm
 
  \begin{prp} \label{below} 
In the definition of the hydrodynamic Green function, the symmetry requirement is redundant. It is a consequence of the normalization iii).
    \end{prp}
    \proofn 
    \[\begin{split}
       &\oint_{\partial \Sigma} G_{\rm hydro}(\cdot;r)\star dG_{\rm hydro}(\cdot;s)-
       \oint_{\partial \Sigma} G_{\rm hydro}(\cdot;s)\star dG_{\rm hydro}(\cdot;r)\\ &
       = \int_{\Sigma} d\Big(G_{\rm hydro}(\cdot;r)\star dG_{\rm hydro}(\cdot;s)\Big)-
       \int_{\Sigma}  d\Big(G_{\rm hydro}(\cdot;s)\star dG_{\rm hydro}(\cdot;r)\Big)\\ &\,\,\,\,\, =
        \int_{\Sigma} dG_{\rm hydro}(\cdot;r)\wedge\star dG_{\rm hydro}(\cdot;s)-
        \underbrace{\int_{\Sigma}  dG_{\rm hydro}(\cdot;s)\wedge\star dG_{\rm hydro}(\cdot;r)}_{ \int_{\Sigma} dG_{\rm hydro}(\cdot;r)\wedge\star dG_{\rm hydro}(\cdot;s)}  \\ & \,\,\,\,\,\, \,\,\,\,\, \,\,\,\,\,\, \,\,\,\,\,  -
        \int_{\Sigma} G_{\rm hydro}(\cdot;r)\delta_s(\cdot)\mu(\cdot)+
        \int_{\Sigma} G_{\rm hydro}(\cdot;s)\delta_r(\cdot)\mu(\cdot)\\ & \,\,\,\,\,\, \,\,\,\,\,  =
       -  G_{\rm hydro}(s;r)+ G_{\rm hydro}(r;s)\,,
    \end{split}
  \]

Therefore, if the normalization condition is verified, then $G_{\rm hydro}$ is symmetric.  \qed

\vspace{2mm}
Conversely,  if a $G_{\rm hydro}$  is symmetric then the normalization  is satisfied up to an overall constant.

\begin{prp} \label{below1} Suppose that $G_{\rm hydro}$ verifies the conditions (i) and (ii),  
  including symmetry. Then the following normalization condition holds
     \begin{equation} \oint_{\partial \Sigma} G_{\rm hydro}(\cdot;r)\star dG_{\rm hydro}(\cdot;s)=c\ \text{ for all}\
    r,s\in \Sigma \,,\label{norm2}\end{equation}
where $c$ is independent of $s$ and $r$. 
 \end{prp} 
 \proofn Let $q_o\in b_o$, $q_1\in b_1\ldots$ be any points in the boundary components of $\Sigma$.
 Since the periods are prescribed $\oint_{b_k} \star dG_{\rm hydro}(\cdot,s)=-p_k, \, 1 \leq k \leq g, $ (condition (ii)) with
 $p_o=1-(p_1+\ldots p_n)$, 
 and $G$ is constant at the boundary components, we have
 \[
   \oint_{\partial \Sigma} G_{\rm hydro}(\cdot;r)\star dG_{\rm hydro}(\cdot;s)=-\sum_{k=0}^g G(q_k,r)p_k,\,\,\,\int_{\partial \Sigma} G_{\rm hydro}(\cdot;s)\star dG_{\rm hydro}(\cdot;r)=-\sum_{k=0}^g G(q_k,s)p_k
\]
The symmetry of $G_{\rm hydro}$ implies that these two last functions must be
the same. The first depends only on $r$ and the second only on $s$ so  neither depend  on  $r$ nor on   $s$.
\qed

 \begin{exa} \label{impulsive} No values of circulations  $p_{\ell}$ seem to be  more natural than  others.  One  choice is 
 when  a flow,  initially at rest,  starts impulsively by releasing   `stirrers'\footnote{`Stirrers', also called `agitators'   appear in studies of chaotic mixing, see eg.  \cite{Aref1984,Khak,Ottino,Ditch}. }  with strengths $\Gamma_k, 1 \leq k \leq N$ at  given  points $s_1(0), \cdots, 
  s_1(0)$.  
   The speed of sound (or information) is infinite in an incompressible fluid:  the boundaries  respond instantaneously. 
Then the inner boundaries circulations will be 
\begin{equation}  \label{constantmotionimp}
p_{\ell} =    - \sum_{k=1}^g \Gamma_k\, \oint_{b_{\ell}} \star dG_{{\rm electro}}( \cdot , s_k(0)) =
\sum_{k=1}^g \Gamma_k\, u_{\ell}( s_k(0)) .   
\end{equation}
$\,$
  \end{exa}

For the proof, consider the following   \textit{gedanken} experiment: 
take a pair  $s_o, \tilde{s}_o$ of image points in $\Hat{\Sigma}$.
During a very short time $\tau$ make
the vorticity $\Gamma(t)$ of the stirrer at  the front face  to  increase steadily from 0 to 1, and its  image decrease from 0 to -1.  They will move according to the time-dependent complete system of vortex equations  (\ref{sflat}, \ref{Pell})). 

We assert: 
making $\tau \rightarrow 0$,  in the limit neither  the pair of vortices nor the harmonic components  will 
  change from their initial values. 
This is simply because   the right hand side of the ODEs are bounded.
Therefore, the initial conditions for the flow starting impulsively will have  in the Schottky double the same original data:
   $s, \tilde{s}_o,\,B=0$,  
and   vortex strengths $\pm 1$.    \qed

\begin{exa}  \label{exaannulus}  Hydrodynamical Green function  for the annulus $A: r \leq |z|\leq R.$    
 The harmonic measure for 
 $\, b_o:  |z|= R $ (counter-clockwise),  $\, b_1: |z| = r $ (run clockwise) is 
 \begin{equation}
u_1(z)= u = \frac{\log R/|z|}{\log R/r}.
\end{equation} 

\noindent Let $c$ be the prescribed circulation on the  inner circle $b_1$ (run clockwise).
Theorem \ref{capacteo} gives:
$$ G_{\rm hydro}(z;z_o) =   G_{\rm electro}(z;z_o) +  Q (u(z) - c)(u(z_o) - c) .
$$
The matrices $P$ and $Q$ have only one entry each,  $Q=1/P$.  A quick calculation gives
$$
P=\frac{1}{2}\int_A du\wedge\star du =\frac{\pi}{\log R/r}.
$$
Modulo the factor $1/2$ in $P$ (which can be viewed as a factor of convention) this agrees with what one 
finds in elementary text books for the electrostatic capacity. The ``hydrostatic'' capacity is
\begin{equation}
Q=\frac{\log R/r}{\pi},
\end{equation} 
The electrostatic   Green function  $G_{\rm electro}(z;z_o) $   for the annulus can be found  in several sources,  e.g., in Courant-Hilbert
\cite{CourantHilbert}  (V, section 15,  art. 7).  Vortex problems on   annular domains have been studied by several authors, see \cite{Nadia, Vaskin, Kurakin}.
\end{exa}

\subsection{About  the three approaches on multiply connected domains} 

As for the comparison between i) Marsden-Weinstein reduction, ii)  C. C. Lin's Kirchhoff-Routh hamiltonian, and iii) the hamiltonian via the hydrodynamic Green function, we wish to point out some highlights of the former.

We believe that the  Marsden-Weinstein's reduction has a conceptual advantage   because
it stems from our general equations
in the beginning of the paper.  Moreover,  the Hamiltonian in (\ref{GLindef})  is presented with a quite visible  structure,
quadratic in the $\Gamma_j$ and the $p_j$ taken together.
 This structure is given in the general
context in Theorems \ref{gus} and \ref{Hamstru}. 

Lin's Green function is a special case of the hydrodynamic Green function, and we think Lin perhaps used it because the  Green function with vanishing  inner circulations 
was well-known at the time of his writing  (via the work of Koebe and others). The circulation of this Green function
around the outer boundary component is then equal to 1,  by necessity.

In the case of  a flow  with zero inner circulation on the inner boundaries one can add up these Lin's Green functions for  the several vortices without problems. \\

   However,  in order to deal with 
general boundary  circulations one has to compensate the stream function adding harmonic flows. The term ``outer agency''
may have some historical interest, but to us  it sounds strange\footnote{Could it be a pun?  There were a lot of spy agents
around in the war time when Lin  was writing the  paper. }.

In the Kirchhoff-Routh  hamiltonian  (\ref{KRHamilt})  the added terms  appearing in Lin's paper  were seemingly not quadratic in the vortex strengths
and circulations. Eventually they are quadratic of course, but that was somewhat hidden. 

In Theorem \ref{Linteo},
there are explicit single factors $\Gamma_i$ in the last term of  (\ref{KRHamilt}), and then there are implicit factors 
${\bf p}$ in the ``outside agency'', as we now exhibit in (\ref{GLindef}).

\subsection{Further comments and related questions}

\noindent i) The choice of inner circulations $p_i=0, \, 1 \leq i \leq g$  does not  seem to represent  any particular insight. Probably Lin simply took  it
by default
from Koebe's theory of conformal mappings.  
It is instructive to take an annular region and  compare the two  cases $p=0$ and $p=1$.
They are equally good and represent all 
circulation going to one and the same component, either the outer  or the inner component, respectively.
In  both cases, $$f(z)=\exp\big(-2\pi(G_{\rm hydro}(z,w)+iG^*_{\rm hydro}(z,w))\big)$$ maps the domain
onto a circular slit unit disk. The component taking the circulation being mapped onto the unit circle
(this because the normalization requires $G_{\rm hydro}$ to vanish on that component). 
And we have $G_{\rm hydro}=G_{\rm Lin}$ only when $p=0$. 
With $p=\frac{1}{2}$, half  the circulation goes to the inner component, half to the outer. 
The function $f$ above is not univalent, not even single-valued.\\

\noindent ii)  About Lin's ``outside agencies".  As we mentioned, the terminology is a bit odd and  somewhat misleading. 
We are working on a fully deterministic problem given by the initial conditions and Euler equations, with  no  external influences
in the flow.  Helmholtz theorem prevents the boundary circulations to change.   Of course, interferences on a flow can (and need) to  be done in practical matters, for example in  boundary layer control of   aerodynamic flows,  by inserting  stirrers and  sinks/sources on the boundaries. 
However, control theory was not what Lin had in mind. He just wanted a mathematical procedure to deal with the more general case of non-zero (but fixed) circulations on the internal boundaries.  
 
 \vspace{1.5mm}

The next comments are meant to be just a glimpse of  themes from  \textit{geometric function theory}  related to vortex motion and several other physical applications.  We plan to prepare a review paper involving classic differentials, Bergman (and other)  kernel functions,  capacities, Robin functions and curvature.  There is a sizeable literature going  back to the 1970's.  Some of the papers (among others) related to  vortex motion  are
 \cite{Gustafsson-1979-b, Bandle, Flucherbook, Richardson}.
 
 \vspace{1.5mm}
 
\noindent iii)  In the expression (\ref{G2}) for the full hydrodynamic Green function, the contribution from the conformal geometry, namely the matrix $Q$  is separated from the circulations $p_j. $
Previously, in \cite{Gustafsson-1979-b, Flucher-Gustafsson-1997-a} these two contributions were tied up with each other. Mysteriously, when taking mixed second derivatives towards Bergman kernels, the dependence of circulations disappeared and only that of $Q$ survived. By Theorem \ref{capacteo}  it is now clear why. \\

\noindent  iv) For critical points of the  electrostatic Green function  $G_{\rm electro}(z, \zeta)$  (with $\zeta$ fixed) of  bounded multiply connected domains, see  e.g \cite{GuSebar, Borah} and references therein.  If the  connectivity is $g + 1$, then  there are $g$
critical points.   This is elementary.: $\partial\, G_{\rm electro}(z,\zeta)/\partial z$ is a meromorphic  differential in the double, so  the relation $\#$
  zeros  $ - \, \#$  poles   = $2g - 2 $ becomes 
$$  2  \times \text{number of critical points for} \,  G_{\rm electro} - 2 = 2g - 2 .$$
Main themes are  the   differential geometry of the level lines, and the  limiting positions 
of the equilibria $z_j(\zeta),\, 1\leq j \leq g$ as the pole $\zeta$ approaches the boundary. For instance, it is known that they stay in a compact set independently of $\zeta$  \cite{GuSebar, Solynin}.
 The same problem    may be formulated
with respect to the hydrodynamic Green function with similar results.\\

\noindent v)    If  $f_{\zeta}(z) = G_{\rm hydro}(z, \zeta)$ has a critical point  at 
 $z_o(\zeta)$  then usually $ f_{z_o(\zeta)}(z) = G_{\rm hydro}(z, z_o(\zeta))$  does not have a critical point at $z = \zeta$.  In some cases this may happen, for instance if there exists an isometry exchanging $\zeta$ and $z_o(\zeta)$.
  If furthermore both $\zeta$ and $z_o(\zeta) $ are  critical points of the Robin function, then  the pair will be an equilibrium for the 2-vortex system no matter the vortex strengths, as noted in Remark \ref{insight}. \\
 
 \noindent vi)  The electrostatic Robin function tends to $-\infty$ the boundary.  This follows from the inequaity
 $$
 \log d(z) \leq  R_{\rm electro}(z)\leq\log d(z) + {\rm const.}
 $$
where $d(z)$ is the distance to the boundary \cite{Gustafsson-1979-b}.  Similar estimates for $R_{\rm hydro}$  follow  immediately, since keeping the circulations fixed the difference between the two Robin functions is bounded.\\

\noindent vii) Therefore one may search for \textit{maxima} inside the domain.  When the domain is convex  (in particular simply connected) there is only one such point for the electrostatic Robin function (which is also the hydrodynamic).  It is shown in \cite{Bandle,Flucherbook,Guconvex} that  $-R_{\rm electro}$ is  convex.  Actually  a strong result holds: the so-called
\textit{mapping radius} function
 $r(z) = \exp(2\pi R_{\rm electro}(z))$ is concave.\\

\noindent viii) 
 We end this subsection with a simple observation.  The single vortex system is governed by
\[
R_{\rm hydro}=  R_{\rm electro}(s)+ \frac{1}{2}\sum\limits_{i,j=1}^g Q_{ij}(u_i(s)-p_i)(u_j(s)-p_j)
\]
The equation for the equilibrium is
\begin{equation*}
 d R_{\rm electro}(s)+ \sum\limits_{j=1}^g du_i(s)Q_{ij}(u_j(s)-p_j)=0 .
\end{equation*} 
  If we fix a point $s \in \Sigma$ and consider $p\in\R^g$ as an unknown, then each of the two components of the differential gives
  a scalar equation for $p$.  More precisely, we obtain the linear system 
  \vspace{1mm}
  \begin{equation*}
    M_{2\times g}\, p_{g\times 1}=b_{2\times 1}
    \label{Mb}
    \end{equation*}
  where one can suppose that  at least generically $M$ has rank 2, 
 \begin{equation*} \begin{split}\,\,\,\,\,\,\,\,\,\,&M_{1,j} =\sum\limits_{i=1}^g \partial_x u_i(s)Q_{ij} \,,\, \,\,
      M_{2,j} =\sum\limits_{i=1}^g \partial_y u_i(s)Q_{ij}\quad (j=1,\ldots,g)\\
      &b_1 =\partial_ x R_{\rm electro}(s)+\sum\limits_{j=1,j}^g \partial_xu_i(s)Q_{ij}u_j(s),\,\,\,\,
            b_2 =\partial_ y R_{\rm electro}(s)+\sum\limits_{j=1,j}^g \partial_y u_i(s)Q_{ij}u_j(s). .
          \end{split}
\end{equation*}
        
 The system in overdetermined, determined, underdetermined if $g$ is 1,2, $>2$,
 respectively. 
 \vspace{1mm}
 
When  $g\ge 2$ we expect to have a set of solutions of dimension $g-2$.
 This means that  for $g\ge 2$  \textit{every}  point of the domain may be a vortex equilibrium if the circulations
 are conveniently chosen.

\subsection{Implementation of  the hydrodynamical Green function}
\vspace{1.5mm}

 D.Crowdy and collaborators use Schottky-Klein prime functions to find the hydrodynamical Green functions   of \textit{canonical} multiply connected   planar domains \cite{Crowdybook,Crowdy0, Crowdy1, Crowdy, Crowdy-prime}.  Here the word canonical means
  domains with slits and circles as boundaries. 
  Conceptually,   if  one knew the  modified  Green functions of a general domain $\Sigma$, then  with $r$ fixed, the function 
$$
f(\cdot; r)=\exp \big(-2\pi G_{\rm hydro}(\cdot ;r)-2\pi \I G_{\rm hydro}^*(\cdot ;r)\big) $$
will map $\Sigma$ conformally onto the unit disk with $g$ circular slits removed, taking $r$ to the origin.

Note that $|f(s)|$ is constant on each boundary component and that the multivaluedness of the harmonic
conjugate $G_{\rm hydro}^*(s;r)$ is swallowed by the exponential.
This is because all periods of $$\star dG_{\rm hydro}(\cdot;r)=dG_{\rm hydro}^*(\cdot;r)$$  
are either one or zero. Thus $f(\cdot;r)$ is single-valued, and an application of the argument principle shows that 
it in addition is univalent, see \cite{Nehari-1952a, Sario-Oikawa-1969a}.
\smallskip

We have an alternative proposal. 
\begin{enumerate}
\item Compute    $G_{{\rm electro}} = - (1/2\pi) \log |z-z_o| + h(z;z_o)$ numerically. This is classical.  
For each $z_o$, the Dirichlet boundary conditions for the  harmonic complement $h(z; z_o)$ are  known.
\item Compute the harmonic measures $u = (u_1, \cdots, u_g)$ (again, these are Dirichlet problems).
\item Compute  the Riemann relations $Q $  for the domain (computationally mature: \cite{Trager, Luo}).
\end{enumerate}

\subsection*{The ``World islands"}
\vspace{1mm} 

A convincing  example  is the  unbounded domain   of connectivity 210, exterior to an artificial archipelago located in the waters of the Arabian Gulf, 4 km off the coast of Dubai,  known as ``The World Islands".   We refer to    Nasser \cite{Nasser, Nasser1}. 
His numerical codes allows to  compute the hydrodynamic Green function for domains with high multiple connectivity,  complex geometry,  with close-to-touching boundaries, and domains with piecewise smooth boundaries.

\begin{figure}[!ht]
\includegraphics[scale=0.55]{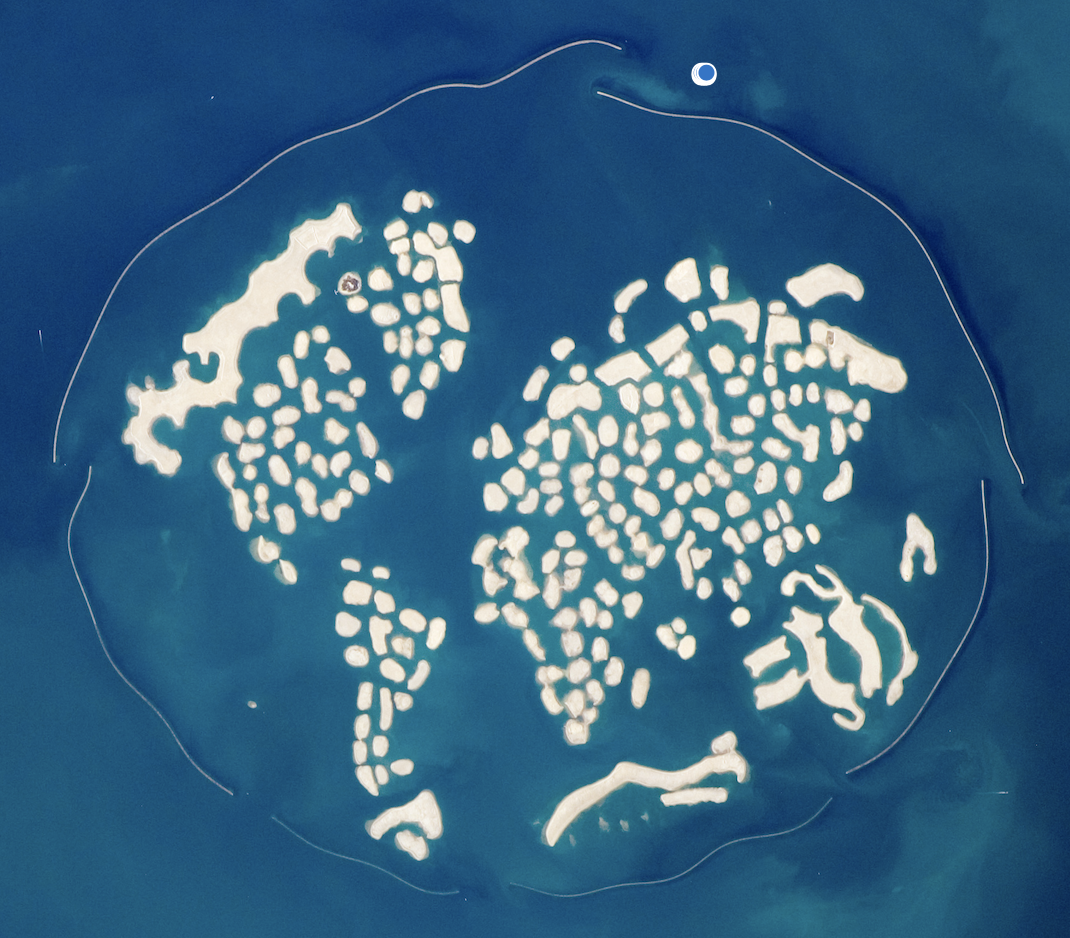}
\caption{`World'Archipelago, Dubai. Crop from \url{https://en.wikipedia.org/wiki/The_World_(archipelago)}}
\end{figure}

\subsection*{Dimensionalization} 

 In order to   ``navigate by instruments" on the  flow  using Euler's equation,  a  simple scaling suffices.  Choose units of length (L) and time (T)  appropriate for the context.    
The other  five fundamental base units  of the SI system are not used, not even mass - the  fluid density does not appear in the equations 
(viscosity  is neglected and the fluid is assumed incompressible).  Keep time   $t$ dimensional. \\

\noindent i)  Compare the  size of the true domain with that  given abstractly by the  Riemannian metric. Suppose,  for instance,  that a  lake has area
  $A \, { \rm m}^2$ ($A$ is a `pure' number), while the Riemannian metric gives  has the value $V$ (a `pure' number also) for the area. \\
  
\noindent    Then, 
if in the  Riemannian metric the distance between $s_1$ and $s_2$ is  $\ell(s_1, s_2)$  (nondimensional), the true distance of the corresponding points in the lake is  $\sqrt{A/V}\, \ell(s_1, s_2)$ (in meters).  
 By the same reason, if  $|\dot{s}|$ is the norm  (using the  inner product) of the tangent vector $\dot{s}  = ds/dt \in T\Sigma$, then  
the true velocity of the stream in the lake will be also dimensionalized by $ \sqrt{A/V} |\dot{s}| $ m/s .  \\

\noindent ii) To solve the ODEs we  must first nondimensionalize  the the vortex strengths   $\mathbf{\Gamma}$,   and when there are boundary curves,
the  circulations    $\mathbf{p}$  around them.  The letters in bold are used to keep in mind the dimension  $L . L/T = L^2/T .$
Therefore
$$     \Gamma^{\rm model} =  (V/A) \mathbf{\Gamma}  \,, \,\,\,  p^{\rm model}  = (V/A) \mathbf{p} \,\,\,\,\,\,\, (\text{having dimension}\,  \, T^{-1})
$$

  The superscript $^{model}$  means ``to insert in the ODEs".
 
Observe that to measure  circulation around a boundary curve,   any  curve  $\gamma$ (not necessarily a stream curve)  drawn 
conveniently away from it could be used.   The only requirement is that there are no vortices   between that curve and the boundary.\\

Fluid metrology is a well developed subject\footnote{\url{https://www.nist.gov/pml/sensor-science/fluid-metrology}}. 
For measurements on  a lab, one uses  particle image velocimetry (PIV).  For large scale phenomena  in the ocean and in the atmosphere
satellite images and balloons/buoys  with meteorological instruments.  

\newpage

\section{COMMENTS AND QUESTIONS} \label{questions}

\vspace{1mm}

 In addition to the questions scattered in the previous sections  in this final  section we (humbly)
  list some  problems that occurred to us while writing this paper. We refer to  Yudovich (\cite{Yudovich2003}) and to  the recent article by Khesin, Misiolek and Shnirelman \cite{Khesinopen}, a   collection  state-of-the-art  open  problems 
in the    hydrodynamical mainstream (no pun intended).

\vspace*{-2mm}

\subsection*{General setting  (arbitrary dimension)}  

\vspace{1mm}
  \begin{problem}  \label{pb1} 
   Find theoretical estimates for the energy interchange between   the vorticity  (${\omega}$)   and    the harmonic ($\eta$) components  (there is a first attempt   in 
 \cite{Iushutin}, but we believe it must be scrutinized). 
 \end{problem}

    \begin{problem} Kelvin-Arnold's stability test when the homology is non trivial.  What does the Hodge decomposition imply?
   \end{problem}
 See  \cite{Davidson}, sections 11.6, 11.7  for an elementary exposition of  Arnold-Kelvin theorem and \cite{Vladimirov} for
  historical background\footnote{Lord Kelvin (\cite{Kelvin}, 1887):  `The condition for steady motion of an incompressible inviscid fluid filling a finite fixed portion of space (that is to say, motion in which the velocity and direction of motion  continue unchanged at every point of space in which the fluid is placed)  is that, with given vorticity, the energy is a thorough maximum, or a thorough minimum, or a minimax. The further condition of \textit{stability}  is secured, by consideration of energy alone, for any case of steady motion for which the energy is a thorough maximum or a thorough minimum; because when the boundary is held fixed the energy is of necessity constant. But the mere consideration of energy does not decide the question of stability for any case of steady motion in which the energy is a minimax."}.``Steady fluid flows are critical points of the restriction of the
Hamiltonian (which is the kinetic energy function defined on the dual space) to the
spaces of isovorticed fields, i.e., sets of fields with diffeomorphic vorticities. If the
restriction of the Hamiltonian functional has a sign-definite (positive or negative)
second variation at the critical point, the corresponding steady flow is Lyapunov
stable."(B. Khesin, \cite{Khesin}). Other recent references are \cite{Izosimov1, Izosimov2, Izosimov3}.

\vspace{2mm}

   \begin{problem}  \label{pb1a}  We called {\it incomplete} the system with the second equation  removed from
(\ref{omegaeta}). 
The harmonic part  $\eta$ is taken as a constant flow (say,  zero). 
  Consider the complete system and suppose  that the flow starts `purely vortical'  at  $t=0$. An harmonic component  will immediately appear  and, in turn, will start to 
 affect the vorticity.  Quantify by how much the incomplete  is off the mark, compared with the evolution of
  the complete system. Simulations in \cite{SanDiego} indicate that the divergence may occur  quickly, but it should depend on the relative initial energies.
 \end{problem}
 \begin{problem} 
 Analogous question, comparing with a flow in the universal cover,  where the initial vorticity  is confined in one of the cells.  For tori this has been addressed in \cite{Peskin}.
 See problems \ref{pb6}, \ref{pb7}.
\end{problem}

\noindent {\bf Comments. }  Flows on bounded multiply connected domains with  non-smooth initial data, specially vortex sheets  were  considered  in \cite{Helena}.  It was  claimed that Kelvin's boundary circulation conservation could be violated. But what would be the physical mechanism?

We saw that for genus $g=1$  the exchange of vortical and
harmonic energy should occur  when the surface is curved. 
In this situation, heuristically, curvature should play  a role similar to topography in lake equations \cite{Deke, grote1999dynamic}, yielding the coupling between the
vortical (the ``small-scale") and the harmonic (the ``mean flow") parts
of the flow.

Long-time numerical simulations in the plane  with periodic boundary conditions and in the sphere
  suggest  that,  starting with random initial vorticities,  after long times the fluid arranges itself  in ``vortex patches" \cite{Modin}.   Curiously, on some occasions, the dynamics seems being attracted to trajectories typical of integrable systems\footnote{See the nice videos in \url{https://www.youtube.com/user/thdrivas}}.
  Other numerical works obtained similar conclusions \cite{Shnirelman}.
For non-flat metrics such  formation of vortex patches could be revisited,  \textit{taking under consideration the interplay with the harmonic fields.}

Problem  \ref{pb1}   belongs to the $L^2$ realm.  Out of our scope are the subtler questions belonging  to $L^{\infty}$ and  higher order Sobolev norms  (were pointed to us by  Edriss Titi). For instance,  double exponential growth in time of
the vorticity gradient - a 2d phenomenon that  takes place near
boundaries\footnote{We just mention (somewhat opportunistically):   simply \textit{taking the Schottky double, these phenomena would also happen on a compact boundaryless surface.} }  found by  Yudovich \cite{Yudovich, Yudovich1}.  See \cite{Kiselev} for the current status of the theme.   Another research line is the spatial spread of vorticity,  starting on a confined region.

   \begin{problem}
  What is the long term effect of viscous dissipation (properly defined) on the coupled system?  Will a steady nonzero harmonic component subsist?
      \end{problem}
  The Hodge Laplacian $\Delta_H  = d\delta + \delta d$  annihilates harmonic forms. In  a flow $\nu = \delta \Psi + \eta $  (Theorem \ref{coupledeq}) a $\Delta_H $-viscous dissipation will
  act only   the vorticity component $\delta \Psi$, but  decomposing the dissipation term  $\Delta_H \delta \Psi =\delta d \delta \Psi $ may introduce an harmonic component. Worse even: in the problem statement  we wrote `properly defined'  because the correct  operator to describe dissipation on curved manifolds  is a matter of dispute \cite{finlandia}.

\subsection *{Dimension 2} \label{relating}   
\begin{problem}
Computation of the Green function $G_{{\rm double}}$ for Schottky doubles of planar domains.
\end{problem}

 Unfortunately, the image method for hydrodynamical problems, conceptually so elegant\footnote{Maxwell's \textit{Treatise on Electricity and Magnetism}
(1873, Vol. 1, Ch. XI, p. 245-283, 3rd Ed.) devotes 40 pages to the reflection techniques.},  seems to be not useful  in practice due to the technical difficulty of getting a hand in $G_{{\rm Schottky}}$.

Coming back to the remark \ref{Gelectroback}, one can extend $G_{{\rm  electro}}$  to $\Hat{\Sigma}$ immediately,  but it has two opposite  poles:  $s$   ($+$) in the front side and  $\tilde{s}$  ($-$) 
on the backside.
In order to  compute $G_{{\rm double}}$  from $G_{\rm electro}$  one would need, so to speak,
`un-balayage' the  concentrated counter-vorticity  at $\tilde{s}$, spreading it  all over $\Hat{\Sigma}$  uniformly.   Recall that the curvature of the Lipschitz metric is concentrated on the doubling boundaries. 
This question is up to grabs.

In section \ref{simplyconnSho} we presented a workable procedure  for simply connected domains.   
 $G_{{\rm double}}(\D) $  is known, and one has (in principle)   the Riemann mapping theorem at one's  disposal.

\subsection*{Numerical experiments for torus metrics (genus=1)}

\begin{problem} \label{pb6}
 Simulations with various torus  metrics to investigate how fast the solutions of the incomplete vs. solutions of the full system diverge.
 For initial  $|\omega |$  `big',   
    supported on a `small' region of a cell of  $\Sigma$ in the universal cover,   how long it takes for the fluid to `perceive'   the topology?
    \end{problem}
    \begin{problem} \label{pb7}
On the other direction,  starting with a  `big'  harmonic flow $| \eta |$, how much is it perturbed by a `small' initial vorticity?
 \end{problem}

\subsection*{Numerical experiments on hyperbolic surfaces (genus $\geq 2$)}

\vspace{1mm}

 There are beautiful Riemann surfaces waiting to be filled with vortex dynamics (generalized  Bolza, Klein, Bring, Macbeath, and many more).  
Theoretical as well as computational tools are already  available for   the numerical computations both of harmonic forms and   Green/Robin functions  on compact surfaces with the constant  negative curvature   metric (\cite{Avelin}, \cite{Jorgenson}, \cite{Strohmaier}).
Codes by Strohmaier and Uskii can be used  for this task\footnote{\url{ http://www1.maths.leeds.ac.uk/~pmtast/hyperbolic-surfaces/hypermodes.html}\\
 \url{http://www1.maths.leeds.ac.uk/~pmtast/publications/eigdata/datafile.html} .}.   
 
The Green function $G_{\Sigma}$ can be obtained  integrating the heat kernel  of $\Sigma$ for $t \in (0,\infty)$.  There are two standard representations of the heat kernel of $\Sigma$, roughly speaking one good for short and the other for long times. For implementations on an ODE solver, one needs to compute values of Green and Robin functions  for  a representative number of points in the fundamental domain $\mathcal{F}_{\Sigma} $.
    For first and second derivatives, numerical differentiation could be done using the tabulated values of $G,\, R$.   Actually,  one can derive the   partial derivatives up to the second order:
     differentiating the  heat kernel increases the decay  in the covering space,  which helps in the sum of the group replicas.

More generally, computer graphics and machine learning communities reached a mature stage in methods to invert  the Laplace-Beltrami operator,  for any  metric on a compact surface\footnote{See e.g.   works by  Yau and collaborators \cite{Yau} and from Desbrun and associates \url{http://fernandodegoes.org/}.}.
Several codes are now readily available  for applications. 
    
    \vspace{1mm}
    \begin{problem} The Robin function of Bolza's surface was numerically computed in \cite{Ragazzo0}.  A natural continuation is to study the complete
  dynamics    (\ref{Guseq1}, \ref{Guseq2}) taking the harmonic part into account. How much would
 the Robin function oscillate?
  \end{problem}
  
\begin{problem} \label{pb9} A vortex dipole can be thought as being in the ground state.
For pair $s_1, s_2$ of opposite intensities, the vortical piece of the energy goes as $\frac{1}{4\pi} \log \ell(s_1, s_2) \sim - \infty $.  
For how long a pair of close by opposite vortices shadows a geodesic? If the initial harmonic component is very large,  can the interchange of energy make
the vortex pair to dissociate?
\end{problem}

\vspace{1mm}

\subsection*{Dimension $\geq 3$ (see Appendix B  for background on  Biot-Savart )}

\vspace{2mm}

Given an exact 2-form $\omega$,    Hodge theorem  guarantees that there exists a 2-form $\Psi$ such that \vspace{1.5mm} 

\centerline{ $d (\delta \Psi) = \omega$
(any $ \Psi + {\rm kernel}(\delta)$ would be fine).}

\begin{problem}  As a  tribute to Jean Baptiste Biot e F\'elix Savart one may consider the following general question: in dimensions $\geq 3$, to recover  the   vortical term
$\delta \Psi$ given the vorticity $\omega$.
\end{problem}
\vspace{-1mm}
 
  \begin{problem} Dynamics of vortex lines in the 3-sphere (see \cite{Dix}).   Here there is no harmonic component.  Use Hopf fibration for a reduction to $S^2$.
  
  \vspace{1mm}
 
\end{problem}
  Biot-Savart formulas    for $S^3$   can be found in
  papers by H. Gluck and associates \cite{Gluck1,Gluck2,Gluck3}.   

  \begin{problem} \label{prob10} Product of spheres and flat Tori.
  Find Biot-Savart formulas for $T^3 (= S^1 \times S^1 \times S^1$),  for $S^2 \times S^1$ and more generally, for  products of spheres of  various dimensions. 
  \end{problem}
    \vspace{1mm}
    
  \begin{problem}
Decomposed Euler equations for them.  
  Special solutions via symmetry reductions.
  \end{problem}
 
   \vspace{1mm}
\begin{problem} Could Hodge splitting  be useful for  topological hydrodynamics in dimension three?
\end{problem}

The famous ABC+G flow (Arnold-Beltrami-Childress-Gromeka) in the 3-torus \cite{ABC, Gromeka}   has been the subject of extensive  numerical studes, see eg. \cite{Dombre, Zhao, Galoway}. From the theoretical aspect, recall that
Beltrami flows are those for which vorticity and velocity  are parallel \cite{Ghrist0, Ghrist00, Ghrist}.  

Around 25 years ago their connection with contact topology and 3d-hydrodynamics  has been shown in a series of papers by  Etnyre and Ghrist,  including results on  the Weinstein  conjecture.
 Beautiful new  results about steady solutions of Euler equations with Beltrami vorticity have been obtained  by   Boris Khesin and associates (already cited: \cite{Khesin3dHodge, helicity3d, Khesin}), and by Eva Miranda and her collaborators,  \cite{Miranda,  PSalas,  Enciso2}. 

\vspace{1mm}

\section*{ACKNOWLEDGMENTS}

\noindent (JK) wishes to thank Vladimir Dragovic, Valery Kozlov, Ivan Mamaev, Zoran Rakic and Dmitry Treschov, organizers of
GDIS 2022, Zlatibor, Serbia, June 5-11 2022,
and the organizers of the conference 
on the memory of Alexey  Borisov,
November 22-December 3, 2021, Steklov   Institute. \\ \\
\noindent (BG)  and (CGR) wish to thank Stefanella Boatto and other organizers of the conference in November 2012
in Rio de Janeiro, ``N-vortex and N-body dynamics: common properties and approaches". This is what made (BG)  return to working on vortex motion after many years of absence from this field. \\ \\
\noindent The three authors thank Boris Khesin and Edriss Titi for references and insights. 
\section*{AUTHORS' CONTRIBUTIONS}
\begin{center} All authors participated in discussing the results and in writing the article. \\ BG was   Salviati, CGR was
Sagredo and JK was
 Simplicio.
 \end{center} 

\vspace{1.5mm}

\begin{center}
APPENDICES 
\end{center}

$\,\,\,\,\,\,\,\,\,\,\,\,\,\,\,\,\,\,\,\,\,\,\,\,\,\,\,\,\,\,\,\,\,\,\,\,\,\,\,\,\,\,\,\,\,\,\,\,\,\,\,\,\,\,\,\,\,\,\,\,\,\,\,\,\,\,\,\,\,\,\,\,\,\,\,$ A: Hodge decomposition 

$\,\,\,\,\,\,\,\,\,\,\,\,\,\,\,\,\,\,\,\,\,\,\,\,\,\,\,\,\,\,\,\,\,\,\,\,\,\,\,\,\,\,\,\,\,\,\,\,\,\,\,\,\,\,\,\,\,\,\,\,\,\,\,\,\,\,\,\,\,\,\,\,\,\,\,$ B: Biot-Savart  

$\,\,\,\,\,\,\,\,\,\,\,\,\,\,\,\,\,\,\,\,\,\,\,\,\,\,\,\,\,\,\,\,\,\,\,\,\,\,\,\,\,\,\,\,\,\,\,\,\,\,\,\,\,\,\,\,\,\,\,\,\,\,\,\,\,\,\,\,\,\,\,\,\,\,\,$ C: Comments on Lemma \ref{lem3}  (`Hat trick')

\newpage

\begin{center}  APPENDIX A: HODGE DECOMPOSITION   \label{Hodge}\\
\end{center}
\setcounter{equation}{0}
\renewcommand{\theequation}{{\rm A}.\arabic{equation}}

\subsection*{ Hodge theory  for compact manifolds $\Sigma^n$ without boundary }  

 See   \cite{Warner} for a complete exposition.  A metric $g$ in $T\Sigma$ extends  to tensors and forms via Gramians,
\begin{equation} \label{Hodgem}  \langle \nu_2, \nu_2 \rangle_{Hodge} :=  \mathlarger{\int}_{\mathsmaller{\Sigma}} \, \nu_1 \wedge \star \nu_2
\end{equation}
 where the operator $\star \nu$ from $k$ to $(n-k)$ forms  such that $(\star)^2= (-1)^{k(n-k)}$ can be defined by 
\begin{equation}   \xi \wedge \star \nu = g(\xi,\nu) \mu \,\,  (\mu =\, \text{volume form})
\end{equation}
\begin{theorem} (Hodge)
 There is an {\it orthogonal decomposition}  of any k-form  $\nu \in  \Omega ^{k}(\Sigma)$  as a sum
$$
   \nu =d\alpha \oplus \delta \Psi  \oplus \eta
$$
\centerline{  $\alpha \in  \Omega ^{k-1}(\Sigma)$, $\Psi \in  \Omega ^{k+1}(\Sigma)$,
$\eta \in  kernel(\Delta)\,\,$}
\begin{enumerate}
\item $\delta:\Omega ^{k+1}(\Sigma) \to \Omega ^{k}(\Sigma)$, given by $\delta = (-1)^{nk+1} \star d \,\star$  is  the  adjoint of $d$: $  \langle d\nu_1, \nu_2 \rangle = \langle \nu_1, \delta \nu_2 \rangle $.           
\item   $\Delta = d\delta + \delta d$   is  an elliptic operator. \\ $\,$\\
 $kernel(\Delta) =  {\rm Harm}^k(\Sigma).\,\,$  is of finite dimension,  and determines  the cohomology. 
\item  $ \Delta \eta = 0$  {\rm iff }\,\, {\rm  both}  $d \eta = 0 $ and {\rm} $\delta \eta = 0\,\,\,$ and
  $\,\,\,\, \left\{ d\alpha  \oplus \delta \omega \right\} ={\rm Image}(\Delta)\,\,\,\,  (\perp \,{\rm to}\,\,  {\rm kernel}(\Delta)  ). $     \end{enumerate}
\end{theorem}
Schematically,
$$
\Omega^k(\Sigma)={\rm im \,}d\oplus {\rm ker\, }\delta= {\rm im \,}\delta\oplus {\rm ker\, }d
={\rm im \,}d\oplus {\rm im\, }\delta \oplus{\rm ker\, \Delta}={\rm im \,}\Delta\oplus {\rm ker\, }\Delta.
$$

\vspace{-1mm}

\subsection*{Manifolds with boundary:
 Hodge-Morrey-Friedrichs  (\cite{Friedrichs-1955-b},\cite{Schwarz-1995}, \cite{Morrey}). \\For  computational implementation,  \cite{Poelke,
 Desbrun}. For application in  blood physiology \cite{Raza, Saqr, Razaf}. }

\vspace{1mm} 
   We did not explore  this situation in this paper.  In order to show why Hodge theorem has to be refined, observe  that if  $\Sigma$ has boundary, then it is possible for a
1-form to be   in the kernel of the Laplacian    without being both closed and co-closed.

A simple example: Let $\Sigma = A$ be the annulus $A: 0 <  a \leq r \leq b$.
The 2-form
$\omega = (1/2)\, \log (x^2+y^2) dx \wedge dy $  is harmonic (since $\log r$ is harmonic in the annulus).
 It is actually also exact:
 $\omega =  d [ \int^r \, r \log r^2) d\theta ] .$
 Notice that
 it is
not  co-closed:
 $   \phi := \delta \omega = (-y dx + x dy)/(x^2 + y^2) 
 $  represents the 1-dimensional cohomology ${\rm Harm}^1(A)$.

\subsection*{Hodge theory in 2d: formalism} 

\vspace{1mm}

 For  short we will  will omit the wedge $\wedge$ and sometimes denote by $\mu $ the area form. The total area will be denoted by $V$.  All  coordinate systems $(x,y)$  will be isothermal, with
  $  ds^2 = \lambda^2 (dx^2 + dy^2) . $

We will  circumvent  the complex notation $dz$ and $d\bar{z}$, and write
 \begin{equation} 
   \nu = \nu_x dx + \nu_y dy  \,\,,  \,\,\,  v = \nu^\sharp = (1/\lambda^2) ( \nu_x \partial_x + \nu_y \partial_y)
\end{equation}

The main  rules  of the game are $\,\, \star \nu = - v_y dx + \nu_x dy \,\,\,\,$ (90 degrees rotation), together with  \\

  \centerline{ $\star 1 = \lambda^2 dx \wedge dy = \mu  \,\,\,\,,\,\,\,\,\,
    \star dx = dy \,\, , \,\, \star dy = -dx ,  \,\,\,\, \star  \mu = 1  $.}
  
\smallskip \smallskip

 \centerline{ $\star\star \nu = - \nu\,\,\, \,\,$ (for 1-forms),  $   \delta = - \star d \star $ (any degree)  $ \,\,\,\, \Delta = d\delta + \delta d $   commutes with $d$ and $\delta$. }
 
 \smallskip
\centerline{
$\Delta f = -\frac{1}{\lambda^2} (f_{xx}+f_{yy}) \,\,\,\,\,\, $ (for functions)}

The following formulas are  useful in computations with 1-forms: 
  \begin{equation} \label{useful} 
   \star  \nu = i_v \mu  \,\,\,,\,\,\,\,  
    L_v = d i_v + i_v d  \,\,\,\, \Rightarrow \,\,\,\,
    d (\star \nu) = d ( i_v \mu) = L_v \mu =: ({\rm div} v) \mu .
 \end{equation}

 \noindent  All our vector fields will be incompressible:   $d \star \nu = 0$. This assumption is used to say that
locally
\begin{equation} \star \nu = d \psi .
\end{equation} 

\begin{proposition}
  Inner product of 1-forms  requires just the  complex structure
   \end{proposition}
\proofn  By definition $     \langle \nu_1 , \nu_2 \rangle =  \int_{\Sigma} \nu_1 \wedge \star \nu_2 .
  $
Locally, $\nu_1 = a_1 dx +  a_2 dy \,,\, \nu_2 = b_1 dx +  b_2 dy$
  then
  $$   \langle \nu_1 , \nu_2 \rangle = \int (a_1 b_1 + a_2 b_2) dx dy $$
 If one goes back to the vector fields,
 $  \nu_1^{\sharp} =  (a_1 \partial_x  +  a_2 \partial_y)/\lambda^2 , \,\, \nu_2^{\sharp} =  (b_1 \partial_x  +  b_2 \partial_y)/\lambda^2
 $
 and  $$ \langle \nu_1,\nu_2 \rangle = (a_1 b_1 + a_2 b_2)\lambda^2/\lambda^4 , $$ because  $ |\partial_x|  =  \lambda$, etc. So
 $$ \langle \nu_1^\sharp, \nu_2^\sharp \rangle = \int \frac{a_1 b_1 + a_2 b_2}{\cancel{\lambda^4}} \, \cancel{\lambda^2 (\lambda^2} dx dy) .
 $$
Thus there is no need to refer to a  specific metric for forms, 
although  for the inner product between vector fields one has to specify the   metric tensor.  
As seen, things miraculously compensate:

\vspace{2mm}

\begin{proposition}
   The harmonicity of 1-forms, namely the two conditions $d\nu = 0, d \star \nu = 0$  does not  depend on the chosen metric in the conformal class. 
\end{proposition}

\newpage

\vspace{1mm}
\begin{center} APPENDIX B:  BIOT-SAVART  \label{biotsavart}\\
\end{center}
\setcounter{equation}{0}
\renewcommand{\theequation}{{\rm B}.\arabic{equation}}

\subsection*{Two dimensions}

Recall  that $\Psi$ is the Hodge star of the Green's function $G^\omega$ (notations  in section \ref{hodgesubsection})
and Green's functions can in principle be constructed by potential theoretic methods (minimization of an energy,  Perron's method of sub-harmonic functions, etc).

 For 2-forms  in dimension two one has  $\Delta = d \delta + \cancel{\delta  d } = d \delta $, since the second term disappears by default: $d$ is applied in the maximal dimension of the manifold.

Given a vorticity 2-form $\omega$, write  Poisson's equation   (for functions),
$\Delta \psi  = \star \omega $
where $\psi$ is a function. By abuse of language one can omit the $\star$ in front of the 2-form $\omega$ (so $\omega$ is interpreted as a density multiplying $\mu$, just to conform with the  fluid mechanics notation).

The requirement  for the solution $\psi$  to exist is   $\omega$  to be exact (it is a vorticity 2-form)

\centerline{ ($\omega$  is exact $ \Leftrightarrow
  \omega$  is $\perp$ to  harmonic 2-forms (constant multiples of $\mu$)  $ \Leftrightarrow
    \mathlarger{\int}_{\mathsmaller{\Sigma}} \omega = 0$.)}

Let  $\Psi = \star \psi = \psi \mu.$  Consider now $\delta \Psi$ (which is pure vortical by definition).
Then $$d (\delta \Psi)  =   \Delta \Psi - \delta \cancel{d \Psi}  = \omega .  $$ 

Note that the  vortical part of the energy can be rewritten as
\begin{equation*}
H_{vort}(\omega):= H(\delta \Psi)  = \frac{1}{2} \langle \delta \Psi , \delta \Psi \rangle
 =\frac{1}{2} \langle d\delta \Psi ,  \Psi \rangle  = \frac{1}{2} \langle \Delta \Psi ,  \Psi \rangle = \frac{1}{2} \langle \omega ,  \Delta^{-1} \omega \rangle \, .
\end{equation*}

Note that $\Psi$ is not unique. 
One can add any $\Psi_o$ such that  $\delta\Psi_o=0$. 
This means that $\Psi_o$ is a constant multiple of $\mu$. One may normalize $\Psi=\Psi_\omega$ 
by requiring that it is orthogonal to all harmonic functions (namely the constants), i.e. by requiring
$$
\int_\Sigma \Psi_\omega=0.
$$
Then, by identification 
$$
\nu=\delta \Psi_\omega+\eta=-\star d\star \Psi_\omega+\eta= -\star dG^\omega+\eta
$$
so the Hodge star of $\Psi_\omega$ is the Green function: 
$G^\omega=\star \Psi_\omega$ with normalization
$$
\int_\Sigma G^\omega \, \mu=0.
$$
Extending the Hodge decomposition to certain singular forms gives, for example,
$$
G(s,r)=G^{\delta_r\mu}(s)=\star \Psi_{\delta_r\mu}(s).
$$

\vspace{3.5mm}

\subsection*{Biot-Savart in $\R^n,  \, n \geq 3$}
\vspace{1mm}

 In dimension greater than two, for 2-forms, the term $\delta  d$ in   $\Delta = d \delta +  \delta  d   $ no longer vanishes, so we cannot use Poisson's equation anymore. The Biot-Savart law in $\R^3$ is taught in basic electromagnetism classes.  For $\R^n, n \geq 4$  we just found the very recent reference \cite{Glotzl}.
 One of the authors (BG) formulated a version of his own in \cite{Gustafsson-1990b}, Lemma~1.3.

Let $f=(f_1,\dots, f_n)$ be  a vector field in $\R^n$, 
 vanishing  sufficiently
fast at infinity (in order for the convolutions to make
sense). No regularity of  $f$ is assumed, we rather think of a distributional setting, allowing situations in which $f$ is singular, such as having support on lower dimensional manifolds (lines, surfaces, etc). Let
\begin{equation}
E(x)=\frac{c_n}{|x|^{n-2}} \quad (n\geq 3) 
\end{equation}
be the standard fundamental solution of the Laplacian, $c_N$  chosen so that $\Delta E=\delta_o$.

\noindent {\bf Proposition.}  With $\star = \star_{\rm conv}$ denoting convolution, we have the Biot-Savart type representation:
\begin{equation}\label{Biot-Savart}
f=-({\rm div\,}f) \star \nabla E-({\rm curl\,} f) \star \nabla E. 
\end{equation}

The component-wise interpretation of this will be clear from the derivation below, where $E_j=\partial E/\partial x_j$ (the components of $\nabla E$).
For each $k=1,\dots,n$ we have, using that the derivative of a convolution can be moved to an arbitrary factor in it, and also that
$\partial E_j/\partial x_k=\partial E_k/\partial x_j$:

\begin{equation}  \label{biotN}
\begin{split}
f_k & =-(\Delta f_k)*E= -\sum_{j=1}^n \frac{\partial^2 f_k}{\partial x_j^2}\star E= -\sum_{j=1}^n \frac{\partial f_k}{\partial x_j}\star E_j \\
& = -\sum_{j=1}^n \frac{\partial}{\partial x_j}(f_k\star E_j)+\sum_{j=1}^n \frac{\partial}{\partial x_k}(f_j\star E_j)-\sum_{j=1}^n \frac{\partial}{\partial x_j}(f_j\star E_k)\\
 & =-\sum_{j=1}^n \frac{\partial f_j}{\partial x_j}\star E_k-\sum_{j=1}^n \Big(\frac{\partial f_k}{\partial x_j}-\frac{\partial f_j}{\partial x_k}\Big)\star E_j.
\end{split}
\end{equation}
The ${\rm curl}$ of the vector field  $f$ is  implicit in the second term of this formula. It is an antisymmetric tensor
corresponding to the exterior differential acting on a one form.

\noindent {\bf Case $n=3$}.  The above  reduces to the ordinary Biot-Savart law.
We have $ f =  \left( f_1, f_2, f_3 \right),$
\begin{equation*}
\begin{split}
 {\rm curl} f =  & \left(  \frac{\partial f}{\partial x_2} - \frac{\partial f}{\partial x_3} ,  \frac{\partial f}{\partial x_3} -  \frac{\partial f}{\partial x_1} ,  \frac{\partial f}{\partial x_1}  - \frac{\partial f}{\partial x_2}        \right)
\\
      {\rm div}  f = &  \frac{\partial f}{\partial x_1} +  \frac{\partial f}{\partial x_2} + \frac{\partial f}{\partial x_3}\,\,\,,\,\,\,\,
\nabla E =   (E_1,E_2,E_3) = - \frac{1}{4\pi} \frac{x}{|x|^3} .
\end{split}
\end{equation*}
as vector and scalar fields in $\R^3$. Spelling out formula (\ref{biotN}) in this vector
analysis language results in
\begin{equation}
f(x) = \frac{1}{4\pi} \int_{\Re^3} ({\rm div} f)(y) \frac{x-y}{|x-y|^3} d^3y + \frac{1}{4\pi} \int_{\Re^3} ({\rm curl} f)(y) \times \frac{x-y}{|x-y|^3} d^3y ,
\end{equation}
where $d^3y = dy_1dy_2dy_3 $ and $\times$ denotes the ordinary vector product. When $f $ is a stationary magnetic field, so that ${\rm div} f = 0 $ and ${\rm curl} f =$ current density, by Maxwell's equations, the above formula agrees with Biot-Savart's law found in textbooks. If  $f $ instead represents a stationary electric field, then $ {\rm curl} f = 0 $ and we obtain Coulomb's law for an extended charge distribution ${\rm div} f.$

\subsection*{ Formulation for one-forms}
\vspace{1mm}

Next we wish to reformulate (\ref{biotN}) in a way which in principle opens up for generalizations to curved manifolds. We then consider $f$ as a differential form rather than a vector field. Thus
$$ f = f_1 dx_1 + \cdots+ f_ n dx_n
$$
 {\bf Proposition.}
We have
\begin{equation}
 f = dE \star_{\rm conv}  \delta f - i_{(dE)^{\sharp}}  \star_{\rm conv} df,
\end{equation}
The second term shall be interpreted as ``interior convolution
product". 
\noindent \proofn Noting that, in our Cartesian context:
\begin{equation}
\begin{split}
 df  & = \sum_{k,j=1}^n  \frac{\partial f_k}{\partial x_j}  dx_j \wedge dx_k = \frac{1}{2} \left(\frac{\partial f_k}{\partial x_j}  - \frac{\partial f_j}{\partial x_k}  \right) dx_j \wedge dx_k  \\
 \delta f  & = - d \star d \star f = -  \sum_{k=1}^n \,  \frac{\partial f_k}{\partial x_k  } = - {\rm div} f \\
 & \hspace{-3mm}  i_{(dE)^{\sharp}} (dx_j \wedge dx_k ) =  E_j dx_k - E_k dx_j .
\end{split}
\end{equation}

We have by (\ref{biotN}):
\begin{equation}
\begin{split}
F = & - \delta f  \star_{\rm conv}  dE - \sum_{k,j=1}^n  \left(\frac{\partial f_k}{\partial x_j}  - \frac{\partial f_j}{\partial x_k}  \right) \star_{\rm conv}  E_j dx_k  \\
= &  \delta f  \star_{\rm conv}  dE  -   \frac{1}{2} \left(\frac{\partial f_k}{\partial x_j}  - \frac{\partial f_j}{\partial x_k} \right) \star_{\rm conv} \left(  E_j dx_k - E_k dx_j \right) = \\
= &   \delta f  \star_{\rm conv}  dE  -   \frac{1}{2} \left(\frac{\partial f_k}{\partial x_j}  - \frac{\partial f_j}{\partial x_k}  \right) \left( i_{(dE)^{\sharp}} \star_{\rm conv}  dx_j \wedge dx_k \right) = \\
= & dE  \star_{\rm conv} \delta f -  i_{(dE)^{\sharp}}  \star_{\rm conv} df .
\end{split}
\end{equation}

For curved manifolds one has to replace the convolutions above, which can be interpreted as different kinds of Newtonian potentials, by expressions involving ``Green potentials. For example, with a $0$-form $\rho, \,E \star_{\rm conv} \rho$ would be written $E^{\rho}$, or $G^{\rho}$ with $G $ some Green function.

\newpage

\begin{center}  APPENDIX C: DIRECT PROOF OF LEMMA \ref{lem3}   \label{hattrickproof}\\
\end{center}
\setcounter{equation}{0}
\renewcommand{\theequation}{{\rm C}.\arabic{equation}}

We first show that the function  
\begin{equation}
  U_\gamma(s)=\oint_\gamma \star_r d_r G(r,s) \quad(\text{integral in\ }r)
\end{equation} 
is harmonic if  $s\not\in \gamma$.  
The Laplace operator with respect to $s$
can be written as $$\Delta_sG(r,s)= - \star_s d_s \star_s d_s G(r,s) 
\quad(\text{see Appendices A and B).}  $$ 

The use of
local isothermal coordinates $s=(x,y)$ implies: \\

\centerline{$\mu_s=\lambda(x,y)dx\wedge dy$, 
  $\star \mu_s=1$,  
$ \,\, \Delta_s G(r,s)= - \lambda(x,y)(\partial_x^2+\partial_y^2)G(r,x,y). $ } 

Clearly,
$$\Delta_s \star_rd_rG(r,s)=\star_rd_r\Delta_s G(r,s) \,\, {\rm  if } \,\,  r\ne s . $$

  The
symmetry of the Green's function $G(r,s)=G(s,r)$ and equation (\ref{geometerdelta})  
implies that  $$\Delta_sG(r,s)= - V^{-1} \,\, {\rm for}   \,\, r\ne s . $$
 Therefore, if $s\not\in\gamma$
\begin{equation}
  \Delta_s U_\gamma(s)=\oint_\gamma \Delta_s\star_r d_r G(r,s) =
  \oint_\gamma\star_r d_r  \Delta_sG(r,s)=  - \oint_\gamma\star_r d_r (1/V)  =0
\end{equation}

We now study  $U_\gamma(s)$ in a neighborhood of a point in $\gamma$. We assume
that $\gamma$ is a real analytic curve. In this case,
any  $s_0\in\gamma$  has a neighborhood $N$ in which
there are isothermal coordinates $(x,y):N\to \C$ such that
$$\gamma\cap N=\{|x|< a,y=0\} \,\,\, {\rm 
and}\,\,  (x,y)(s_0)=(0,0). $$  

The orientation of $\gamma$ coincides with the orientation
of the $x$-axis. Let $s$ be a point in $N$ with coordinates $(x,y)(s)=(\xi,\eta)$
such that 
\[
  U_\gamma(s)=\int_{\gamma\cap N}
  \star_r d_r G(r,s)
  +\int_{\gamma-\gamma\cap N} \star_r d_r G(r,s)
\]
Since  $G(r,s)$ is analytic for
$r\in\{\gamma-\gamma\cap N\}$ and $s\in N$,
the second integral is an analytic function of  $s$.
Therefore, the singular behavior of $ U_\gamma(s)$ is
determined by
\[
  \int_{\gamma\cap N}
 \star_r d_rG(r,s)=
 \int_{-a}^a- \partial_yG(r,s)dx.
\]
In the isothermal coordinates
\[ G(r,s)=-\frac{1}{4\pi}\log\big((x-\xi)^2+(y-\eta)^2\big) +h (x,y,\xi,\eta)\,
  \]
  where $h$ is an analytic function on $N$.
  The singular behavior of $ U_\gamma(s)$ is, therefore,
  determined by the limit
  {\small\[\begin{split} &
        \frac{1}{4\pi} \int_{-a}^a \partial_y
\log\big((x-\xi)^2+(y-\eta)^2\big)
   \Big|_{y=0}dx= \\ & \,\,\,\,\,\,\, = 
   - \frac{1}{2\pi} \int_{-a}^a\frac{\eta }{(x-\xi)^2+\eta^2}dx\\ & \,\,\,\,\,\,\, =
   -\frac{1}{2\pi}\bigg\{ \arctan\left(\frac{a-\xi}{\eta}\right)+
   \arctan \left(\frac{a+\xi}{\eta}\right)\bigg\} \,.
   \end{split}
    \]}
  We conclude that for $s\in N$
  \begin{equation}
    U_\gamma(\xi,\eta)=
    -\frac{1}{2\pi}\bigg\{ \arctan\left(\frac{a-\xi}{\eta}\right)+
    \arctan \left(\frac{a+\xi}{\eta}\right)\bigg\}+U_R(\xi,\eta)\,,
    \label{Uxieta}
    \end{equation}
  where $U_R$ is an analytic function.\\

 Let $(\xi,0)$ be  a point in $\gamma \cap N$, which implies $|\xi|<a$, and make
  $(\xi,\eta)\to (\xi,0_-)$ from the right-hand side of $\gamma$, namely
  with $\eta<0$,
  then
 \begin{equation}
  U(\xi,0_-)=  \lim_{\eta\to 0_-} U_\gamma(\xi,\eta)=
    \frac{1}{2}+U_R(\xi,0)
    \end{equation}
  If the limit is taken from the left-hand side of $\gamma$, namely with $\eta>0$,
  then
  \begin{equation}
 U(\xi,0_+)=   \lim_{\eta\to 0_+} U_\gamma(\xi,\eta)=
    -\frac{1}{2}+U_R(\xi,0)
    \end{equation}
    
  Therefore $U(\xi,0_-)-U(\xi,0_+)=1$
  and  the function $U_\gamma(s)$ jumps by one when $\gamma$ is crossed from the
  left to the right.
  The differential of $U_\gamma$ as given in equation (\ref{Uxieta}) is
  $dU_R$ plus
  \[
    -2a\frac{2  \xi \eta d\xi+ \left(a^2-\xi^2+\eta^2\right)d\eta}
    {\left((a-\xi)^2+\eta^2\right) \left((a+\xi)^2+\eta^2\right)}\,,
  \]
  which is analytic  over $\gamma$. Therefore,
  $dU_\gamma$ is a harmonic differential that is regular over $\gamma$.  \qed

\bigskip
  
 This  was presented  just as a guide for a the non-expert.  In different degrees of sophistication such an analysis can be found for example in Hermann Weyl's classical  1913 treatise \cite{Weyl}, and it is also similar to jump formulas for Cauchy integrals \cite{Young, Solom}. The Cauchy integral of an analytic $f $ in a domain D equals $f $ inside the domain and zero outside, hence the jump across the boundary is exactly $f$. With $f=1$ we are integrating just the Cauchy kernel, and the singularity of this is of the same type as that of $dG$ (or rather $dG+i \star dG$). In complex analysis, for the Cauchy integral, the jump formula
often goes under the name Sokhotski-Plemelj jump formula (tracing back to 1868).


\newpage






\begin{thebibliography}{99}



\bibitem{Gustafsson}
Gustafsson, B., Vortex pairs and dipoles on closed surfaces,
\textit{ J. Nonlinear Sci.},  2022, vol. 32, 62.

\bibitem{Ragazzo} Grotta-Ragazzo, C.,
Errata and Addenda to: ``Hydrodynamic Vortex on Surfaces"
and ``The Motion of a Vortex on a Closed Surface of
Constant Negative Curvature",
\textit{ J. Nonlinear Sci.},  2022, vol. 32, 63.

\bibitem{Ragazzo0} Grotta-Ragazzo, C., The motion of a vortex on a closed surface of constant negative curvature. \textit{Proc. R.
Soc. A Math. Phys. Eng. Sci.}, 2017, vol. 473: 20170447.20170447.

\bibitem{Bogatskii1}   Bogatskii,  A., Vortex flows on closed surfaces, \textit{J. Phys. A: Math. Theor.}, 2019,  vol. 52,  475501.

\bibitem{Bogatskii} A. Bogatskii, A., Vortex flows on surfaces and their anomalous hydrodynamics,
\textit{Ph.D. thesis}, Department of Physics, University of Chicago, 2021.

\bibitem{MarsdenWeinstein}  Marsden, J.,  Weinstein, A., Coadjoint orbits, vortices, and Clebsch variables for incompressible fluids, \textit{Physica D: Nonlinear Phenomena}, 1983, vol.  7, no. 1-3, pp.  305-323

\bibitem{Hodge} Hodge, W. V. D., \textit{The Theory and Applications of Harmonic Integrals}, Cambridge University Press, 1941.

\bibitem{KoillerBoatto}  Boatto, S., Koiller, J. Vortices on Closed Surfaces. in: Chang, D., Holm, D., Patrick, G., Ratiu, T. (eds), \textit{Geometry, Mechanics, and Dynamics: The Legacy of Jerry Marsden},  Fields Institute Communications, vol 73. Springer, New York, NY, 2015.

\bibitem{Lin-a} Lin, C.C., On the motion of vortices in two dimensions. I. Existence of the Kirchhoff-Routh function, \textit{Proc.
Nat. Acad. Sci. U.S.A.}, 1941, vol. 27, pp 570--575,  (see also the second part of the paper in the same volume pp.
575-577), and \textit{On the Motion of Vortices in Two Dimensions}, University
of Toronto Studies, Applied Mathematics Series, no. 5, University of
Toronto Press,    1943.

\bibitem{Gustafsson-1979-b}  Gustafsson, B., On the motion of a vortex in two-dimensional flow of an
ideal fluid in simply and multiply connected domains, \textit{Bulletin TRITA-MAT-1979-7}, Royal Institute of
Technology Research Bulletins,  1979, pp. 1-109.

\bibitem{Flucher-Gustafsson-1997-a} Flucher, M., Gustafsson, B.,  Vortex motion in two-dimensional
hydrodynamics, \textit{Bulletin TRITA-MAT-1997-MA-02},  Royal Institute of Technology Research Bulletins, 1979, pp. 1-24.

\bibitem{Flucher-1999-a}  Flucher, M., Vortex Motion in Two Dimensional Hydrodynamics, Chapter 15 of   \textit{ Variational problems with concentration}, Progress in Nonlinear Differential Equations and their Application, vol. 36, Birkh\"auser Verlag, Basel, 1999.

\bibitem{Marsden}   Marsden,J.,  Weinstein, A.,  Reduction of symplectic manifolds with symmetry, \textit{Reports on Math. Phys.}, 1974, vol. 5, , 121-130.

 \bibitem{Friedrichs-1955-a} Friedrichs, K. O.,   Differential forms on Riemannian manifolds. \textit{Comm. Pure
Appl. Math.}, 1955,  vol. 8, pp. 551-590.

\bibitem{Schwarz-1995-a} Schwarz, G., \textit{Hodge Decomposition - A Method for Solving Boundary Value Problems}, 
Springer Lect. Notes in Mathematics, vol. 1607, 1995. 

\bibitem{Morrey-a} Morrey, C.B.,  A variational method in the theory of harmonic integrals II,
\textit{Amer. J. Math.}, 1956, vol. 78, pp. 137-170.

 \bibitem{Raza-a}   Razafindrazaka, F.,   Poelke,  K.,  Polthier,  K.,   Goubergrits, L., A Consistent Discrete 3D Hodge-type
Decomposition: implementation and practical
evaluation, \textsf{arXiv:1911.12173} (16 Dec 2019).

 \bibitem{Saqr-a} Saqr, K.M., Tupin, S., Rashad, S. et al. Physiologic blood flow is turbulent, \textit{Scientific Reports (Nature)}, 2020, vol. 10, 15492. 

 \bibitem{Razaf-a} Razafindrazaka F.H,
Yevtushenko, P., Poelke, K., Polthier, K., Goubergrits
L., Hodge decomposition of wall shear stress
vector fields characterizing biological flows,  2019,  \textit{R.
Soc. Open Sci.}, vol. 6,  181970.  

\bibitem{Poelke-a}  Poelke, K., Polthier, K., Boundary-aware Hodge decompositions for piecewise constant vector fields, \textit{Computer-Aided Design}, 2016, vol. 78,  pp. 126-136.

 \bibitem{Desbrun-a} Zhao, R., Debrun, M., Wei, G., Tong, Y., 3d Hodge decompositions of edge and face based vector fields,
\textit{ACMTransactions On Graphics}, 2019, vol. 38, pp. 1-13.

\bibitem{SanDiego}  Yin, H.,  Nabizadeh, M.,   Wu, B., Wang, S.,  Chern, A.,   Fluid Cohomology, \textit{ACM Trans. Graph.}, 2023, vol. 42, no. 4, Article 126. 

\bibitem{Arnold}  Arnold, V. I., Sur la g\'eom\'etrie diff\'erentielle des groupes de Lie de dimension
infinie et ses applications \`a l'hydrodynamique des fluides parfaits, \textit{Ann.
Inst. Fourier}, 1966 vol. 16, pp. 319-361.

\bibitem{Modin1}  Modin. K., Geometric Hydrodynamics: from Euler, to Poincar\'e, to Arnold, in:
\textit{13th Young Researchers Workshop on Geometry, Mechanics and Control: Three Mini-courses}, Textos de Matematica 48, Universidade de Coimbra, 2019.

\bibitem{ArnoldKhesin}  Arnold, V.I. and B. A. Khesin, B.A., \textit{Topological Methods in Hydrodynamics},
Applied Mathematical Sciences, vol 125, New York: Springer-Verlag,  1998.

\bibitem{Chorin} Chorin, A. J. (1968), Numerical Solution of the Navier-Stokes Equations, \textit{Math. Comp.}., 1968, vol. 22,  no. 104, 745-762. 

\bibitem{Kuch}  K\"uchemann, D., Report on the I.U.T.A.M. Symposium on concentrated vortex
motions in fluids, \textit{J. Fluid Mech.}, 1965,  vol. 21, pp. 1-20.

\bibitem{Saffman} Saffman, P. G., {\it Vortex dynamics}, Cambridge University Press, 1992.

\bibitem{Khesin3dHodge}   Khesin, B.,   Kuksin,  S.,  Peralta-Salas, D.,
KAM theory and the 3D Euler equation, \textit{Advances in Mathematics}, 2014, vol.  267, pp. 498-522.

\bibitem{helicity3d} Khesin, B., Peralta-Salas, D., Yang, C.,
The helicity uniqueness conjecture in 3D hydrodynamics. \textit{Trans. of the Amer. Math. Soc.}, 2022, vol. 375, no.2,   909-924.

\bibitem{Peskin}  Bao, Y.,  Donev, A., Griffith, B.,  McQueen, D.,  Peskin, C.,  An Immersed Boundary method with divergence-free velocity interpolation and force spreading, 2017,  \textit{J. Comput. Physics}, vol. 347, pp. 183-206.

\bibitem{Joseph}  Joseph, D. D., Helmholtz decomposition coupling rotational to irrotational flow of a viscous fluid, \textit{Proc. Natl. Acad. Sci. USA}, 2006,  vol. 103, no. 39, pp. 14272-14277.

\bibitem{Bhatia}  Bhatia, H.,  Norgard, G., Pascucci, V., Bremmer, P.,   Helmholtz-Hodge Decomposition - A Survey, \textit{IEEE Trans.  Vis.  Comp. Graph.}, vol. 19, no. 8, 1386-1404 (2012) 

\bibitem{Lefevre}  Lef\`evre, J., Leroy, F., Khan, S., Dubois, J., Huppi, P., Baillet, S., Mangin, J.,   Identification of Growth Seeds in the Neonate Brain through Surfacic Helmholtz Decomposition, International Conference on Information Processing in Medical Imaging, Williamsburg, in \textit{Lecture Notes in Computer Science}, vol. 5636,   Jerry L. Prince,
  Dzung L. Pham,  and  Kyle J. Myers (Eds.),  2009, pp. 252-263. 



\bibitem{Marchioro} Marchioro, C. ,   Pulvirenti, M.,  \textit{Mathematical theory of incompressible nonviscous fluids},  Springer-Verlag, New York, 1994. 



\bibitem{Weis}  Weis-Fogh, T., Quick estimates of flight fitness in hovering animals, including novel mechanisms for lift production, \textit{J. Exp. Biol.}, 1974, vol. 59, pp. 169-230.

\bibitem{Lighthill-a}  Lighthill, M. J., On the Weis-Fogh mechanism of lift generation, \textit{J. Fluid Mech.}, vol. 60,  1973, 1-17.

\bibitem{Kol} Kolomenskiy, D., Moffat, H., Farge, M., Schneider, K.  (2011). The Lighthill-Weis-Fogh clap-fling-sweep mechanism revisited, \textit{J. Fluid Mech.}, 2011,  vol. 676,  pp. 572-606. 

\bibitem{Cheng}   Cheng, X., Sun, M., Revisiting the clap-and-fling mechanism in small wasp Encarsia formosa using quantitative measurements of the wing motion,
\textit{Physics of Fluids}, 2019, vol. 31, 101903. 

\bibitem{Helmholtz} Helmholtz, H., \"Uber integrale der hydrodynamischen gleichungen
   welche den Wirbelbewegungen entsprechen, \textit{Journal f\"ur reine und angew. Mathematik}, 1858, Vol. 55, 25-55. 
A translation to English was approved by Helmholtz:  \textit{On integrals of the hydrodynamic equations which express vortex motions}, published by P.G. Tait
in Philosophical Magazine and Journal of Science, Vol. 33, no. 4, 485-512, 1867. 

\bibitem{Telionis}  Telionis , D.P.,  Impulsive Motion. In: \textit{Unsteady Viscous Flows}, Springer Series in Computational Physics. Springer, Berlin, Heidelberg, 1981,  pp. 79-153. 
 
\bibitem{BKT}
Kosterlitz, J. M. and J. Thouless, D.J.,  Early Work on Defect Driven Phase Transitions,
in \textit{40 Years of Berezinskii-Kosterlitz-Thouless Theory}, J. V, Jos\'e (Ed.), World Scientific, 2013.

\bibitem{Moffatt}  Moffatt, H. K., Singularities in fluid mechanics,  \textit{Phys. Rev. Fluids}, 2019, vol. 4, 110502.

\bibitem{FarkasKra} Kra, I.,  Farkas, M.,  \textit{Riemann Surfaces},  Graduate Texts in Mathematics vol. 71,  Springer,
2nd edition, 1992.

\bibitem{Chai} Chai, C.L., The period matrices and theta functions of Riemann, in \textit{The Legacy of Bernhard Riemann after One Hundred and Fifty Years}, eds. L. Ji, F. Oort and S.-T. Yau, International Press and Higher Education Press, 2016, pp. 79-106.

\bibitem{Okikiolu} Okikiolu, K., A Negative Mass Theorem for the 2-Torus, \textit{Commun. Math. Phys.}, 2008, vol. 284, 775-802.

 \bibitem{Gustafsson-2019a} Gustafsson, B., Vortex motion and geometric function theory: the role of connections, \textit{Philos. Trans. Roy. Soc. A}, 2019, vol. 377,   20180341 . 

\bibitem{Kleina} Klein, F.,  On Riemann's Theory of Algebraic Functions and their Integrals  - A Supplement to the Usual Treatises, Cambridge: Macmillan and Bowes, 1893.     German original: \"Uber Riemanns Theorie der algebraischen Funktionen und ihrer Integrale, Leipzig, 1882.
     
 \bibitem{Miranda0} Guillemin V., Miranda E., Pires A.R., Symplectic and Poisson geometry on b-manifolds, \textit{Adv. Math.}, 2014,  vol. 264, pp. 864-896.

\bibitem{Zambon}  Geudens, S.,  Zambon, M.,, Deformations of Lagrangian submanifolds in log-symplectic manifolds, textit{Adv. Math.}, 2022,  vol. 397, 108202.

\bibitem{Kimura} Kimura, Y., Vortex Motion on Surfaces with Constant Curvature, \textit{R. Soc. Lond. Proc. Ser. A Math.
Phys. Eng. Sci.}, 1999, vol. 455, no. 1981, pp. 245-259.

\bibitem{RagVig} Grotta-Ragazzo, C., Viglioni, H.H.B., Hydrodynamic vortex on surfaces, \textit{J. Nonlinear Sci.}, 2017, vol. 27, no. 5, 1609-1640.

\bibitem{HolmanSchuss0}   Holcman, D., Schuss, Z. , Escape Through a Small Opening: Receptor Trafficking in a Synaptic Membrane, \textit{Journal of Statistical Physics}  2004, vol. 117, no. 5-6, pp. 975-1014.

\bibitem{Schuss} Schuss, Z. The Narrow Escape Problem - a  short review of recent results, \textit{Journal of Scientific Computing}, 2012,  vol. 53, no. 1, pp. 194-210
.

\bibitem{HolmanSchuss}   Holcman,  D.,  Schuss,Z., The narrow escape problem, \textit{SIAM Review}, 2014,  vol. 56, no. 2,  pp. 213-257.

\bibitem{SteinerDoyle2009} Doyle, P. G., Steiner, J.,  Spectral invariants and playing hide-and-seek on surfaces, ArXiv:1710.09957v1

 \bibitem{Morpurgo-a} Morpurgo, C., Zeta functions on $S^2$, in  J.R. Quine, P. Sarnak, eds., \textit{Extremal Riemann Surfaces}, Contemporary Mathematics 201, 1997, pp.   213-226.

 \bibitem{Steiner-a}  Steiner, J., A geometrical mass and its extremal properties for metrics on $S^2$,
\textit{Duke Math. J.}, 2005, vol. 129, no. 1), 63-86.

\bibitem{ragsub-a} Grotta-Ragazzo, C.,  Vortex on surfaces and  Brownian-motion in higher dimensions: special metrics,
{\it Submitted}.

\bibitem{Wiener} Wiener, N., Differential space. \textit{J. Math. Phys. Mass. Inst. Tech.}, 1923, vol. 2 , 131--174;.

\bibitem{doylesnell}   Doyle, P.,   Snell, J., 
\textit{Random Walks and Electrical Networks }, Carus Mathematical Monographs,  Mathematical Association of America,  1984.

\bibitem{stolar} Stolarksy, K. B.,  Review on  Random Walks and Electric Networks. \textit{The American Mathematical Monthly}, 1987, vol. 94, no. 2, pp. 202-205. 

\bibitem{Lighthill} Lighthill, J.,  Real and Ideal Fluids,    in: L. Rosenhead (ed.), \textit{Laminar Boundary Layers},  Oxford Univ. Press,  1963.

 \bibitem{Howe} Howe, M., Vorticity and the theory of aerodynamic sound, \textit{J. of Engineering Math.}, 2001, vol. 41, pp. 367-400. 



\bibitem{Tkachenko} Tkachenko, V. K., Stability of Vortex Lattices, \textit{Soviet Physics JETP}, 1966, vol. 23,
no. 6, pp. 1049-1056. 

 \bibitem{Neil} O'Neil, K. A., On the Hamiltonian Dynamics of Vortex Lattices, \textit{J. Math. Phys.}, 1989, vol. 30, no. 6, pp. 1373-1379.
 
  \bibitem{StremlerAref} Stremler, M., Aref, H., Motion of Three Point Vortices in a Periodic Parallelogram, \textit{J. Fluid Mech.}, 1999, vol. 392, pp. 101-128.

 \bibitem{Stremler} Stremler, M., On relative equilibria and integrable dynamics of point vortices in periodic domains, \textit{Theoretical and Computational Fluid Dynamics}, 2010, vol.  24., no. 1, pp. 25-37.

\bibitem{crowdyrect} Crowdy, D., On rectangular vortex lattices, \textit{Applied Mathematics Letters}, 2010,  vol. 23, no. 1, pp. 34-38.

\bibitem{KilinArtemova} Kilin, A.A., Artemova, E.M., Integrability and Chaos in Vortex Lattice Dynamics, \textit{Regul. Chaotic Dyn.,} 2019, vol. 24, no. 1, pp. 101-113.

\bibitem{Green}  Green,  C.,  Marshall J.,   Green's function for the Laplace-Beltrami operator on a toroidal surface, \textit{Proc. R. Soc. A.}, 2013, vol. 469, 2012047920120479

 \bibitem{Sakajo} Sakajo, T.,  Shimizu, Y., Point Vortex Interactions on a Toroidal Surface, \textit{Proc. Roy. Soc. London Ser. A}, 2016, vol. 472, no. 2191, 20160271 . 
 
\bibitem{Sakajo-a} Sakajo T., Vortex crystals on the surface of a torus, \textit{Phil. Trans. R. Soc. A }, vol. 377, 2019, 20180344.

 \bibitem{guenther}  Guenther, N.,   Massignan, P.,    Fetter, A., Superfluid vortex dynamics on a torus and other toroidal surfaces of revolution, \textit{Physical Review A}, 2020, 053606. 
 
 \bibitem{Arefob} Borisov, A., Meleshko, V., Stremler, M., van der Heist, G.,  In Memoriam: Hassan Aref  (1950-2011), \textit{Regular and Chaotic Dynamics}, 2011, vol. 16, no. 6, pp. 671-684.
  
\bibitem{chineses} Lin, C., Wang, C., Elliptic functions, Green functions and the mean
field equations on tori, \textit{Annals of Mathematics},  2010,  vol. 172, pp. 911-954.

\bibitem{Willmore} Willmore, T., Surfaces in Conformal Geometry, \textit{Annals of Global Analysis and Geometry}, 2000, vol. 18, 255-264.

\bibitem{Coda} Marques, F.C., Neves, A., Min-Max theory and the Willmore conjecture, 
 \textit{Annals of Math.}, 2014,   vol. 179 , no. 2,  pp.  683-782. 
See also: The Willmore conjecture, arXiv:1409.7664 .


\bibitem{Pinkall}  Pinkall, U.,   Sterling, I., Willmore surfaces, \textit{Math. Intelligencer}, 1987,vol. 9, no.2,
pp. 38-43. 

\bibitem{Magdalena} Heller, B., Pedit, F., 
Towards a Constrained Willmore Conjecture, in 
Toda, M. D., (ed.), \textit{Willmore Energy and Willmore Conjecture}, Chapman and Hall/CRC, 2017. 


\bibitem{Barros2001}  Barros, M., Equivariant tori which are critical points of the conformal total tension functional, \textit{Rev. R. Acad. Cien. Serie A. Mat. Spain}, 2001, vol. 95, no.2 pp. 249-258.
 
\bibitem{Barros2014} Barros, M.,  Ferr\'andez, A.,  Garay, O., 
Equivariant Willmore surfaces in conformal homogeneous three spaces,
\textit{J. Mathematical Analysis and Applications}, 2014, 
vol. 409, no. 1, pp.  459-477. 

\bibitem{Wente}   Wente, H., Counterexample to a conjecture of H. Hopf, \textit{Pacific J. Math.}, 1986, vol. 121,  no. 1, 193-243.

\bibitem{Abresch}  Abresch, U., Constant mean curvature tori in terms of elliptic functions,  \textit{J. Reine Angew. Math. }, 1987, vol. 374, 169-192.

 \bibitem{Andrews} Andrews, B.,  Li, H., Embedded constant mean curvature tori in the three-sphere, 
\textit{J. Differential Geom.}, 2015, vol. 99, no. 2, 169-189

\bibitem{Hauswirth} Hauswirth, L., Kilian, M., Schmidt, M.U., Mean-convex Alexandrov embedded constant mean curvature tori in the 3-sphere, \textit{Proceedings of the London Mathematical Society}, 2016, vol. 112, 588-622.

 \bibitem{lawson}          Lawson, H., Complete Minimal Surfaces in $S^3$, 
\textit{Annals of Math.}, 1970, vol. 92, no. 3,  pp. 335-374.

\bibitem{Penskoi} Penskoi, A.V. Generalized Lawson Tori and Klein Bottles, \textit{J. Geom. Anal.}, 2015, vol. 25, pp. 2645-2666.  

\bibitem{Pinkall1} Pinkall, U., Hopf tori in $S^3$,  \textit{nvent. math.}, 1985, vol. 81, 379-386.


\bibitem{Mironov}  Mironov, A., On a Family of Conformally Flat Minimal Lagrangian Tori in $\C P^3$,  \textit{Mathematical Notes (Matematicheskie Zametki)}, 2007, vol. 81, no. 3, pp. 329-337. 

\bibitem{Arefcrystals}   Aref, H.,    Newton, P. K.,    Stremler, M.,   Tokieda, T.,  Vainchtein, D., Vortex Crystals, \textit{Advances in Applied Mechanics}, 2003, vol. 39, 1-79.

 \bibitem{KMontaldi}   Koiller,  J.,  Getting into the vortex: On the contributions of James Montaldi, \textit{J.  Geometric Mechanics}, 2020, vol. 12, no. 3,  507-523. 
  
  
 \bibitem{Bolzapaper} Bolza, O.  On Binary Sextics with Linear Transformations into Themselves, \textit{American Journal of Mathematics}, 1887, vol, 10, no. 1, 47-70. 
  
 \bibitem{Schwarztriangle}   Magnus, W., \textit{Noneuclidean Tesselations and Their Groups}, Pure and Applied Mathematics series vol. 61, Elsevier, 1974.

\bibitem{Balazs}  Balazs, N.L., Voros A.,  Chaos on the pseudosphere,
  \textit{Phys. Rep.}, 1986, vol. 143, pp. 109-240.

\bibitem{Gilman}  Gilman, J., Compact Riemann Surfaces with Conformal Involutions, \textit{Proc.   American Math. Society}, 1973,
vol. 37, no. 1,  pp. 105-107.

\bibitem{Schaller} P. S. Schaller, Involutions and simple closed geodesics on Riemann surfaces, \textit{Annales Academiae Scientiarum Fennicae,
Mathematica}, 2000, vol.
25,  91-100.

\bibitem{Haas} Haas, A., Susskind, P., The geometry of the hyperelliptic involution in genus two,
\textit{Proc. of the Amer. Math. Soc.}, 1989, col. 105, no. 1, 159-165.


\bibitem{Parlier}  Costa, A. F.,  Parlier, H., A geometric characterization of orientation-reversing involutions, \textit{J.  London Math. Society}, 2008, vol. 77, no. 2, pp. 287-298. 


\bibitem{Schottky-1877} Schottky, F., Ueber die conforme Abbildung mehrfach zusammenh\"angender ebener Fl\"achen, \textit{J. Reine Angew. Math.}, 1877, vol. 83,
pp. 300-351.



\bibitem{Schiffer-Spencer-1954-a} Schiffer, M. and Spencer, D. C., \textit{Functionals of finite Riemann surfaces}, Princeton University Press, Princeton, N. J., 1954.

 \bibitem{Hawley-Schiffer-1967}
  Hawley, N.  and  Schiffer, M.,  Riemann surfaces which are doubles of plane domains,
\textit{Pacific J. Math}, 1967,  vol. 20, no. 2,  pp, 217-222

\bibitem{Davis-1974} Davis, P. J., \textit{The Schwarz function and its applications},
Carus Mathematical
Monographs, No. 17., The Mathematical
Association of America, Buffalo, N. Y., 1974.


\bibitem{Cohn-1980-a} Cohn, H., \textit{Conformal mapping on Riemann surfaces}, Dover Publications,
Inc., New York, 1980. Reprint of the 1967 edition.



\bibitem{Gustafsson-Roos-2018} Gustafsson, B.,   Roos, J.,  Partial balayage on Riemannian manifolds,  \textit{J. Math. Pures Appl.}, 2018,  vol. 118, 82-127.


 \bibitem{RKCRCD}  Regis, A., Castilho, C., Koiller, J., On the Linear Stability of a Vortex Pair Equilibrium
on a Riemann Surface of Genus Zero, \textit{Reg. Chaotic Dynamics}, 2022, vol. 27, no. 5, 498-524.


 \bibitem{Alling} Alling, N. L.,  Greenleaf, N., \textit{Foundations of the theory of Klein
surfaces},  Lecture Notes in Mathematics, vol. 219. Springer-Verlag,
Berlin-New York, 1971. 

\bibitem{Vanneste}Vanneste , J., Vortex dynamics on a M\"obius strip, \textit{Journal of Fluid Mechanics}, 2021, vol. 923, A12.



\bibitem{Balabanova}  Balabanova, N., Algebraic and Geometric Methods in Mechanics,
\textit{Ph.D. thesis}, Mathematics Department, The University of Manchester, 2022.

 \bibitem{Gustafsson-Tkachev-2011} Gustafsson, B., Tkachev, V. G.,  On the exponential transform
of multi-sheeted algebraic domains, \textit{Comput. Methods Funct. Theory}, 2011, vol. pp. 591-615.

\bibitem{GuSebar} Gustafsson, B., Sebbar, A., Critical Points of Green's Function and
Geometric Function Theory, \textit{Indiana University Math. J.}, 2012,
vol. 61, no. 3, 939-1017.

\bibitem{Krichever}  Krichever, I.,  Marshakov, A., Zabrodin, A., Integrable Structure of the Dirichlet Boundary Problem in Multiply-Connected Domains, \textit{Commun. Math. Phys.},  2005, vol. 259, pp. 1-44.


\bibitem{Yamada}  Yamada, A., Positive Differentials, Theta Functions and Hardy $H^2$ Kernels,
\textit{Proceedings of the AMS}, 1999, vol. 127, no. 5, pp. 1399-1408.

\bibitem{GGreen}  Green G, \textit{Essay on the Application of Mathematical Analysis to
the Theory of Electricity and Magnetism}, 1828,  reprinted in Green, G. , Mathematical Papers of the Late George Green (Cambridge Library Collection - Mathematics) (N. Ferrers, Ed.), Cambridge University Press , 2014. 

 \bibitem{Crowdy} Crowdy, D., Marshall, J., Green's functions for Laplace's equation in multiply connected domains, \textit{IMA J.  Applied Math.},  2007, vol.  72, pp. 278-301.
        
\bibitem{Crowdybook} Crowdy, D., \textit{Solving Problems in
Multiply Connected
Domains}, SIAM, 2020.

\bibitem{Crowdy0} Crowdy, D.,   Marshall, J., Analytical formulae for the Kirchhoff-Routh
path function in multiply connected domains, \textit{Proc. R. Soc. Lond. A}, 2005, vol. 461, pp. 2477-2501.

 \bibitem{Crowdy1}   Crowdy, D.,   Marshall, J., The motion of a point vortex around multiple circular islands, \textit{Phys. Fluids}, 2005, 056602

 \bibitem{Crowdy-prime}  Crowdy, D.,  The Schottky-Klein Prime Function on the Schottky Double of Planar Domains, \textit{Computational Methods and Function Theory}, 2010,  vol10,  no. 2, 501-517

\bibitem{Koebe} Koebe, P.,  Abhandlungen zur Theorie der konformen Abbildung: IV. Abbildung mehrfach zusammenh\"angender schlichter Bereiche auf Schlitzbereiche, 
 \textit{Acta Math.}, 1916, vol. 41, pp. 305-344.
 
\bibitem{Koebe1} Koebe, P.,  Abhandlungen zur Theorie der konformen Abbildung, \textit{Math. Z.,}  1918, vol. 2, pp. 198-236.

\bibitem{Bandle} Bandle, C,    Flucher, M., Harmonic Radius and Concentration of Energy; Hyperbolic Radius and Liouville's Equations $\Delta U = e^U $ and $\Delta U = U^{\frac{{n + 2}}{{n - 2}}}$,
\textit{SIAM Review},  1996, vol.  38, no. 2, 191-238.
 
 \bibitem{Ahlfors-1973} Ahlfors, L., \textit{Conformal invariants: topics in geometric function theory}, McGraw-Hill Series in Higher Mathematics, McGraw-Hill Book Co., 1973. 

\bibitem{Aref1984} Aref H., Stirring by chaotic advection, \textit{J. Fluid Mech.}, 1984, vol. 143, pp 1- 21.

\bibitem{Ottino} Ottino, J.,  \textit{The Kinematics of Mixing: Stretching, Chaos and Transport}, Cambridge University Press, 1989.  

\bibitem{Ditch} Ditch, A., T\'el, T., Dynamics of blinking vortices, \textit{Physica. Rev. E}, 2009, vol. 79, 016210. 

\bibitem{Khak} Khakhar, D., Rising, H.,  Ottino, J.,Analysis of chaotic mixing in two model systems, \textit{J. of Fluid Mechanics}, 1986, vol.172, pp. 419-451. 

\bibitem{CourantHilbert} Courant, R., Hilbert. D. , \textit{Methods of Mathematical Physics}, Wiley, 1989.

    \bibitem{Vaskin}	Vaskin V. V., Erdakova N. N.,
On the dynamics of two point vortices in an annular region, \textit{Nonlinear dynamics},
2010,  vol. 6, no. 3, pp.  531-547.

\bibitem{Kurakin}  Kurakin, L. G., Influence of annular boundaries on Thomson's vortex polygon stability, \textit{Chaos}, 2014, vol. 24, 023105 . 


\bibitem{Nadia}  Erdakova, N.,  Mamaev, I.,
On the dynamics of point vortices in an annular region, 2014
\textit{Fluid Dyn. Res.}, 2014, vol. 46, 031420


\bibitem{Flucherbook} Flucher, M., 
\textit{Variational Problems with Concentration}, Springer, 1991.



\bibitem{Richardson} 
Richardson, S., \textit{Vortices, Liouville's equation and the Bergman kernel function}, Mathematika,  1980, vol. 27. no. 2 , 321-334. 


\bibitem{Borah} Borah, D., Haridas, P., Verma, K. Comments on the Green's function of a planar domain, \textit{Anal.Math.Phys.}, 2018, vol. 8, pp. 383-414. 

\bibitem{Solynin} Solynin, A.,  A note on equilibrium points of Green's function, \textit{Proc. Amer. Math. Soc.}, 2008, vol. 136, 
pp. 1019-1021.

\bibitem{Guconvex} Gustafsson, B., On the convexity of a solution of Liouville's equation,
\textit{Duke Math. J.}, 1990, vol. 60, no. 2, pp. 303-311.


 \bibitem{Nehari-1952a} Nehari, Z., \textit{Conformal  mapping}, McGraw-Hill Book Co., 1952.

\bibitem{Sario-Oikawa-1969a} Sario, L. and Oikawa, K.,  \textit{Capacity functions}, Die Grundlehren
der mathematischen Wissenschaften, Band 149, Springer-Verlag New
York Inc.,  1969.

\bibitem{Trager} Gianni, P.;, Sepp\"al\"a, M., Silhol, R., Trager, B., Riemann surfaces, plane algebraic curves and their period matrices, \textit{J. Symb. Comput.}, 1998, vol. 26, no.6, 789-803.  


\bibitem{Luo} Luo, W., Error estimates for discrete harmonic 1-forms over Riemann surfaces,
\textit{Communications in analysis and geometry}, 2006, vol. 14, no. 5, 1027-1035.


\bibitem{Nasser} Nasser, M., Fast Computation of Hydrodynamic Green's Function, \textit{Revista Cubana de F\'isica}, 2015, vol.  32, no. 1, pp. 26-32.

\bibitem{Nasser1} Nasser, M.,  Fast solution of boundary integral equations with the generalized Neumann kernel, 
\textit{Electronic Transactions on Numerical Analysis}, 2015, vol. 44, pp. 189-229, 2015. 


\bibitem{Yudovich2003} Yudovich,  V.,  Eleven Great Problems of Mathematical Hydrodynamics, \textit{Moscow Mathematical Journal}, 2003,  vol. 3, no. 2, pp. 711-737.


 \bibitem{Khesinopen}  Khesin,  B., Misiolek, G. ,  Shnirelman,  A.,  Geometric Hydrodynamics in Open Problems,
\textit{Arch. Rational Mech. Anal.}, 2023, vol. 247, 15.


\bibitem{Iushutin}  Yushutin, V.,   On stability of Euler flows on closed surfaces of positive genus,
\textsf{arXiv:1812.08959v2} [math.AP] (24 Dec 2019).


\bibitem{Davidson}  Davidson, P.A., \textit{Incompressible Fluid Dynamics}, Oxford University Press,  2022.


\bibitem{Vladimirov} Vladimirov, V., Ilin, K., On Arnold's variational principles in fluid mechanics,
In: E. Bierstone, B. Khesin, A. Khovanskii, J.E. Marsden, Eds., \textit{The Arnoldfest: Proceedings of a Conference in Honour of V. I. Arnold for his Sixtieth Birthday}, Fields Institute Communications, vol. 24, 1999,  pp. 471-496.

\bibitem{Kelvin} Kelvin, Lord, 1887, On the stability of steady and of periodic fluid motion. Maximum and minimum
energy in vortex motion,  \textit{Phil. Mag.}, 1887, vol. 23,  pp. 529-539.


\bibitem{Khesin} B. Khesin, Symplectic structures and dynamics on vortex membranes, \textit{Moscow Math. J.}, 2012,
 vol. 12, no. 2,  pp. 413-434.

 \bibitem{Izosimov1}   Izosimov,  A.,  Khesin,  B., Characterization of Steady Solutions to the 2D Euler Equation, \textit{International Mathematics Research Notices}, 2017,  vol. 2017,  no. 24,  pp. 7459-7503.

\bibitem{Izosimov2}   Izosimov,  A.,  Khesin,  B., Mousavi, M.,
Coadjoint orbits of symplectic diffeomorphisms of surfaces and ideal
hydrodynamics, \textit{ Ann. Inst. Fourier}, 2016, vol.
66, no.  6, pp.  2385-2433.

\bibitem{Izosimov3}   Izosimov,  A.,  Khesin,  B., Classification of Casimirs in 2d hydrodynamics,
\textit{Moscow Math. J.}, 2017,
vol. 17,  no.  4,  pp. 699-716

\bibitem{Helena} D. Iftimie, D;,   Lopes Filho, M, Nussenzveig Lopes, H., Weak vorticity formulation of the incompressible
2D Euler equations in bounded domains, \textit{Commun.   Partial Diff. Equations}, 2020,
vol. 45, pp. 109-145.

\bibitem{Deke}
Dekeyser, J.,  Van Schaftingen, J., Vortex Motion for the Lake Equations, \textit{Commun. Math. Phys.}, 2020, vol. 375, 1459-1501.

 \bibitem{grote1999dynamic} Grote, M. J., Majda, A. J., Grotta-Ragazzo,
C., Dynamic mean flow and small-scale interaction through
topographic stress, \textit{J.   Nonlinear Science}, 1999, vol.9, no. 1, 89-130.

\bibitem{Modin}    Modin, K. , Viviani, M.,  A Casimir preserving scheme for long-time simulation of spherical ideal hydrodynamics, \textit{J.  Fluid Mechanics}, 2020,
   vol. 884, A22. 

 \bibitem{Shnirelman}   Shnirelman,  A.,  On the long time behavior of fluid flows, \textit{Procedia IUTAM}, 2013,  vol. 7, pp. 151-160.

 \bibitem{Yudovich}   Yudovich, V. I., On the loss of smoothness of the solutions of the Euler equations and the inherent
instability of flows of an ideal fluid,
\textit{Chaos}, 2000,  vol. 10, 705-719. 

 \bibitem{Yudovich1}
 Morgulis, A.,  Shnirelman, A.,    Yudovich, V., Loss of smoothness and inherent instability of 2D inviscid fluid flows, \textit{Comm. Partial Differential Equations}, 2008, vol. 33, pp. 943-968.

 \bibitem{Kiselev}    Kiselev, A.,  Vladimir Sver\'ak, V., Small scale creation for solutions of the incompressible two-dimensional Euler equation,
 \textit{Annals of Mathematics}, 2014, vol. 180, no. 3,   pp. 1205-1220.

\bibitem{finlandia} Samavaki, M.,  Tuomela, J., Navier-Stokes equations on Riemannian manifolds, \textit{J. of Geometry and Physics}, 2020, vol. 148, 103543.

\bibitem{Avelin}  Avelin, H.,   Computations of Green's Function and its Fourier Coefficients on Fuchsian Groups, \textit{Experiment. Math.}, 2010, vol. 19, no. 3, pp. 317-334.

\bibitem{Jorgenson} Jorgenson, J. and Kramer J.,  Bounds on canonical Green's functions. \textit{Compositio Math.}, 2006, vol. 142, pp. 679-700.


 \bibitem{Strohmaier}    Strohmaier, A. and   Uski, V.,  An Algorithm for the Computation of Eigenvalues, Spectral Zeta Functions and Zeta-Determinants on Hyperbolic Surfaces,  
\textit{Commun. Math. Phys.}, 2013,  vol. 317,   827-869.  


\bibitem{Yau} Gu, X-D., Yau, S-T., \textit{Computational Conformal Geometry}, Interntional Press, 2008.

\bibitem{Dix}  Dix, O. M.   and   Zieve, R.J.,  Vortex simulations on a 3-sphere, \textit{Physical Review Research}, 2019, vol. 1, 033201.

\bibitem{Gluck1}   DeTurck, D.   and H. Gluck, H.,  Linking integrals in the $n$-sphere, \textit{Matem\'atica Contempor\^anea}, 2008, vol. 34, pp. 233-249.

\bibitem{Gluck2} DeTurck, D.   and H. Gluck, H., Electrodynamics and the Gauss linking integral on the 3-sphere and in hyperbolic 3-space, \textit{J. Math. Phys.}, 2008, vol. 49, 023504.

\bibitem{Gluck3} Parsley, R.J., The Biot-Savart operator and electrodynamics on bounded subdomains of the three-sphere,  \textit{Ph.D. thesis}, Mathematics Department,  University of Pennsylvania, 2004.

\bibitem{ABC}  Arnold, V. I.,  Sur la topologie des \'ecoulements stationnaires des fluides parfaits, \textit{C. R. Acad. Sci. Paris}, 1965, vol. 261, pp. 17-20.

\bibitem{Gromeka} Gromeka, I.,  Some cases of incompressible fluid motion, \textit{Scientific notes of the Kazan University},1881.


\bibitem{Dombre} Dombre, T.,    Frisch, U.,  Greene,   J. M.,   H\'enon, M.,   Mehr, A.,    Soward. A.,  Chaotic streamlines in the ABC flows, \textit{J.Fluid Mechanics}, 1986,  167, pp. 353-391.

 \bibitem{Zhao} Zhao, X.,   Kwek,K.,    Li, J.,  Huang. K., Chaotic and Resonant Streamlines in the ABC Flow, \textit{SIAM J.   Applied Math.}, 1993,  vol. 53, no. 1, pp. 71-77.

\bibitem{Galoway} David Galloway, D., ABC flows then and now,
\textit{Geophysical $\&$ Astrophysical Fluid Dynamics}, 2012,
vol. 106,  no. 4-5, pp. 450-467.

    \bibitem{Ghrist0} Etnyre, J. and   Ghrist, R.,  Contact topology and hydrodynamics I: Beltrami fields and
the Seifert conjecture, \textit{Nonlinearity}, 2000, vol. 13, pp. 441-458.

\bibitem{Ghrist00} Etnyre, J. and   Ghrist, R., Stratified integrals and unknots in inviscid flows, \textit{Contemp. Math.}, 1999, vol.
246, pp. 99-111.

\bibitem{Ghrist} J. Etnyre and R. Ghrist,
Contact topology and hydrodynamics III:
Knotted orbits, \textit{Trans. Amer. Math. Soc.}, 2000, vol. 352,, pp. 5781-5794.

\bibitem{Miranda} R. Cardona, Miranda, E., Peralta-Salas, D.,  Computability and Beltrami fields in Euclidean space,
\textit{Journal de Math\'ematiques Pures et Appliqu\'ees}, 2023, 
vol.  169, pp.  50-81.


\bibitem{PSalas} Robert Cardona, Eva Miranda, Daniel Peralta-Salas, Francisco Presas, Constructing Turing complete Euler flows in dimension 3, Proc. Natl. Acad. Sci 118 (19) e2026818118

\bibitem{Enciso2} Enciso, A.,   Peralta-Salas, D.,
Knots and links in steady solutions of the Euler equation
\textit{Annals of Math.}, 2012, vol. 175,  pp. 345-367.  See also \textit{Procedia IUTAM}, 2013,
vol. 7,
pp. 13-20.


\end{thebibliography}

\begin{thebibliography}{99}

\bibitem{Warner} F. Warner, \textit{Foundations of Differentiable Manifolds and Lie Groups}, Springer Graduate Texts in Mathematics vol. 94, 1983.

 \bibitem{Friedrichs-1955-b} Friedrichs, K. O.,   Differential forms on Riemannian manifolds. \textit{Comm. Pure
Appl. Math.}, 1955,  vol. 8, pp. 551-590.

\bibitem{Schwarz-1995} Schwarz, G., \textit{Hodge Decomposition - A Method for Solving Boundary Value Problems}, Springer Lect. Notes in Mathematics, vol. 1607, 1995. 

 \bibitem{Morrey} C. B. Morrey, C.B.,  A variational method in the theory of harmonic integrals II,
\textit{Amer. J. Math.}, 1956, vol. 78, pp. 137-170.

\bibitem{Poelke}  Poelke, K. and Polthier, K., Boundary-aware Hodge decompositions for piecewise constant vector fields, \textit{Computer-Aided Design}, 2016, vol.
 78,  pp. 126-136.

 \bibitem{Desbrun} Zhao, R., Debrun, M., Wei, G., Tong, Y., 3d Hodge decompositions of edge and face based vector fields,
\textit{ACMTransactions On Graphics}, 2019, vol. 38, pp. 1-13.

 \bibitem{Raza}  Faniry H. Razafindrazaka, Konstantin Poelke, Konrad Polthier, Leonid Goubergrits, A Consistent Discrete 3D Hodge-type
Decomposition: implementation and practical
evaluation, \textsf{arXiv:1911.12173} (16 Dec 2019).

 \bibitem{Saqr} Saqr,  K.M., Tupin, S., Rashad, S. et al. Physiologic blood flow is turbulent, \textit{Scientific Reports (Nature)}, 2020, vol. 10, 15492. https://doi.org/10.1038/s41598-020-72309-8

 \bibitem{Razaf} Razafindrazaka F.H,
Yevtushenko, P., Poelke, K., Polthier, K., Goubergrits
L., Hodge decomposition of wall shear stress
vector fields characterizing biological flows,  2019,  \textit{R.
Soc. Open Sci.}, vol. 6,  181970. \\
\url{http://dx.doi.org/10.1098/rsos.181970}

\end{thebibliography}

\begin{thebibliography}{99}


 \bibitem{Glotzl}  Gl\"otzl, E.,   Richters, O.,
Helmholtz decomposition and potential functions for n-dimensional analytic vector fields,
\textit{J. of Mathematical Analysis and Applications}, 2023,
Vol. 525, no. 2, 127138,

\bibitem{Gustafsson-1990b} Gustafsson, B.,   On quadrature domains and an inverse problem in potential theory, \textit{J. Analyse Math.}, 1990, vol. 55, pp. 172-216.
\end{thebibliography}

\begin{thebibliography}{99}
  
  
 \bibitem{Weyl}Weyl, H., \textit{Die Idee der Riemannsche Fl\"ache} , Teubner,1913; new ed. by R. Remmert,  Teubner, 1997.
  
\bibitem{Young} Young,  J., On the Cauchy Integral and Jump Decomposition, 
arXiv:2301.12287 [math.CV]


\bibitem{Solom}  Solomentsev, E., Cauchy integral. Encyclopedia of Mathematics. \url{http://encyclopediaofmath.org/index.php?title=Cauchy_integral&oldid=52051}

  
  \end{thebibliography}
\end{document}